\documentclass[
journal=jacsat, 
manuscript=article]{achemso}

\usepackage[version=3]{mhchem} 

\author{Jonantan Kallus}
\affiliation[MS]{Mathematical Sciences, University of Gothenburg and Chalmers University of Technology}
\email{kallus@chalmers.se}
\author{Jos\'e~S\'anchez}
\affiliation[AZ1]{Discovery Sciences, AstraZeneca Gothenburg}
\author{Alexandra Jauhiainen}
\affiliation[AZ2]
{Early Clinical Biometrics, AstraZeneca Gothenburg}
\author{Sven Nelander}
\affiliation[UU]
{Department of Immunology, Genetics and Pathology, Uppsala
      University}
\author{Rebecka J\"ornsten}
\affiliation[MS]
{Mathematical Sciences, University of Gothenburg and Chalmers University of Technology}
\email{jornsten@chalmers.se}

\title[ROPE]
{ROPE: high-dimensional network modeling \\ with robust control of edge FDR}

\begin{document}

\begin{abstract}
\noindent \textbf{Motivation:}
Network modeling has become increasingly popular for analyzing genomic data, to aid in
the interpretation and discovery of possible mechanistic components and therapeutic targets.
However, genomic-scale networks are high-dimensional models and are usually estimated from a relatively small
number of samples. Therefore, their usefulness is hampered by estimation instability. In addition,
the complexity of the models is controlled by one or more penalization (tuning) parameters where small changes to these can lead to vastly different networks, thus making interpretation of models difficult.
This necessitates the development of techniques to produce robust network models accompanied by estimation quality assessments. \\
\textbf{Results:}
We introduce Resampling of Penalized Estimates (ROPE): a novel statistical method for robust network modeling. The method utilizes resampling-based network estimation
and integrates results from several levels of penalization through a constrained, over-dispersed beta-binomial mixture model. ROPE provides robust False Discovery Rate (FDR) control
of network estimates and each edge is assigned a measure of validity, the q-value, corresponding to the FDR-level for which the edge would be included in the network model.
We apply ROPE to several simulated data sets as well as genomic data from The Cancer Genome Atlas.  We show that ROPE outperforms state-of-the-art methods in terms of FDR
control and robust performance across data sets. We illustrate
how to use ROPE to make a principled model selection for which genomic associations to study
further. ROPE is available as an R package on CRAN. \\
\textbf{Availability and implementation:}
The proposed method has been implemented in the R package rope available on
CRAN.
\end{abstract}

\section{Introduction}

%
%
Large-scale network modeling has the potential to increase our understanding of complex genomic data structures.
However, the interpretability of such high-dimensional models are limited by their estimation instability and
sensitivity to model tuning  parameters. Network modeling is often a preliminary step toward
identifying biomarkers for disease stratification or therapeutic targets \citep[e.g.][]{PeerHacohen2011}. It is therefore essential that
network modeling is accompanied by reliable measures of validity, e.g. false discovery rate of detected edges.
Here, we focus on the network modeling of gene expression data, but the methodology is generally applicable to other
genomic data sets \citep{KlingNelander2015}. Transcriptional network models aim to identify genes (transcripts) that are directly
connected. How connectivity is defined depends on the method utilized. For instance, in \emph{graphical lasso} \citep{FriedmanTibshirani2008}
a network model is obtained through a penalized Gaussian likelihood
estimate of the precision matrix (the inverse covariance matrix). Non-zero entries of this matrix identify
directly connected genes as those for which the estimated partial
correlation exceeds a penalization threshold.
Methods like WGCNA \citep{LangfelderHorvath2008} or ARACNE \citep{MargolinCalifano2006} similarly
identify connections as those for which a metric of gene-gene association
(correlation for WGCNA, mutual information for ARACNE) exceeds a certain
penalization threshold.
Thus, common to all these methods, the complexity of the estimated network is controlled by a
penalization parameter, $\lambda$, regulating the sparsity of the estimates.
For graphical lasso, much work has focused on estimating the proper penalization for
asymptotically consistent selection or optimal bias variance trade off
\citep{MeinshausenBuhlmann2010,LiuWasserman2010}.
Specifically, stability selection \citep{MeinshausenBuhlmann2010} performs model selection
based on many subsamples of the data and with different levels of penalization. The
method addresses selection of high-dimensional models in general and can readily
be applied for selection of network models. An upper bound for the expected
number of falsely selected variables (edges), family wise error rate (FWER), is derived.
In practice, the estimated bound depends on the range of used penalization
levels.
Alternatively, one can approach the problem of proper
penalization in terms of controlling false discovery rate (FDR) using subsampling or bootstrapping.
Bootstrap inference for network construction (BINCO) \citep{LiWang2013} models
the bootstrap selection frequency for spurious edges, to estimate FDR.

Other methods for selection includes StARS (stability approach to regularization selection) \citep{LiuWasserman2010}
which estimates the expected probability of edges to be
selected in one subsample and not in another, as a
function of the penalization level. This estimate, denoted the
instability of variable selection, cannot
 trivially be extended to control FDR.
Bolasso \citep{Bach2008} was the first method to combine bootstrapping and the
lasso for variable selection and retains variables consistently
selected for all bootstrap samples. Results focus on selection accuracy rather
than false discovery control.

Here, we introduce Resampling of Penalized Estimates (ROPE) to provide robust FDR control
for edge selection accompanied by a measure of validity for each edge: \emph{q-values} \citep{StoreyTibshirani2003}.
q-values are  assigned to each edge so that if all edges
with $q<\alpha$ were retained, an FDR of $\alpha$ would be achieved. Thus,
q-values have the same relation to FDR as p-values have to false positive rate.
This results in a highly interpretable representation where the inferred
network is visualized with edge widths corresponding to edge  q-value.
We show that ROPE outperforms state-of-the-art FDR-controlling methods through comprehensive simulation
studies and application to RNA-seq expression data from the Cancer Genome Atlas \citep{TCGAStuart2013}.
An easy-to-use R package is provided through CRAN.

\begin{figure*}[!tpb]
\centerline{\includegraphics[width=0.95\textwidth]{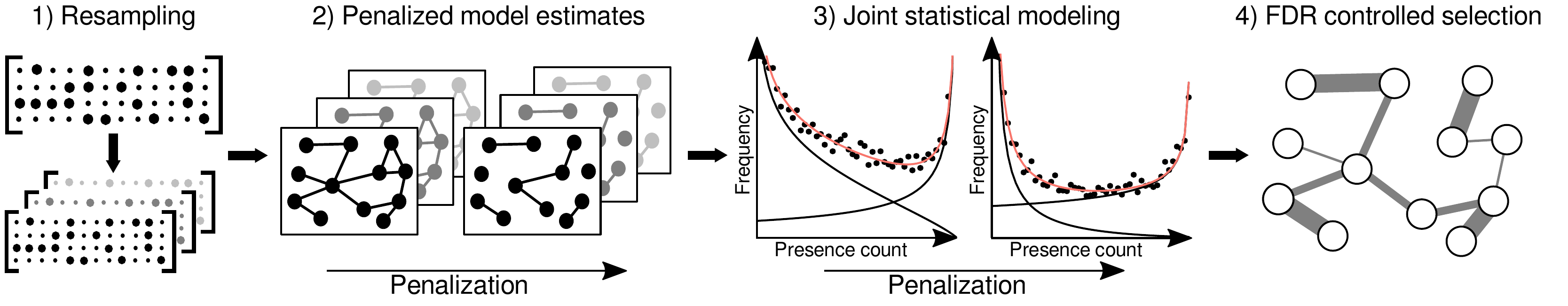}}
\caption{Summary of ROPE (resampling of penalized estimates) for network
modeling with control of the rate of falsely discovered edges (FDR).
1)~The input data is resampled.
2)~For each resample, network models are estimated with varying penalization.
3)~The number of resamples in which edges are present is modeled as a mixture
of ``spurious'' and ``relevant'' edges with the mixture proportion jointly estimated across
penalization levels. 4)~From the mixture model, each edge is assigned a q-value, the minimal FDR
target for which the edge is included.}\label{summary}
\end{figure*}



This article is structured as follows. This section has introduced the problem
at hand.
Section~\ref{methods} provides a detailed description of our method and a comparison
with the state-of-the-art.
Section~\ref{results} evaluates the method with
comprehensive simulation studies and includes method comparisons
on genomic data from glioblastoma tumors in TCGA.
Our method finds several hub genes known to have glioblastoma associated functions, and estimates the validity of each of their connections.
Section~\ref{discussion} concludes with the authors' thoughts on the significance
of this work and directions for future research.

\section{Methods}
\label{methods}
Variable selection is central to the understanding
of high-dimensional data. In network modeling of genomic data, variable
selection takes the form of selecting which gene-gene direct interactions (edges)
to include.
Traditional methods for model selection, e.g.\ cross validation, are
unsatisfactory for high-dimensional problems, due to their tendency to
overfit \citep{JornstenNelander2011}. Furthermore, measurement errors are
expected in genomics data and high-dimensionality makes erroneous
observations both influential and hard to filter. Therefore, single model
estimates  are not informative and resample based methods are
needed.

In this article we use neighborhood selection
\citep{MeinshausenBuhlmann2006} for network modeling. However, we emphasize that ROPE is applicable
to any network modeling where sparsity is controlled by a tuning parameter.
Neighborhood selection provides a good approximation of graphical lasso and is computationally faster. It models interactions of a gene $j$ to
other genes via the lasso.
\begin{equation*}
\beta^j=\arg\min_{\{\beta :
\beta_j=0\}}\frac{1}{n}||X_j-X\beta||_2^2+\lambda||\beta||_1
\end{equation*}
where $X$ is a matrix of $n$ rows (observations) times $d$ columns (genes).
The parameter $\lambda$ is the amount of sparsity inducing penalization.
The set
$\{(i,j):\beta^i_j\neq0\vee\beta^j_i\neq0\}$ is the edge set of the inferred
network.
Note that in network modeling of $d$ dimensional data, the network model
consists of $p=d(d-1)/2$ potential edges.

Due to estimation instability, single network estimates have limited interpretability. Therefore,
it is advisable to repeat network estimation on resampled data and utilize an estimation aggregate for
inference. Here, we use resampling of \emph{randomized lasso} estimates which
randomizes the amount of penalization for each individual parameter in different resamples in order to
break correlations between variables. Randomized lasso in combination with
resampling weakens the so-called irrepresentability
conditions that data need to adhere to for
consistent selection \citep{ZhaoYu2006}.
The amount of randomization in Randomized lasso is controlled by a
weakness parameter. Weakness 1 corresponds to
no randomization, while a lower weakness trades signal strength in data for a
lower risk of selecting irrelevant variables \citep{MeinshausenBuhlmann2010}.

Introducing some notation, let $R_i$ be a
realization of any uniform resampling procedure, most commonly subsampling with
sample size $m<n$ or bootstrap, so that $R_i(X)$ is the resampled data set. Let
$\hat{S}^\lambda$ be any penalized method for variable selection ($\hat{S}^\lambda(X)$ is
the set of variables selected by $\hat{S}^\lambda$ given $X$).
Let $\hat{S}_i^\lambda$ be randomization $i$ of
penalization in $\hat{S}^\lambda$. The main algorithmic input of ROPE, stability
selection and BINCO is variable selection counts
\begin{equation}
\label{W}
W^\lambda_{j}=\sum_{i=1}^B 1[j\in \hat{S}_i^\lambda(R_i(X))]\in\{0,\ldots,B\}
\end{equation}
for variable (edge) $j$ over $B$ resamples.

We now present a detailed review of the state-of-the-art FDR-controlling methods BINCO and Stability Selection.
BINCO, proposed in \cite{LiWang2013}, selects edges with frequency counts $W^\lambda_j$ exceeding a threshold $t$.
Parameters $\lambda$ and $t$
are chosen to maximize power while controlling FDR. For each $\lambda$,
$W^\lambda$ corresponds to a histogram
$h^\lambda(w)=\sum_j 1(W^\lambda_j=w)$
(Figure~\ref{summary}.3).
Ideally, this histogram should have two clear modes: at count 0 for spurious (null) edges and count $B$ for the relevant (non-null) edges.
For reasonable levels of regularization, $h^\lambda(w)$ is thus "U-shaped".
In BINCO, the null model is estimated by fitting a powered
beta-binomial distribution
to $h^\lambda$ in the range where $h^\lambda$ is decreasing in $w$ (defined in Equation~\ref{pbb}, Section~\ref{rope-method}). By
extrapolation of this null into the range of large frequency counts (dominated by non-null edges), $t$ can be chosen for each $\lambda$ to control
FDR.  In practice, the authors found this results in an overly liberal selection and therefore also propose
a conservative modification.
In conservative BINCO, the density function of the powered
beta-binomial distribution is modified to be constant, instead of decreasing, to
the right of the estimated minimum of the $h^\lambda$-model. This
results in a larger $t$ for a given target FDR, thus selecting fewer
edges.


Stability selection  \citep{MeinshausenBuhlmann2010} selects variables with
$\max_{\lambda\in\Lambda}W^\lambda_j>t$ for some threshold $t$. That is, as long as an edge $j$ has a frequency count exceeding
threshold $t$ for any penalization $\lambda \in \Lambda$, it is included in the model.
An upper bound
on the expected number of falsely selected variables, $F$, when $t>B/2$ is derived
for $\hat{S}^\lambda_i$ randomized lasso and $R_i$ subsampling with sample size
$\lfloor n/2\rfloor$:
\begin{equation*}
E(F)\leq\frac{q^2_\Lambda}{(2\frac{t}{B}-1)p},
\end{equation*}
where $p$ is the number of variables and the expected number of selected
variables $q_\Lambda$ is estimated by
$|\Lambda|^{-1}\sum_{\lambda\in\Lambda}\sum_jW^\lambda_j$. In \cite{LiWang2013}, an FDR bound
is derived from this by dividing both sides
by the number of selected variables
$\sum_j 1(\max_{\lambda\in\Lambda}W^\lambda_j\geq t)$.
This estimate depends, not only on the threshold $t$, but also on the investigated range of penalization. In
\cite{LiWang2013}, the combination of $t$ and $\Lambda$
that selects the maximum number of edges while controlling FDR at the
desired level is used.


It is a necessary condition for the applicability of both BINCO (and our method, ROPE) that
the histogram $h^\lambda$ is approximately U-shaped for some $\lambda$.
\cite{LiWang2013} connect this condition to the irrepresentable condition,
showing that satisfaction of the latter leads to U-shaped histograms. In
practice, however, the BINCO procedure is sensitive to the histogram shape.
First, it is sensitive to correctly estimating the end points
of the decreasing range of $h^\lambda$, from which the null distribution is estimated.
Second, the estimated null distribution is extrapolated into the
increasing range of $h^\lambda$, where any relevant FDR controlling threshold will be.
This extrapolation leads to an unnecessarily large variance for the selected threshold.
Third, non-uniform presence of the alternative
population (relevant edges) in the decreasing range of the histogram will cause a bias in the estimate of the null
distribution.
Forth, the authors warn that the method makes a too liberal selection when the
minimum of the histogram is to the right of $0.8B$, which easily happens in problems
that are sufficiently sparse.
Stability selection, while not having the issue of sensitivity to histogram
shape, has the limitation that it focuses on a worst-case guarantee, rather than
an estimate of the number of false positives.

\subsection{ROPE: joint model for resampled, penalized estimates}
\label{rope-method}

Recognizing the above limitations of state-of-the-art procedures, we here
introduce ROPE, a novel joint modeling of edge presence counts across multiple
penalization levels.
Figure~\ref{summary} summarizes the method. Specifically,
%
\begin{enumerate}
\item{{\bf Resampling of input data. }}
$B$ resamples are
created by resampling $n$ observations with replacement.
\item{{\bf Generation of edge presence counts. }}
Edge presence counts are collected for several levels of
penalization, $\lambda_j \in \Lambda$ (Equation~\ref{W}). Here, we illustrate ROPE for
neighborhood selection in combination with randomized lasso but, as mentioned above, other sparse network models
can be used.
\item{{\bf Modeling of edge presence counts for each $\lambda$, and joint modeling across multiple $\lambda$s. }}
We model $W^\lambda_i$, for each $\lambda$, as coming from a mixture of
overdispersed beta-binomial distributions (Equation~\ref{distribution}).
For improved robustness and accuracy, we leverage the fact that the mixture proportion of null to non-null edges is constant across $\lambda$.
\item{{\bf q-value assessment and selection of final model. }}
Integrating information from $\lambda$s where the modeled null and alternative
populations are most separated (Equation~\ref{g}), q-values are estimated for each edge. FDR is
estimated by the probability mass of the null component to the right of
threshold divided by mass of the total empirical density to the right of
threshold (Equation~\ref{fdr}).
\end{enumerate}

Specifically, edge presence counts are modeled as coming from a mixture of overdispersed
beta-binomial distributions.
Edge selection probabilities depend not only on them being null
or alternative but also on, at least, the strength of the dependence between the
nodes they connect. This warrants the use of a beta-binomial distribution for each mixture component,
where parameters $\mu$ represent mean edge selection probability within each component (null/alternative),
and $\sigma$ the variation of dependence strengths within components:
\begin{equation*}
f_\text{BB}(w)=\binom{B}{w}
\frac{\beta(w+\frac{\mu}{\sigma},B-w+\frac{1-\mu}{\sigma})}
{\beta(\frac{\mu}{\sigma},\frac{1-\mu}{\sigma})},
\end{equation*}
where $\beta$ is the beta function.

For large and sparse graphs, each edge frequency count can be assumed to be independent of
most other edges.
(Locally, however, edge frequency counts can of course be highly correlated.)
Still, the edge count histograms indicate the presence of overdispersion,
likely
caused by unobserved covariates,
hidden correlations (not accounted for in the theoretical null distribution)
and the existence of many real but uninterestingly small effects
\citep{Efron2004}. We account for overdispersion with
inflation components and modifications of the beta-binomial components.
 Inflation
is added for both low and maximum selection counts.  Since graphs are assumed to
be sparse, most edges will have low selection counts. These edges are easily
classified as belonging to the null so a good model fit is not important in that range.  Therefore, the
beta-binomial distribution that captures null edges is inflated in the range
$\{0,\ldots,c^\lambda\}$ where $c^\lambda$ is chosen so that 75\% of edges has selection
count $c^\lambda$ or less. The method is not sensitive to the exact proportion of
edges captured by this inflation.
The distribution for alternative edges is only
inflated at the maximum count $B$. Further overdispersion is added by raising
the beta-binomial density function corresponding to the null population by an
exponent $\gamma$ and renormalizing, in the same vein as BINCO, yielding the
density function
\begin{equation}
\label{pbb}
f_\text{null}(w)=\frac{f_\text{BB}(w)^{\gamma}}{\sum_{k=0}^Bf_\text{BB}(k)^{\gamma}}.
\end{equation}
The
beta-binomial density function corresponding to the alternative population is
modified to have zero mass in $\{0,\ldots,c^\lambda\}$ but still be continuous
\begin{equation*}
f_\text{alt}(w)=\frac{(f_\text{BB}(w)-f_\text{BB}(c^\lambda))_+}
{\sum_{k=0}^B(f_\text{BB}(k)-f_\text{BB}(c^\lambda))_+}.
\end{equation*}
The modification fits better with observed distributions from simulations and
leads to a more conservative edge selection. Thus, $W^\lambda_i$ is modeled as coming from a distribution defined by the
density function
\begin{equation}
\label{distribution}
f(w)=(1-\pi)f_1(w)+\pi f_2(w),
\end{equation}
\begin{equation*}
f_1(w)=\tau_1\frac{1(w\in\{0,\ldots,c^\lambda\})}{c^\lambda+1}+
(1-\tau_1)f_\text{null}(w),
\end{equation*}
\begin{equation*}
f_2(w)=\tau_2 1(w=B)+(1-\tau_2)f_\text{alt}(w).
\end{equation*}
We impose two constraints in order to make parameters identifiable. First,
the null component, $f_1(w)$, is constrained to be decreasing in its right-most part (corresponding
to $\mu_1+\sigma_1<1$). Secondly, the non-null, $f_2$, is constrained to be convex and
increasing (corresponding to $\mu_2=\sigma_2>0.5$).
Data in $\{c^\lambda+1,\ldots,B-1\}$ is described by five parameters
$\theta=(\pi',\mu_1,\sigma_1,\gamma,\mu_2=\sigma_2)$, where $\pi'$ captures the
component sizes within the range.
These are estimated with numerical
maximization of the log-likelihood function
\begin{equation*}
l(\theta)=\sum_{w=c^\lambda+1}^{B-1}h^\lambda(w)\log\left((1-\pi')f_\text{null}(w;\theta)+\pi'
f_\text{alt}(w;\theta)\right),
\end{equation*}
under the two constraints just  mentioned, as well as the constraints implied by
density parametrizations.
Remaining parameters $\pi,\tau_1,\tau_2$ are then given by
the estimated parameters and the data $h^\lambda$.

We have described the method for a given level of penalization $\lambda$.
The choice of range of
penalization $\Lambda$ to fit the model for, and the unification of fits for
different penalizations, remain.
We propose to use selection counts from different levels of penalization
$\lambda$ simultaneously, in order to decrease variance in estimates of model
parameters. The unknown true $\pi$,
the proportion of alternative edges, is of course constant in $\lambda$.
Nevertheless, we can expect $\hat{\pi}$ to have an upward bias for small
$\lambda$: with too little penalization null
edges will be falsely captured by the alternative mixture component.
Conversely, a large penalization will push the
distribution of selection counts for alternative edges leftwards into the
distribution for null edges. We assume the alternative distribution to have its
mode at $B$. Thus the upper end of $\Lambda$ is the maximal penalization for
which $h^\lambda$ is significantly increasing in the proximity of $B$, i.e.
$h^\lambda$ is approximately U-shaped. We have
included a heuristic algorithm to help identify this point in the software
package.
We are interested in which $\lambda$ that best separates the
null and alternative mixture components
and for which we can thus weigh together the evidence of edge presence together across $\lambda$
for better accuracy and FDR control.
We define the \emph{separation} of
mixture components, for a $\lambda$, as the difference of
the amount of correctly and incorrectly selected edges based on the model fit:
\begin{equation}
\label{g}
g(\lambda)=p\sum_{w=0}^B(\pi f_2(w)-(1-\pi)f_1(w))_+.
\end{equation}

Let
$\lambda_a$ be the upper end of an approximate 0.95 bootstrap confidence interval for the
location of the maximum of $g(\lambda)$. Let $\pi^*=\hat{\pi}(\lambda_a)$, i.e.
a conservative estimate of the proportion of alternative edges. Next, we update the
model fit for each $\lambda$ with the additional constraint $\pi\leq\pi^*$, in
order to incorporate the joint estimate of the proportion of alternative edges.
Lastly, let $\lambda_b$ be the lower end of an approximate 0.95 confidence
interval for the location of the maximum of $g(\lambda)$ for the new model fits.
Using a low estimate of $\lambda$ yields a conservative edge selection since
constraint on $\pi$ is in stronger effect there. The model fitted to selection
counts for penalization $\lambda_b$, constrained to $\pi\leq\pi^*$ is used for
final edge classification. A simulation presented in the next section
illustrates how the simultaneous use of counts from different levels of
penalization results in lower bias and lower variance
(Figure~\ref{only-firststep}).

The classification threshold $t^\lambda$ for the given
FDR target is found from the fitted model. For $t^\lambda\in\{0,\ldots,B\}$ the
estimated FDR is given by
\begin{equation}
\label{fdr}
\widehat{\text{FDR}}(t^\lambda)=\frac{p\sum_{w=t^\lambda}^B
(1-\pi)f_1(w)}{\sum_{w=t^\lambda}^Bh^\lambda(w)}.
\end{equation}
where $p$ is the number of potential edges.
The final step of ROPE assigns a q-value to each edge. Given fitted parameters
at the selected penalization, the q-value $q_i$ of an edge $i$ is
$\widehat{\text{FDR}}(W^\lambda_i)$.
We use the upper limit of a confidence interval for $q_i$
in order to ensure conservative estimates. Under our model, the number of type I
errors approximately follows a binomial distribution with $\sum_{w=t^\lambda}^Bh^\lambda(w)$
experiments and $\widehat{\text{FDR}}(W^\lambda_i)$ success probability.
Using the normal approximation of the binomial distribution, the
upper 0.95 confidence bound for $q_i$ is given by
\begin{equation*}
\widehat{\text{FDR}}(W^\lambda_i)+z_{0.975}
\sqrt{\frac{
\widehat{\text{FDR}}(W^\lambda_i)(1-\widehat{\text{FDR}}(W^\lambda_i))}
{\sum_{w=t^\lambda}^Bh^\lambda(w)}}.
\end{equation*}



To conclude this section, we emphasize the methodological differences between
ROPE and BINCO. First, ROPE uses a mixture model that captures both null and
alternative edges, while BINCO models only the null distribution. In practice,
the threshold corresponding to any relevant FDR target will be in a part of the
domain where the population of alternative edges dominates. This leads to the
estimation of BINCO to be based on an extrapolation, resulting, as the next
section will show, in a lower stability of estimates. Furthermore, to estimate a
model that only captures the null population, BINCO is forced to select a
subset of data where the null population is most prevalent. This intermediate
range selection contributes to the lower stability of estimates. In contrast, by
modeling both null and alternative edge selection counts, ROPE can use the most
relevant subset of data to fit its model parameters. Thus, extrapolation is
avoided and the parameter estimates are insensitive to the exact end points of
the subset range. Second,  while ROPE simultaneously
uses counts from different levels of penalization where the overlap of null and
alternative populations is small, BINCO selects the level of regularization that selects
the most edges while estimating an FDR below target. This results in lower stability of BINCO's
estimates, since the selection may change due to small perturbations
of the data, and in a bias of BINCO to underestimate FDR, since models with
underestimated FDR tends to select more edges at a fixed FDR target. Third,
overdispersion is a main modeling difficulty addressed by BINCO and ROPE. Our
richer model, with greater ability to capture overdispersion, results in ROPE
having a more accurate FDR control than BINCO.

\section{Results}
\label{results}
\begin{figure}[!tpb]
\centerline{\includegraphics[width=\linewidth]{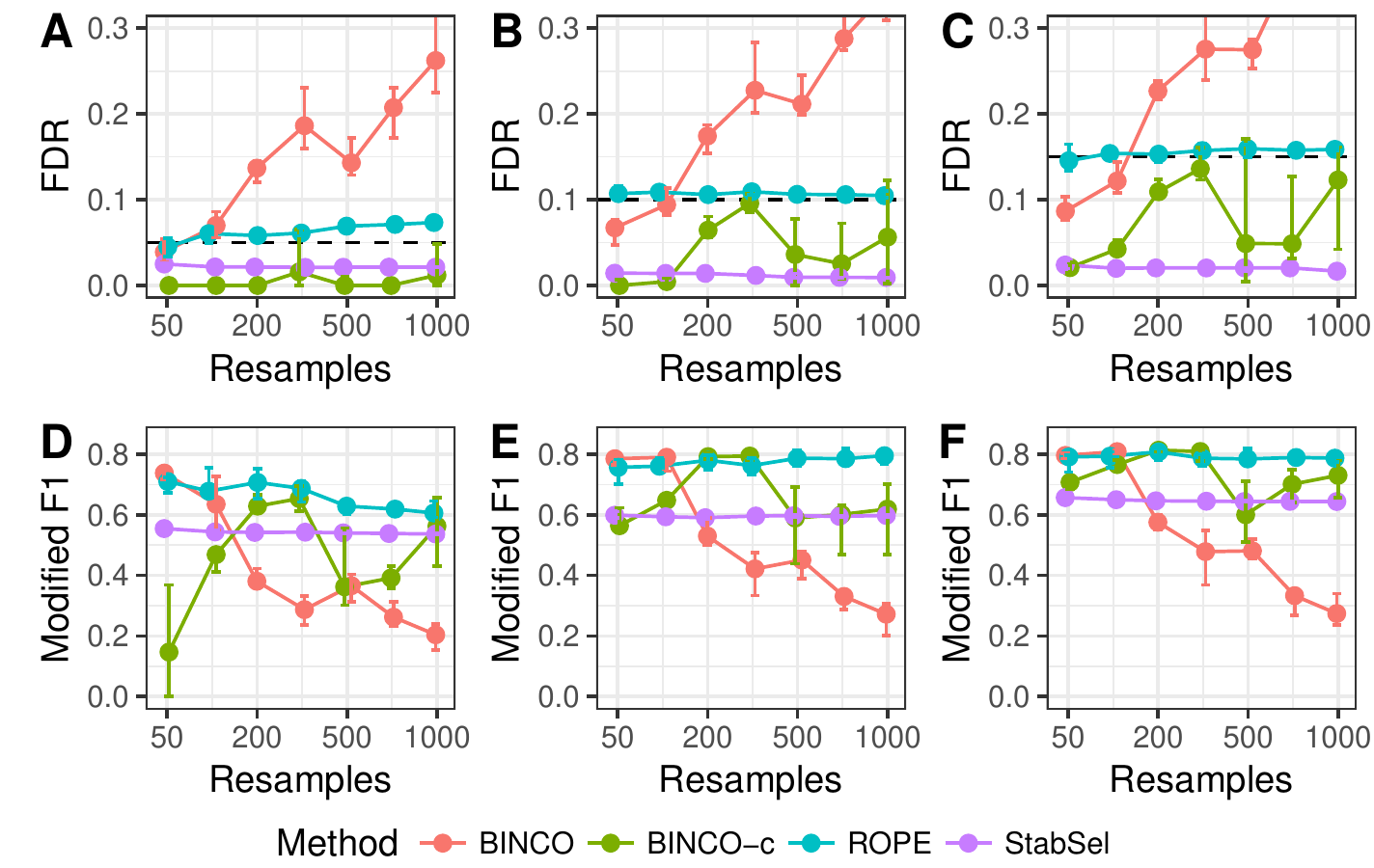}}
\caption{Validation of the proposed method on simulated data. Four methods are compared: BINCO, conservative BINCO
(BINCO-c), ROPE and stability selection (StabSel). Each method has been applied for three FDR
targets. Columns A-D, B-E and C-F
show results for target FDR 0.05, 0.1 and 0.15, respectively. Panels A, B and C compare FDR with target FDR. ROPE achieves an FDR
closest to the target. BINCO tends to make an increasingly liberal
selection as the number of resamples increases. Stability selection is consistently too conservative.
Panels D, E and F show the corresponding
modified F1 score. ROPE scores highest overall.
Points show median result (20 simulations) and whiskers represent
1.5 times IQR.}\label{simres}
\end{figure}
\begin{figure}[!tpb]
\centerline{\includegraphics[width=\linewidth]{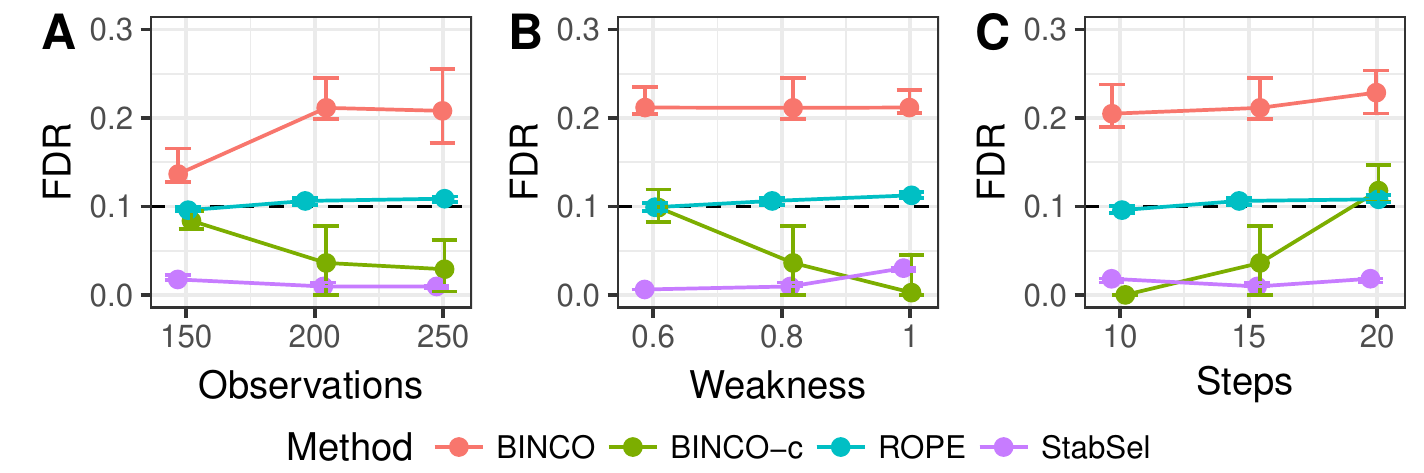}}
\caption{Examination of parameter sensitivity for the same simulated data as in
Figure~\ref{simres}. The number of observations,
weakness and number of steps in the penalization set $\Lambda$ is varied,
in panels A, B and C, respectively. The figure shows that ROPE performs well and
gives consistent results in this parameter subspace, while stability selection is consistently
conservative and BINCO and conservative BINCO give less consistent results. In
general, BINCO is too liberal and conservative BINCO is too conservative.
This figure shows results for a target FDR of 0.1.
Results for other target FDR and settings can be found in the
supplement and are in agreement with our findings here.}\label{simres2}
\end{figure}

We present a comprehensive simulation study to assess the performance of ROPE and
compare it with two state-of-the-art methods: BINCO and stability selection. We also
present an application of the methods to gene expression data from glioblastoma cancer
patients, and compare results. An application of ROPE to variable selection for a non-graphical
model is provided in the supplement.

\subsection{Comparison of accuracy and robustness of FDR control on simulated
data}
Our simulation experiment consists of data from 500-node networks of three topologies:
scale-free, hubby and chain graphs.
We sample standard normal data from
covariance matrices corresponding to the network topologies. The signal strength
is either strong (mean and standard deviation of covariances between connected
nodes is 0.32 respectively 0.13) or weak (mean and standard deviation is 0.25
respectively 0.09). The scale-free networks have 495, 49 (sparse) or 990 (dense)
edges. The hubby network has 20 hub nodes, each connected to between 92 and 4
other nodes. The chain network connects its 500 nodes into one chain
of length 500. In all, this constitutes seven simulated model selection
problems: three topologies, five variations of the scale-free topology. Two of
these are identical to those in \citet{LiWang2013}.

We generate edge presence count matrices $W_j^\lambda$ for each problem by
taking $B$ bootstrap samples, and select edges for each sample using randomized
neighborhood selection with penalization ranging from 0.02 to 0.3. The settings
for $W_j^\lambda$, i.e. $B$, number of steps in $\Lambda$, weakness and $n$, are varied in order
to assess the methods' sensitivity. We compare the methods' selections for three
target FDR levels: 0.05, 0.1 and 0.15. Each combination of settings is rerun
20 times in order to assess sensitivity to randomness in subsampling. We
compare target FDR with achieved FDR and score each selection with a modified F1
score
\begin{equation*}
\text{F1}_\text{m}=2\frac{(1-\text{FDR})\text{TPR}}{m(\text{FDR})+\text{TPR}},~m(\text{FDR})=\begin{cases}
    1-\text{FDR}, \text{if } \text{FDR}\leq\text{FDR}^*\\
    \frac{\text{FDR}}{\text{FDR}^*}-\text{FDR}^*, \text{otherwise}
\end{cases}
\end{equation*}
where FDR$^*$ is the target FDR. The denominator is modified to ensure that
scores are decreasing with FDR when FDR is above target.

Results for the
scale-free
network with 500 nodes, 495 edges and strong signal is presented in
Figures~\ref{simres} and \ref{simres2}. In Figure~\ref{simres}, $B$ is varied,
while $n=200$, weakness is 0.8 and $\Lambda$ consists of 15 steps.
In Figure~\ref{simres2}, $B=500$, while $n$, weakness and number of steps in $\Lambda$ is
varied. Results for remaining
topologies and parameter combinations are presented in the supplement.
\begin{figure}[!tpb]
\centerline{\includegraphics[width=0.71\linewidth]{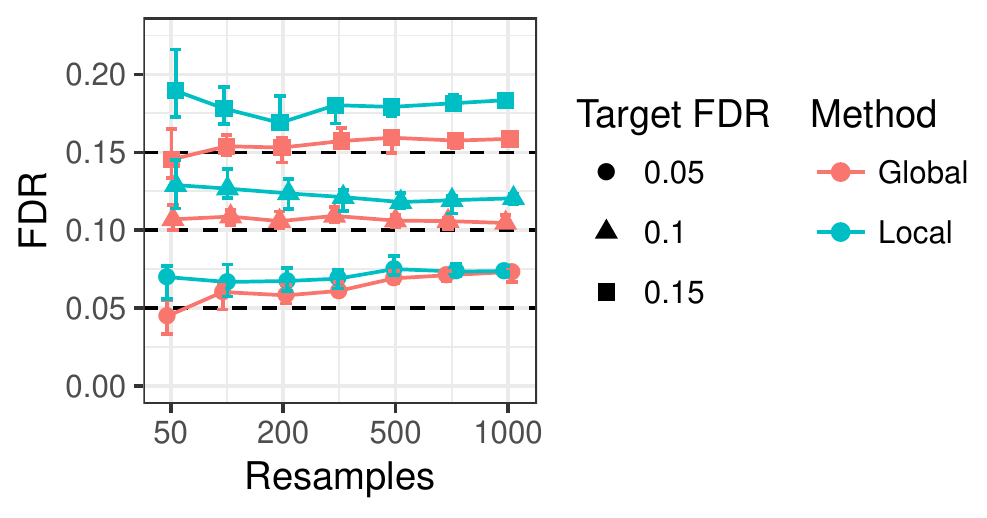}}
\caption{Comparison of ROPE with and without joint modeling of counts from
  different penalization levels. When counts from different levels are used
  (Global) the estimated FDR is closer to the target FDR and the variance
  between simulations is lower.
}\label{only-firststep}
\end{figure}

Results show that ROPE performs best in terms of modified F1, FDR and stability
for the scale-free, dense scale-free, small scale-free and weak signal
  scale-free
networks. In the chain network and the sparse scale-free network ROPE and
stability selection perform similarly. Stability selection makes the most stable
selections, but is generally too conservative, which is to be expected
since the method is based on a bound. For the weak signal scale-free network and
with a target FDR of 0.05, stability selection is too conservative to select any
edge at all. BINCO and conservative BINCO both make far less stable selections
than ROPE and stability selection.
Furthermore, both
BINCO methods are sensitive to the number of bootstraps.
Logically, selection should improve when the number of bootstraps is
increased. Instead, BINCO makes an increasingly more liberal selection.
Similarly disconcerting, BINCO performance worsens
with increased signal strength (number of observations) (Fig. 3A).
Without access to the true model, it would be
difficult to know how many bootstraps that should be performed to get a correct
FDR control. This strong dependency between number of bootstraps, signal strength and achieved
FDR makes BINCO hard to use in practice.
The hubby
network is the one setting where stability selection performs better than ROPE.
There, ROPE makes no selection since the selection count histograms are not
U-shaped. In order to examine how ROPE would perform for the hubby network if
the signal were stronger, we generated additional observations, increasing the
examined range from 150-500 observations to 150-1250 observations. For more than
500 observations, ROPE again yielded the highest modified F1 and the FDR closest
to target.

ROPE uses selection counts from several penalization levels and, as can be seen in Figure~\ref{only-firststep}, this avoids a too
liberal selection and increases stability.  In addition, the Figure indicates that ROPE outperforms
BINCO even
without the joint modeling, which emphasizes the need to model both the null and non-null edge populations as done in ROPE.

In terms of computation time, BINCO and ROPE are slower than stability
selection. At each level of penalization, ROPE fits a five parameter model,
while BINCO estimates the end points of an approximately decreasing range and
then fits three parameters.
Both take only a few seconds per penalization level on a standard desktop computer.
Increasing size of networks or the number of observations does not increase
computation time, since these methods use summary statistics --- the number of
variables having a selection count $w$, for each $w\in\{0,\ldots,B\}$. The computation time of stability
selection, BINCO and ROPE is small compared to the time needed for resampled variable
selection.

\subsection{FDR controlled edge selection for a graphical model of gene
expressions in the PI3K/Akt pathway of glioblastoma cancer patients}
\begin{figure}[!tpb]
\centerline{\includegraphics[width=\linewidth]{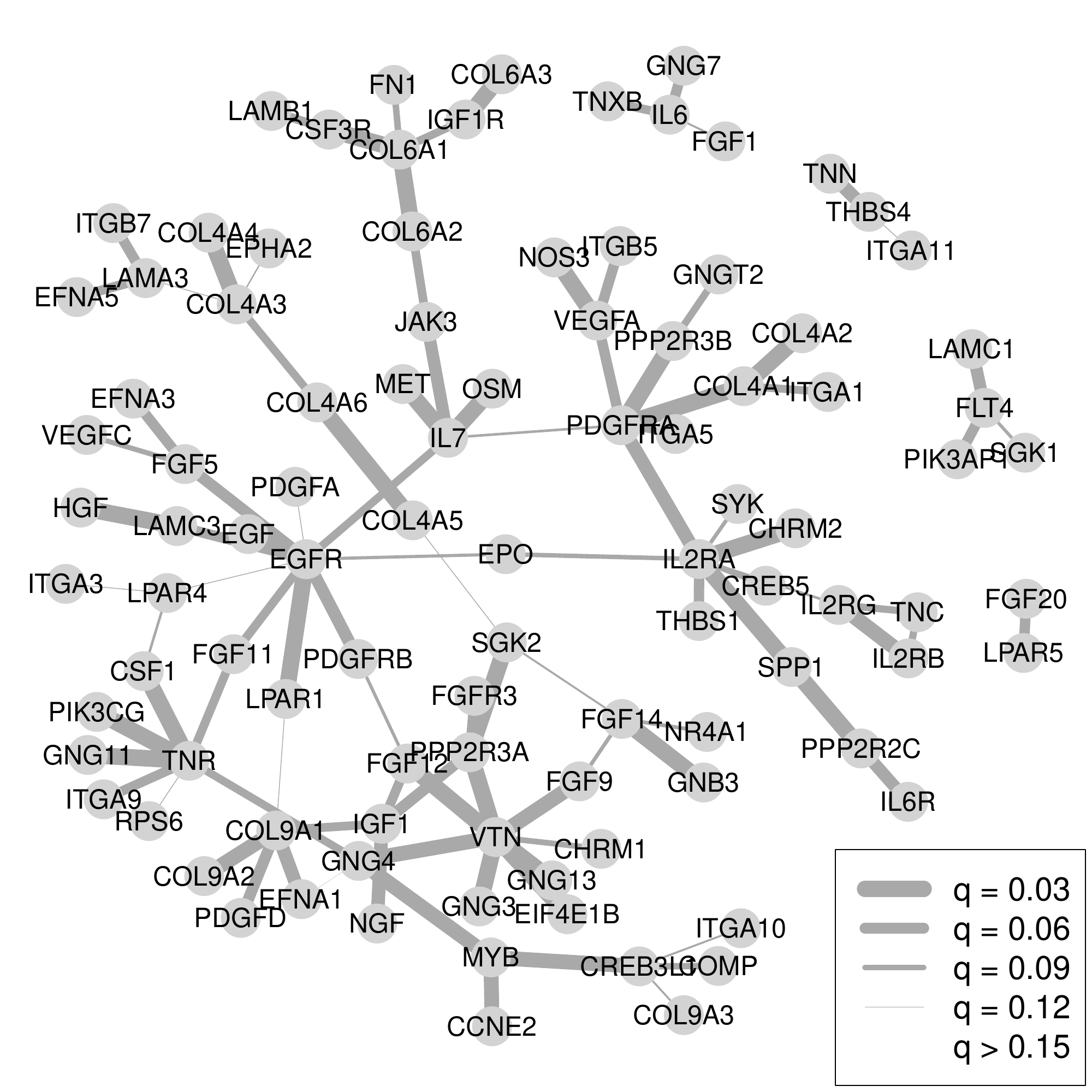}}
\caption{ROPE selection of gene connections in the PI3K/Akt pathway based on gene
expressions from glioblastoma cancer patients in TCGA. Widths of edges
correspond to q-values. Highly connected nodes are known to have functions associated with invasiveness and/or tumor growth in glioblastoma.}\label{network}
\end{figure}
\begin{figure}[!tpb]
\centerline{\includegraphics[width=\linewidth]{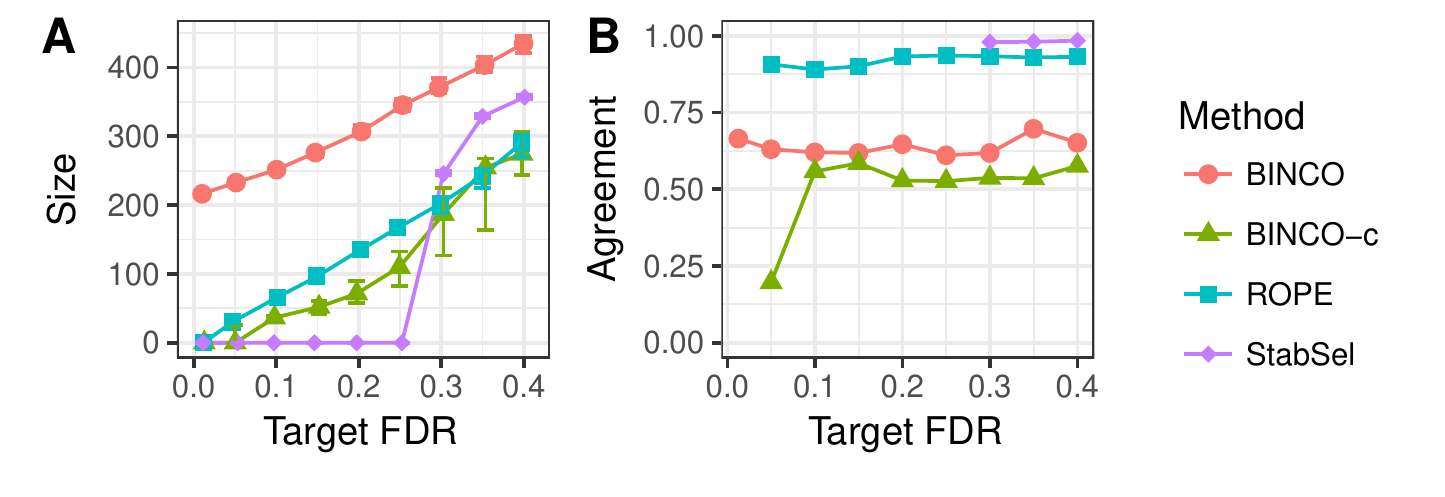}}
\caption{Method comparison on gene expressions in the PI3K/Akt pathway of
glioblastoma cancer patients. Panel A shows the number of selected edges by each
method for a range of target FDR. While the achieved FDR is unknown, we note
that BINCO is liberal enough to select more than 200 edges even at a low target
FDR of 0.0125. As expected, stability selection is conservative producing empty networks
for target FDR 0.25 and below.
Conservative BINCO exhibits substantial variability in network size.
Panel B shows agreement within each
method across 20 subsamples of $W$ as measured by Fleiss' $\kappa$. BINCO and
conservative BINCO are less stable than ROPE and stability selection. The lack of agreement for BINCO at low targets combined with a large selection
size, makes it unlikely that FDR is controlled. Fleiss' $\kappa$ is not defined
for empty selections produced by stability selection below FDR 0.25.}\label{pathway}
\end{figure}
In this section, we apply ROPE to gene expression data and study the selected
network. We also compare ROPE, BINCO and stability selection in terms of size
of FDR controlled selections and stability.
We downloaded RNA-Seq gene
expressions for 172 glioblastoma multiforme cancer patients from the USCS Cancer
Genomics Browser \citep{GoldmanZhu2014}. The data comes from TCGA and had been
normalized across all TCGA cohorts and log transformed. It contains measurements
for {20,530} genes. We downloaded a list of genes in the PI3K/Akt signaling
pathway from KEGG \citep{KanehisaGoto2000}. 337 genes in the gene expression
data set were found in the PI3K/Akt gene list. We discarded half of the genes
with lowest median absolute deviation (MAD) of expression. Remaining genes were
scaled to have MAD 1.
We bootstrapped the data 500 times and estimated graphical models
with 12 different levels of penalization for each bootstrap sample. The weakness
in randomized lasso was set to 0.8. Figure~\ref{network} shows a visualization
of the final network estimated with ROPE. In the visualization we have kept all
edges with an estimated q-value below 0.15, i.e. we expect that 15\% of the
depicted edges are false discoveries. The edge widths correspond to estimated
edge q-value. Zero degree nodes are not shown.
Highly connected network nodes were the epidermal growth factor receptor (EGFR,
8 links), the platelet-derived growth factor receptor alpha (PDGRA, 6 links),
components of the IL2 receptor (IL2RA and IL2RG, with 7 and 3 links),
vitronectin (VTN, 7 links) and tenascin R (6 links). Of these, EGFR and PDGFRA
are well established glioblastoma oncogenes. TNR is a tenascin with neural
restricted expression, and is likely a negative marker of glioma invasiveness
\citep{BrosickeFaissner2015}. By contrast VTN, which is connected to several FGF
and FGFR isoforms in our network, is a pro-migratory/invasion factor
\citep{OhnishiHayakawa1998}. IL2, finally, has been suggested to promote growth
of glioma cells \citep{CapelliNano1999}. Our network may thus serve to
prioritize hub genes for further study, as well as their functionally associated
genes. Edge q-values, along with properties of methodology for subsequent analysis, may facilitate the choice of how many associations to study further.

While the correct network model of the pathway is, of course, unknown, a
comparison of methods on this real data shows relevant differences. We
subsampled the 500 selected models 20 times without replacement. Each subsample
consists of 400 selected models. Counting edge selections within each subsample
gives 20 subsampled $W$. Figure~\ref{pathway} shows a comparison of size of
FDR controlled selections and of stability of selections between subsamples.
BINCO selects more than 200 edges already at a target FDR of 0.0125. Stability
selection selects the empty model for target FDR 0.25 and below, in agreement with
the conservative behaviour observed in the simulations. BINCO and
conservative BINCO show more variation between selected models for different
subsamples, than ROPE and stability selection. The
liberal selection by BINCO agrees with simulation results, suggesting a failure to control FDR.
BINCO's lack of agreement between
selections at low target FDR also suggests a failure to control FDR. The higher
variability in BINCO and conservative BINCO also agrees with simulation results.
We have used Fleiss' $\kappa$, an index of inter-rater agreement among many
raters \citep{Fleiss1971}, to measure agreement between selections across
subsamples.

\section{Discussion}
\label{discussion}
The problem of FDR control in high-dimensional variable selection
problems is of great relevance for interpreting data from molecular biology and
other fields with an abundance of complex high-dimensional data. Many
methods for variable selection in high-dimensional problems exist, but they
suffer from the need to tune intermediate parameters of little scientific
relevance. We have introduced a method for false discovery control in
network models, and presented results showing that this method outperforms
existing alternatives.
With the method and software package
presented here, which achieve accurate and robust FDR control,
we have made possible a principled selection of relevant interactions.




We did consider an alternative statistical model for selection counts where the populations
of alternative and null edges were further stratified into sub populations,
based on their strength or the structure of their neighborhood in the graph.
We did not find such a richer model to be worth the additional cost and estimation
variability. Moreover, such a
model poses the additional challenge of classifying each sub population
as belonging to either the null or the alternative population. We also
considered strengthening the connection between statistical models across all levels of
penalization. Power and stability could potentially be increased by enforcing
smoothness of all model parameters across levels of penalization. But the large
number of edges that are represented in each histogram suggests that
improvements would be small. Furthermore, the numerical fitting of such a global
model is challenging.


ROPE, BINCO and stability selection use only summary statistics, proportions of
variables with each selection count. Thus, their computational complexity is not affected by
an increase in the number of network nodes. Computational time is completely dominated by
the preceding step of resample based estimation. However, resampling based estimation is necessary to
stabilize model selection and this process is parallelizable.

Recently, methods for assigning p-values to variables in high-dimensional
linear models have been proposed. See \cite{DezeureMeinshausen2015} for a review
and comparison.
P-values can be used to approximate q-values
\citep{StoreyTibshirani2003}, and thus to control FDR.
Nevertheless, due to the high
instability of estimated p-values (the so called ``p-value lottery'') resampling
is needed when applying the reviewed methods in practice
\citep{DezeureMeinshausen2015}. The application of this approach to graphical
models is studied in \cite{JankovaGeer2015}. \cite{DezeureZhang2016}
proposes p-value estimation for linear models based a combination
of the de-sparsified lasso and bootstrap. Here, the bootstrap is not used to aggregate many,
unstable estimates but to improve on p-value estimates that relied on asymptotic arguments.
The dependency on a penalization parameter remains (current implementation uses a fixed
penalization chosen via cross-validation).
ROPE can be applied to any resampling based network selection method, including resampling of p-value based
selection, and could thus improve de-sparsified lasso estimates by utilizing
multiple levels of penalization.

Here, ROPE was used for FDR controlled edge selection in a single penalization parameter setting.
An interesting direction for future work would be to generalize ROPE to more complex modeling settings, e.g.
comparative network modeling, with multiple tuning parameters. One could approach this problem in either a sequential fashion (across tuning parameters)
or generalize the distribution mixture modeling to a higher-dimensional parameter space.

Lastly, the use of richer summaries of $W$ than histograms $h^\lambda$ may
improve model selection. One way is to view edge presence counts $W^\lambda_i$
as functional data $W_i(\lambda)$. We have observed that
these functions behave quite differently for different edges. The location and
magnitude of $\min_\lambda \frac{d}{d\lambda}W_i(\lambda)$ are two examples of
quantities that may facilitate edge selection. Another way is to consider
correlation between edges. Edges can compete to explain the node correlation
structure in a network neighborhood. Therefore, selection correlation between
pairs of edges over resamples may also facilitate edge selection. Although
computationally infeasible to estimate in full, the possibility to limit focus
to edge pairs that are, in some sense, closely located in the network makes
this an interesting direction of future research.



%


\acknowledgement

{RJ and JK are supported in part by the Swedish Research Council. RJ is also supported in part by the Wallenberg foundation. SN is supported by the
Swedish Research Council, Swedish Cancer Society and Swedish Childhood Cancer Foundation. AJ and JS are fulltime employees of Astra-Zeneca.}

\emph{Conflict of interest}: none declared.

\bibliography{report}

\providecommand{\latin}[1]{#1}
\makeatletter
\providecommand{\doi}
  {\begingroup\let\do\@makeother\dospecials
  \catcode`\{=1 \catcode`\}=2\doi@aux}
\providecommand{\doi@aux}[1]{\endgroup\texttt{#1}}
\makeatother
\providecommand*\mcitethebibliography{\thebibliography}
\csname @ifundefined\endcsname{endmcitethebibliography}
  {\let\endmcitethebibliography\endthebibliography}{}
\begin{mcitethebibliography}{25}
\providecommand*\natexlab[1]{#1}
\providecommand*\mciteSetBstSublistMode[1]{}
\providecommand*\mciteSetBstMaxWidthForm[2]{}
\providecommand*\mciteBstWouldAddEndPuncttrue
  {\def\EndOfBibitem{\unskip.}}
\providecommand*\mciteBstWouldAddEndPunctfalse
  {\let\EndOfBibitem\relax}
\providecommand*\mciteSetBstMidEndSepPunct[3]{}
\providecommand*\mciteSetBstSublistLabelBeginEnd[3]{}
\providecommand*\EndOfBibitem{}
\mciteSetBstSublistMode{f}
\mciteSetBstMaxWidthForm{subitem}{(\alph{mcitesubitemcount})}
\mciteSetBstSublistLabelBeginEnd
  {\mcitemaxwidthsubitemform\space}
  {\relax}
  {\relax}

\bibitem[Pe'er and Hacohen(2011)Pe'er, and Hacohen]{PeerHacohen2011}
Pe'er,~D.; Hacohen,~N. \emph{{C}ell} \textbf{2011}, \emph{144}, 864--873\relax
\mciteBstWouldAddEndPuncttrue
\mciteSetBstMidEndSepPunct{\mcitedefaultmidpunct}
{\mcitedefaultendpunct}{\mcitedefaultseppunct}\relax
\EndOfBibitem
\bibitem[Kling \latin{et~al.}(2015)Kling, Johansson, S\'anchez, Marinescu,
  J\"ornsten, and Nelander]{KlingNelander2015}
Kling,~T.; Johansson,~P.; S\'anchez,~J.; Marinescu,~V.~D.; J\"ornsten,~R.;
  Nelander,~S. \emph{Nucleic Acids Research} \textbf{2015}, \relax
\mciteBstWouldAddEndPunctfalse
\mciteSetBstMidEndSepPunct{\mcitedefaultmidpunct}
{}{\mcitedefaultseppunct}\relax
\EndOfBibitem
\bibitem[Friedman \latin{et~al.}(2008)Friedman, Hastie, and
  Tibshirani]{FriedmanTibshirani2008}
Friedman,~J.; Hastie,~T.; Tibshirani,~R. \emph{Biostatistics} \textbf{2008},
  \emph{9}, 432--441\relax
\mciteBstWouldAddEndPuncttrue
\mciteSetBstMidEndSepPunct{\mcitedefaultmidpunct}
{\mcitedefaultendpunct}{\mcitedefaultseppunct}\relax
\EndOfBibitem
\bibitem[Langfelder and Horvath(2008)Langfelder, and
  Horvath]{LangfelderHorvath2008}
Langfelder,~P.; Horvath,~S. \emph{BMC Bioinformatics} \textbf{2008}, \emph{9},
  559\relax
\mciteBstWouldAddEndPuncttrue
\mciteSetBstMidEndSepPunct{\mcitedefaultmidpunct}
{\mcitedefaultendpunct}{\mcitedefaultseppunct}\relax
\EndOfBibitem
\bibitem[Margolin \latin{et~al.}(2006)Margolin, Nemenman, Basso, Wiggins,
  Stolovitzky, Favera, and Califano]{MargolinCalifano2006}
Margolin,~A.~A.; Nemenman,~I.; Basso,~K.; Wiggins,~C.; Stolovitzky,~G.;
  Favera,~R.~D.; Califano,~A. \emph{BMC Bioinformatics} \textbf{2006},
  \emph{7}, S7\relax
\mciteBstWouldAddEndPuncttrue
\mciteSetBstMidEndSepPunct{\mcitedefaultmidpunct}
{\mcitedefaultendpunct}{\mcitedefaultseppunct}\relax
\EndOfBibitem
\bibitem[Meinshausen and B\"uhlmann(2010)Meinshausen, and
  B\"uhlmann]{MeinshausenBuhlmann2010}
Meinshausen,~N.; B\"uhlmann,~P. \emph{Journal of the Royal Statistical Society:
  Series B (Statistical Methodology)} \textbf{2010}, \emph{72}, 417--473\relax
\mciteBstWouldAddEndPuncttrue
\mciteSetBstMidEndSepPunct{\mcitedefaultmidpunct}
{\mcitedefaultendpunct}{\mcitedefaultseppunct}\relax
\EndOfBibitem
\bibitem[Liu \latin{et~al.}(2010)Liu, Roeder, and Wasserman]{LiuWasserman2010}
Liu,~H.; Roeder,~K.; Wasserman,~L. \emph{Advances in Neural Information
  Processing Systems 23}; 2010; pp 1432--1440\relax
\mciteBstWouldAddEndPuncttrue
\mciteSetBstMidEndSepPunct{\mcitedefaultmidpunct}
{\mcitedefaultendpunct}{\mcitedefaultseppunct}\relax
\EndOfBibitem
\bibitem[Li \latin{et~al.}(2013)Li, Hsu, Peng, and Wang]{LiWang2013}
Li,~S.; Hsu,~L.; Peng,~J.; Wang,~P. \emph{Ann. Appl. Stat.} \textbf{2013},
  \emph{7}, 391--417\relax
\mciteBstWouldAddEndPuncttrue
\mciteSetBstMidEndSepPunct{\mcitedefaultmidpunct}
{\mcitedefaultendpunct}{\mcitedefaultseppunct}\relax
\EndOfBibitem
\bibitem[Bach(2008)]{Bach2008}
Bach,~F.~R. Bolasso: model consistent lasso estimation through the bootstrap.
  Proc. of the 25th intl. conference on Machine learning. 2008; pp 33--40\relax
\mciteBstWouldAddEndPuncttrue
\mciteSetBstMidEndSepPunct{\mcitedefaultmidpunct}
{\mcitedefaultendpunct}{\mcitedefaultseppunct}\relax
\EndOfBibitem
\bibitem[Storey and Tibshirani(2003)Storey, and
  Tibshirani]{StoreyTibshirani2003}
Storey,~J.; Tibshirani,~R. \emph{Proceedings of the National Academy of
  Sciences} \textbf{2003}, \emph{100}, 9440--9445\relax
\mciteBstWouldAddEndPuncttrue
\mciteSetBstMidEndSepPunct{\mcitedefaultmidpunct}
{\mcitedefaultendpunct}{\mcitedefaultseppunct}\relax
\EndOfBibitem
\bibitem[{The Cancer Genome Atlas Research Network} \latin{et~al.}(2013){The
  Cancer Genome Atlas Research Network}, Weinstein, Collisson, Mills, Shaw,
  Ozenberger, Ellrott, Shmulevich, Sander, and Stuart]{TCGAStuart2013}
{The Cancer Genome Atlas Research Network},; Weinstein,~J.; Collisson,~E.;
  Mills,~G.; Shaw,~K.~M.; Ozenberger,~B.; Ellrott,~K.; Shmulevich,~I.;
  Sander,~C.; Stuart,~J. \emph{Nat Genet} \textbf{2013}, \emph{45},
  1113--1120\relax
\mciteBstWouldAddEndPuncttrue
\mciteSetBstMidEndSepPunct{\mcitedefaultmidpunct}
{\mcitedefaultendpunct}{\mcitedefaultseppunct}\relax
\EndOfBibitem
\bibitem[J\"ornsten \latin{et~al.}(2011)J\"ornsten, Abenius, Kling, Schmidt,
  Johansson, Nordling, Nordlander, Sander, Gennemark, Funa, Nilsson, Lindahl,
  and Nelander]{JornstenNelander2011}
J\"ornsten,~R.; Abenius,~T.; Kling,~T.; Schmidt,~L.; Johansson,~E.;
  Nordling,~T. E.~M.; Nordlander,~B.; Sander,~C.; Gennemark,~P.; Funa,~K.;
  Nilsson,~B.; Lindahl,~L.; Nelander,~S. \emph{Molecular Systems Biology}
  \textbf{2011}, \emph{7}, 486\relax
\mciteBstWouldAddEndPuncttrue
\mciteSetBstMidEndSepPunct{\mcitedefaultmidpunct}
{\mcitedefaultendpunct}{\mcitedefaultseppunct}\relax
\EndOfBibitem
\bibitem[Meinshausen and B\"uhlmann(2006)Meinshausen, and
  B\"uhlmann]{MeinshausenBuhlmann2006}
Meinshausen,~N.; B\"uhlmann,~P. \emph{Ann. Statist.} \textbf{2006}, \emph{34},
  1436--1462\relax
\mciteBstWouldAddEndPuncttrue
\mciteSetBstMidEndSepPunct{\mcitedefaultmidpunct}
{\mcitedefaultendpunct}{\mcitedefaultseppunct}\relax
\EndOfBibitem
\bibitem[Zhao and Yu(2006)Zhao, and Yu]{ZhaoYu2006}
Zhao,~P.; Yu,~B. \emph{J. Mach. Learn. Res.} \textbf{2006}, \emph{7},
  2541--2563\relax
\mciteBstWouldAddEndPuncttrue
\mciteSetBstMidEndSepPunct{\mcitedefaultmidpunct}
{\mcitedefaultendpunct}{\mcitedefaultseppunct}\relax
\EndOfBibitem
\bibitem[Efron(2004)]{Efron2004}
Efron,~B. \emph{Journal of the American Statistical Association} \textbf{2004},
  \emph{99}, 96--104\relax
\mciteBstWouldAddEndPuncttrue
\mciteSetBstMidEndSepPunct{\mcitedefaultmidpunct}
{\mcitedefaultendpunct}{\mcitedefaultseppunct}\relax
\EndOfBibitem
\bibitem[Goldman \latin{et~al.}(2014)Goldman, Craft, Swatloski, Cline,
  Morozova, Diekhans, Haussler, and Zhu]{GoldmanZhu2014}
Goldman,~M.; Craft,~B.; Swatloski,~T.; Cline,~M.; Morozova,~O.; Diekhans,~M.;
  Haussler,~D.; Zhu,~J. \emph{Nucleic Acids Research} \textbf{2014}, \relax
\mciteBstWouldAddEndPunctfalse
\mciteSetBstMidEndSepPunct{\mcitedefaultmidpunct}
{}{\mcitedefaultseppunct}\relax
\EndOfBibitem
\bibitem[Kanehisa and Goto(2000)Kanehisa, and Goto]{KanehisaGoto2000}
Kanehisa,~M.; Goto,~S. \emph{Nucleic acids research} \textbf{2000}, \emph{28},
  27--30\relax
\mciteBstWouldAddEndPuncttrue
\mciteSetBstMidEndSepPunct{\mcitedefaultmidpunct}
{\mcitedefaultendpunct}{\mcitedefaultseppunct}\relax
\EndOfBibitem
\bibitem[Br{\"o}sicke and Faissner(2015)Br{\"o}sicke, and
  Faissner]{BrosickeFaissner2015}
Br{\"o}sicke,~N.; Faissner,~A. \emph{Cell adhesion \& migration} \textbf{2015},
  \emph{9}, 131--140\relax
\mciteBstWouldAddEndPuncttrue
\mciteSetBstMidEndSepPunct{\mcitedefaultmidpunct}
{\mcitedefaultendpunct}{\mcitedefaultseppunct}\relax
\EndOfBibitem
\bibitem[Ohnishi \latin{et~al.}(1998)Ohnishi, Hiraga, Izumoto, Matsumura,
  Kanemura, Arita, and Hayakawa]{OhnishiHayakawa1998}
Ohnishi,~T.; Hiraga,~S.; Izumoto,~S.; Matsumura,~H.; Kanemura,~Y.; Arita,~N.;
  Hayakawa,~T. \emph{Clinical \& experimental metastasis} \textbf{1998},
  \emph{16}, 729--741\relax
\mciteBstWouldAddEndPuncttrue
\mciteSetBstMidEndSepPunct{\mcitedefaultmidpunct}
{\mcitedefaultendpunct}{\mcitedefaultseppunct}\relax
\EndOfBibitem
\bibitem[Capelli \latin{et~al.}(1999)Capelli, Civallero, Barni, Ceroni, and
  Nano]{CapelliNano1999}
Capelli,~E.; Civallero,~M.; Barni,~S.; Ceroni,~M.; Nano,~R. \emph{Anticancer
  research} \textbf{1999}, \emph{19}, 3147\relax
\mciteBstWouldAddEndPuncttrue
\mciteSetBstMidEndSepPunct{\mcitedefaultmidpunct}
{\mcitedefaultendpunct}{\mcitedefaultseppunct}\relax
\EndOfBibitem
\bibitem[Fleiss(1971)]{Fleiss1971}
Fleiss,~J.~L. \emph{Psychological bulletin} \textbf{1971}, \emph{76}, 378\relax
\mciteBstWouldAddEndPuncttrue
\mciteSetBstMidEndSepPunct{\mcitedefaultmidpunct}
{\mcitedefaultendpunct}{\mcitedefaultseppunct}\relax
\EndOfBibitem
\bibitem[Dezeure \latin{et~al.}(2015)Dezeure, B\"uhlmann, Meier, and
  Meinshausen]{DezeureMeinshausen2015}
Dezeure,~R.; B\"uhlmann,~P.; Meier,~L.; Meinshausen,~N. \emph{Statist. Sci.}
  \textbf{2015}, \emph{30}, 533--558\relax
\mciteBstWouldAddEndPuncttrue
\mciteSetBstMidEndSepPunct{\mcitedefaultmidpunct}
{\mcitedefaultendpunct}{\mcitedefaultseppunct}\relax
\EndOfBibitem
\bibitem[Jankov\'a and van~de Geer(2015)Jankov\'a, and van~de
  Geer]{JankovaGeer2015}
Jankov\'a,~J.; van~de Geer,~S. \emph{Electron. J. Statist.} \textbf{2015},
  \emph{9}, 1205--1229\relax
\mciteBstWouldAddEndPuncttrue
\mciteSetBstMidEndSepPunct{\mcitedefaultmidpunct}
{\mcitedefaultendpunct}{\mcitedefaultseppunct}\relax
\EndOfBibitem
\bibitem[Dezeure \latin{et~al.}(2016)Dezeure, B\"uhlmann, and
  Zhang]{DezeureZhang2016}
Dezeure,~R.; B\"uhlmann,~P.; Zhang,~C.-H. \emph{Preprint arXiv:1606.03940}
  \textbf{2016}, \relax
\mciteBstWouldAddEndPunctfalse
\mciteSetBstMidEndSepPunct{\mcitedefaultmidpunct}
{}{\mcitedefaultseppunct}\relax
\EndOfBibitem
\end{mcitethebibliography}

\section*{Supplementary Materials}

\subsection*{FDR controlled variable selection for a multinomial logistic
regression classifier of gene expression profiles}
In our final experiment, we apply ROPE to model selection for a non-graphical
model. In particular, we demonstrate the use of ROPE for a multinomial logistic
regression classifier for classifying the primary cancer type of a gene
expression profile. We downloaded RNA-Seq gene expression profiles consisting of
measurements of {20,530} genes for {9,755} cancer patients from the USCS Cancer
Genomics Browser. The data comes from TCGA. We removed profiles corresponding to
cancer types for which less than 100 observations were present in the data set,
in order to reduce the chance of drawing bootstrap samples without all classes
represented. The resulting data set consists of {9,256} observations and
{20,530} variables. Each observation is classified as having one of 24 primary
cancer types. We drew 100 bootstrap samples and fitted generalized linear models
with lasso penalization and multinomial response to each bootstrap sample. We
used grouped lasso penalization so that each variable is either selected for all
classes or excluded entirely. For each bootstrap sample, one model was fitted
for each of 22 levels of penalization, ranging from 0.015 to 0.039. Lower
penalization resulted in non-convergence when fitting the model and higher
penalization resulted in histograms not being U-shaped. The resulting matrix $W$
of 22 times {20,530} variable inclusion counts was used with ROPE to make an FDR
controlled selection of genes whose expression level is predictive of primary
cancer type. 86, 118 and 133 genes were selected at the 0.05, 0.1 and 0.15 FDR
level, respectively. The selected genes are presented in Table~\ref{mlntable}. This experiment shows that ROPE can be
applied to some variable selection problems other than edge selection in graphical
models.

\subsection*{Additional simulation results}

Figures below show results from all simulations. For each simulation setting, four parameters are
varied one by one (number of bootstraps $B$, number of penalization levels,
number of observations $n$ and weakness in randomized lasso). For each varied
parameter, FDR and modified F1 are shown for each method and three target FDR:
0.05, 0.1 and 0.15. A detailed description of simulation settings and
interpretation of results is given in the main article.

\begin{table}[ht]
\centering
\begin{tabular}{rlrrlrrlr}
  \hline
 & gene & q-value &  & gene & q-value &  & gene & q-value \\
  \hline
1 & ATP5EP2 & 0.025 &   46 & SFTA3 & 0.025 &   91 & KRT74 & 0.051 \\
  2 & AZGP1 & 0.025 &   47 & SFTPA1 & 0.025 &   92 & LYPLAL1 & 0.051 \\
  3 & BCL2L15 & 0.025 &   48 & SFTPB & 0.025 &   93 & MSX1 & 0.051 \\
  4 & C10orf27 & 0.025 &   49 & SLC6A3 & 0.025 &   94 & MUC5B & 0.051 \\
  5 & C14orf105 & 0.025 &   50 & SOX17 & 0.025 &   95 & PTGER3 & 0.051 \\
  6 & C8orf85 & 0.025 &   51 & SPRYD5 & 0.025 &   96 & RNF212 & 0.051 \\
  7 & CALML3 & 0.025 &   52 & ST6GALNAC1 & 0.025 &   97 & SLC5A6 & 0.051 \\
  8 & CDH16 & 0.025 &   53 & TBX5 & 0.025 &   98 & SLCO1A2 & 0.051 \\
  9 & CDHR1 & 0.025 &   54 & TCF21 & 0.025 &   99 & C6orf223 & 0.058 \\
  10 & CDX1 & 0.025 &   55 & TFRC & 0.025 &  100 & ERBB3 & 0.058 \\
  11 & CFHR2 & 0.025 &   56 & TG & 0.025 &  101 & FOXF1 & 0.058 \\
  12 & DPPA3 & 0.025 &   57 & TMEFF2 & 0.025 &  102 & IRX1 & 0.058 \\
  13 & DSG3 & 0.025 &   58 & TPO & 0.025 &  103 & NACAP1 & 0.058 \\
  14 & EBF2 & 0.025 &   59 & TRPS1 & 0.025 &  104 & PHOX2A & 0.058 \\
  15 & EMX2 & 0.025 &   60 & TSIX & 0.025 &  105 & C2orf80 & 0.065 \\
  16 & FLJ45983 & 0.025 &   61 & TYR & 0.025 &  106 & MMD2 & 0.065 \\
  17 & FOXE1 & 0.025 &   62 & UPK1B & 0.025 &  107 & SLC22A2 & 0.065 \\
  18 & FTHL3 & 0.025 &   63 & UPK2 & 0.025 &  108 & APCS & 0.071 \\
  19 & FUNDC2P2 & 0.025 &   64 & ZNF134 & 0.025 &  109 & GJB1 & 0.071 \\
  20 & FXYD2 & 0.025 &   65 & ZNF280B & 0.025 &  110 & LOC285740 & 0.071 \\
  21 & HAND2 & 0.025 &   66 & FABP7 & 0.035 &  111 & BCAR1 & 0.078 \\
  22 & HOXA9 & 0.025 &   67 & HOXC8 & 0.035 &  112 & ACTC1 & 0.084 \\
  23 & INS & 0.025 &   68 & KRT20 & 0.035 &  113 & CTAGE1 & 0.091 \\
  24 & IRX2 & 0.025 &   69 & MAP7 & 0.035 &  114 & ESR1 & 0.091 \\
  25 & IRX5 & 0.025 &   70 & MS4A3 & 0.035 &  115 & GFAP & 0.091 \\
  26 & ITGA3 & 0.025 &   71 & MUC16 & 0.035 &  116 & HKDC1 & 0.091 \\
  27 & KIAA1543 & 0.025 &   72 & NOX1 & 0.035 &  117 & PLA2G2F & 0.091 \\
  28 & KLK2 & 0.025 &   73 & NTRK2 & 0.035 &  118 & SOX10 & 0.091 \\
  29 & LGSN & 0.025 &   74 & PAX3 & 0.035 &  119 & PPARG & 0.103 \\
  30 & LOC407835 & 0.025 &   75 & PRO1768 & 0.035 &  120 & C21orf131 & 0.109 \\
  31 & LOC643387 & 0.025 &   76 & SERPINB3 & 0.035 &  121 & DLX6 & 0.109 \\
  32 & MAB21L2 & 0.025 &   77 & SYCP2 & 0.035 &  122 & GAL3ST3 & 0.109 \\
  33 & NACA2 & 0.025 &   78 & C14orf115 & 0.043 &  123 & HNF1B & 0.109 \\
  34 & NDUFA4L2 & 0.025 &   79 & C14orf19 & 0.043 &  124 & KRT5 & 0.109 \\
  35 & PA2G4P4 & 0.025 &   80 & C1orf172 & 0.043 &  125 & SPINK1 & 0.109 \\
  36 & PAX8 & 0.025 &   81 & FGL1 & 0.043 &  126 & ARHGEF33 & 0.115 \\
  37 & PHOX2B & 0.025 &   82 & GATA3 & 0.043 &  127 & C1orf14 & 0.115 \\
  38 & POU3F3 & 0.025 &   83 & HOXA11 & 0.043 &  128 & APOA2 & 0.121 \\
  39 & PRAC & 0.025 &   84 & KRT7 & 0.043 &  129 & LRRN4 & 0.121 \\
  40 & RFX4 & 0.025 &   85 & PRHOXNB & 0.043 &  130 & SOX2 & 0.121 \\
  41 & RPL17 & 0.025 &   86 & SCGB2A1 & 0.043 &  131 & WNT3A & 0.127 \\
  42 & RPL39L & 0.025 &   87 & FLJ32063 & 0.051 &  132 & GJB7 & 0.133 \\
  43 & RPS4Y1 & 0.025 &   88 & FOXA2 & 0.051 &  133 & NASP & 0.144 \\
  44 & SCGB2A2 & 0.025 &   89 & HECW2 & 0.051 &  134 & ATCAY & 0.150 \\
  45 & SERPINB13 & 0.025 &   90 & KLK3 & 0.051 &  135 & DDR1 & 0.150 \\
   \hline
\end{tabular}
\caption{The 135 transcripts with lowest
q-value as selected with ROPE for a multinomial logistic classifier of
expression profiles by cancer
type.}
\label{mlntable}
\end{table}

\clearpage

\begin{figure}[p]
\centerline{\includegraphics[width=0.9\linewidth]{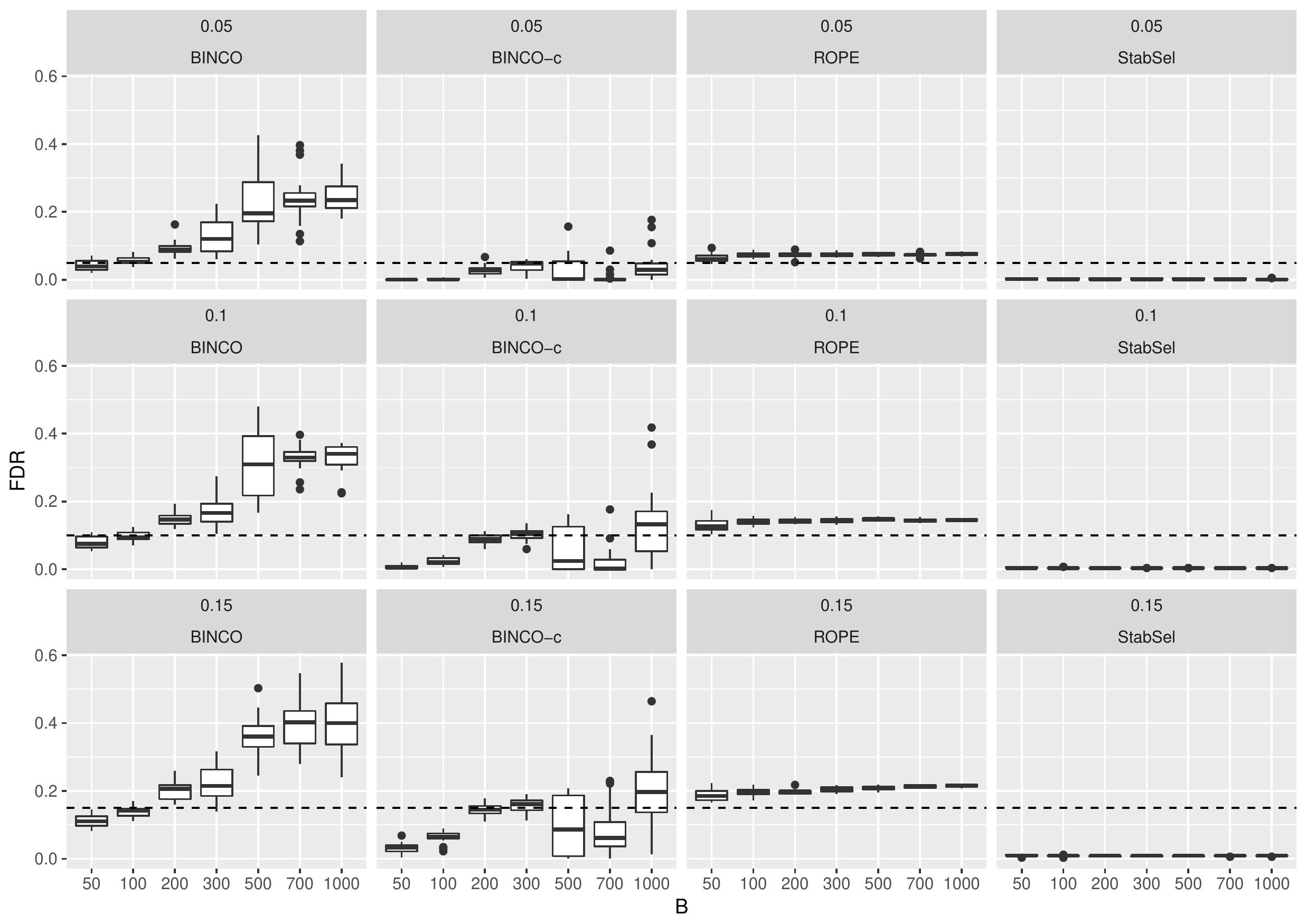}}
\caption{Network topology: chain, steps: 15, $n=200$, weakness: 0.8, facet titles: target FDR and method.}
\label{sim1}
\end{figure}

\begin{figure}[p]
\centerline{\includegraphics[width=0.9\linewidth]{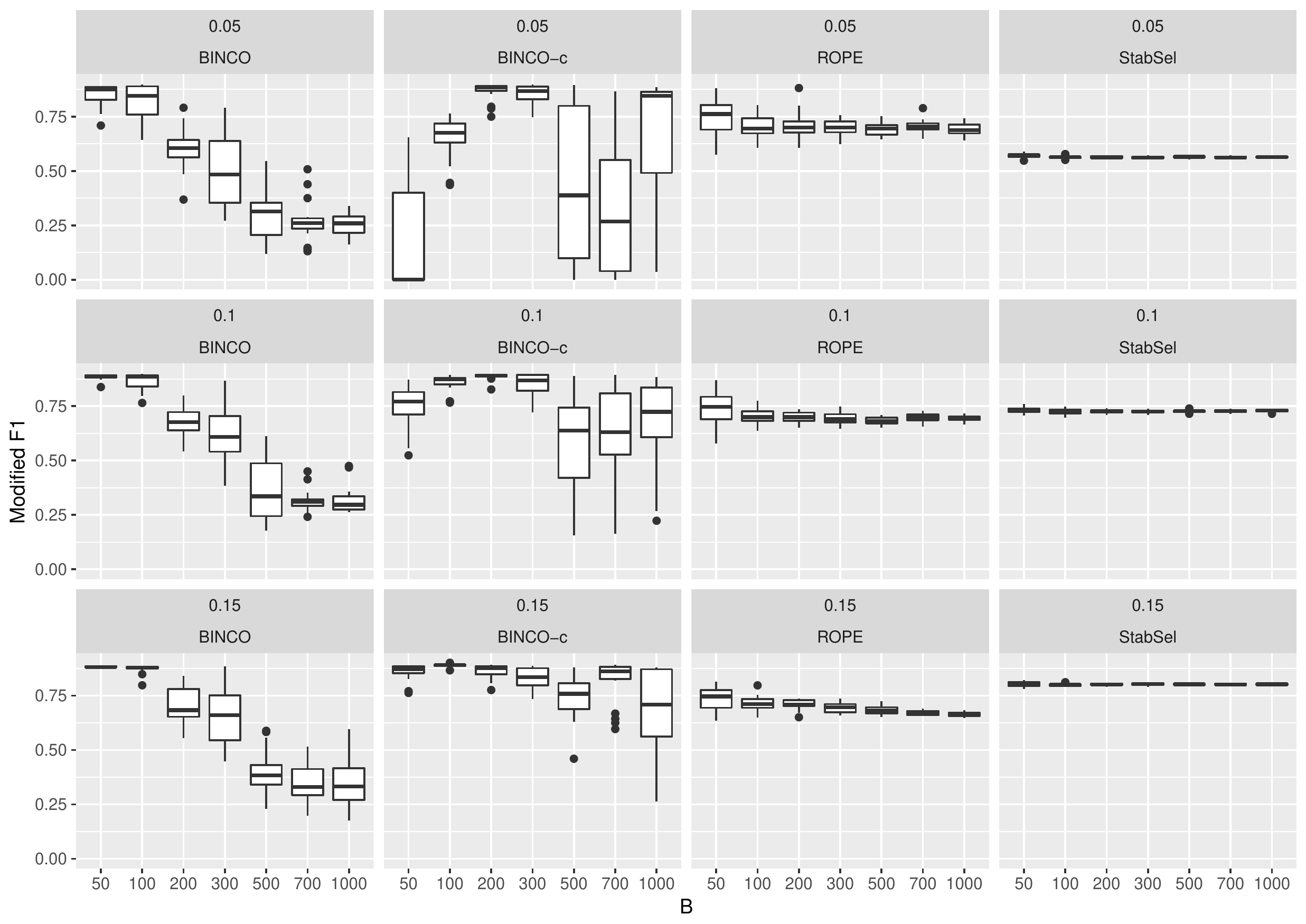}}
\caption{Network topology: chain, steps: 15, $n=200$, weakness: 0.8, facet titles: target FDR and method.}
\label{sim2}
\end{figure}

\begin{figure}[p]
\centerline{\includegraphics[width=0.9\linewidth]{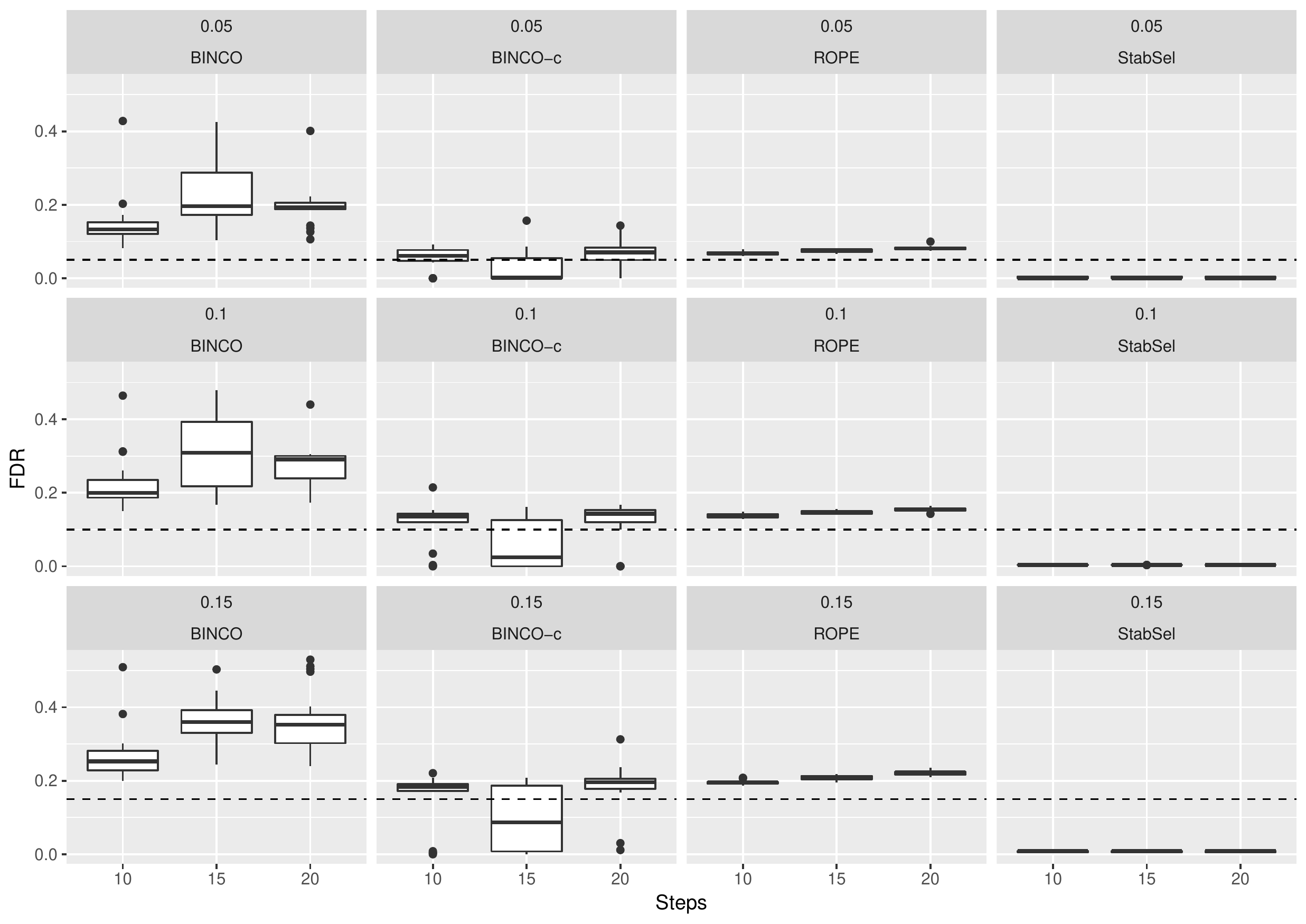}}
\caption{Network topology: chain, $B=500$, $n=200$, weakness: 0.8, facet titles: target FDR and method.}
\label{sim3}
\end{figure}

\begin{figure}[p]
\centerline{\includegraphics[width=0.9\linewidth]{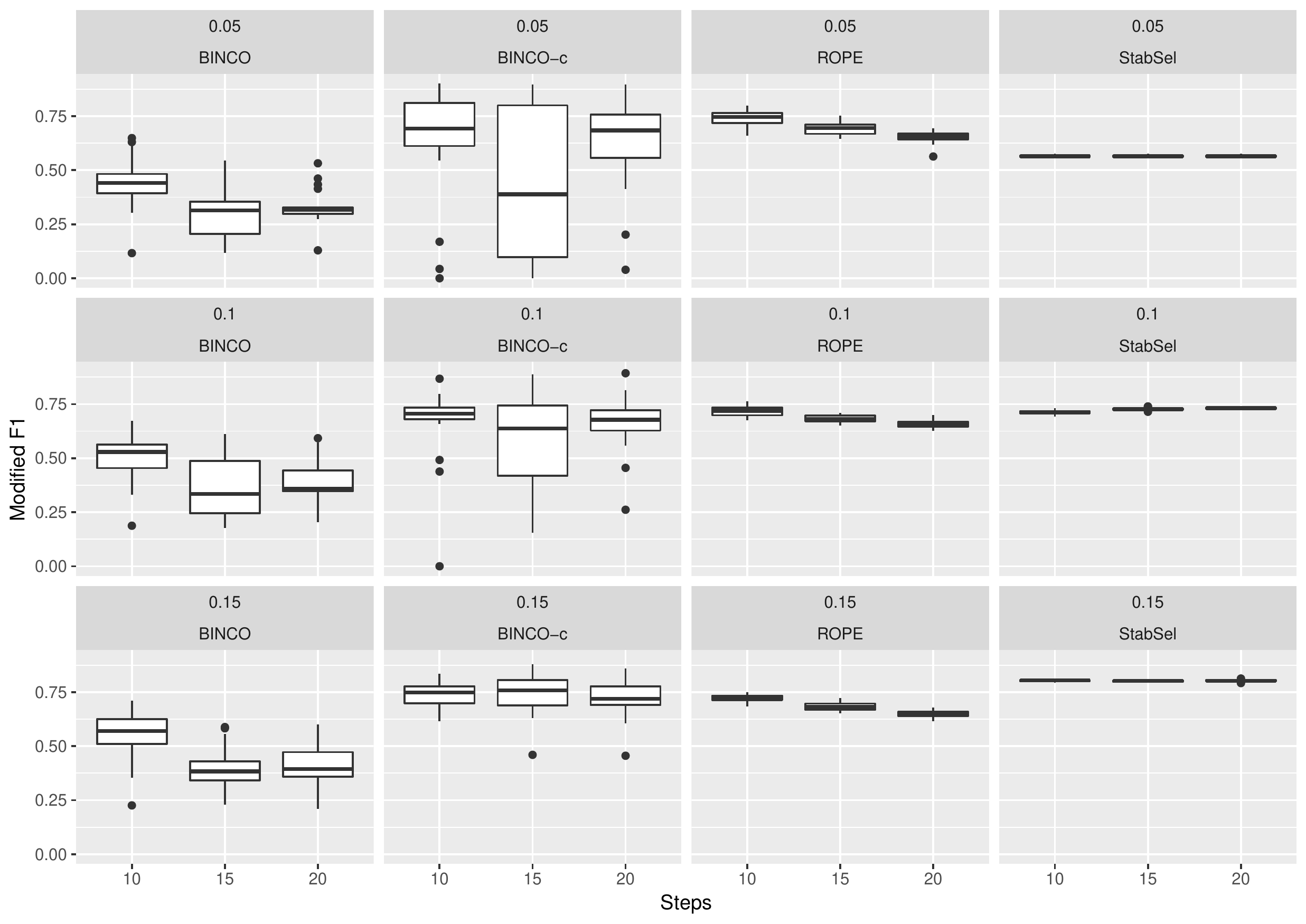}}
\caption{Network topology: chain, $B=500$, $n=200$, weakness: 0.8, facet titles: target FDR and method.}
\label{sim4}
\end{figure}

\begin{figure}[p]
\centerline{\includegraphics[width=0.9\linewidth]{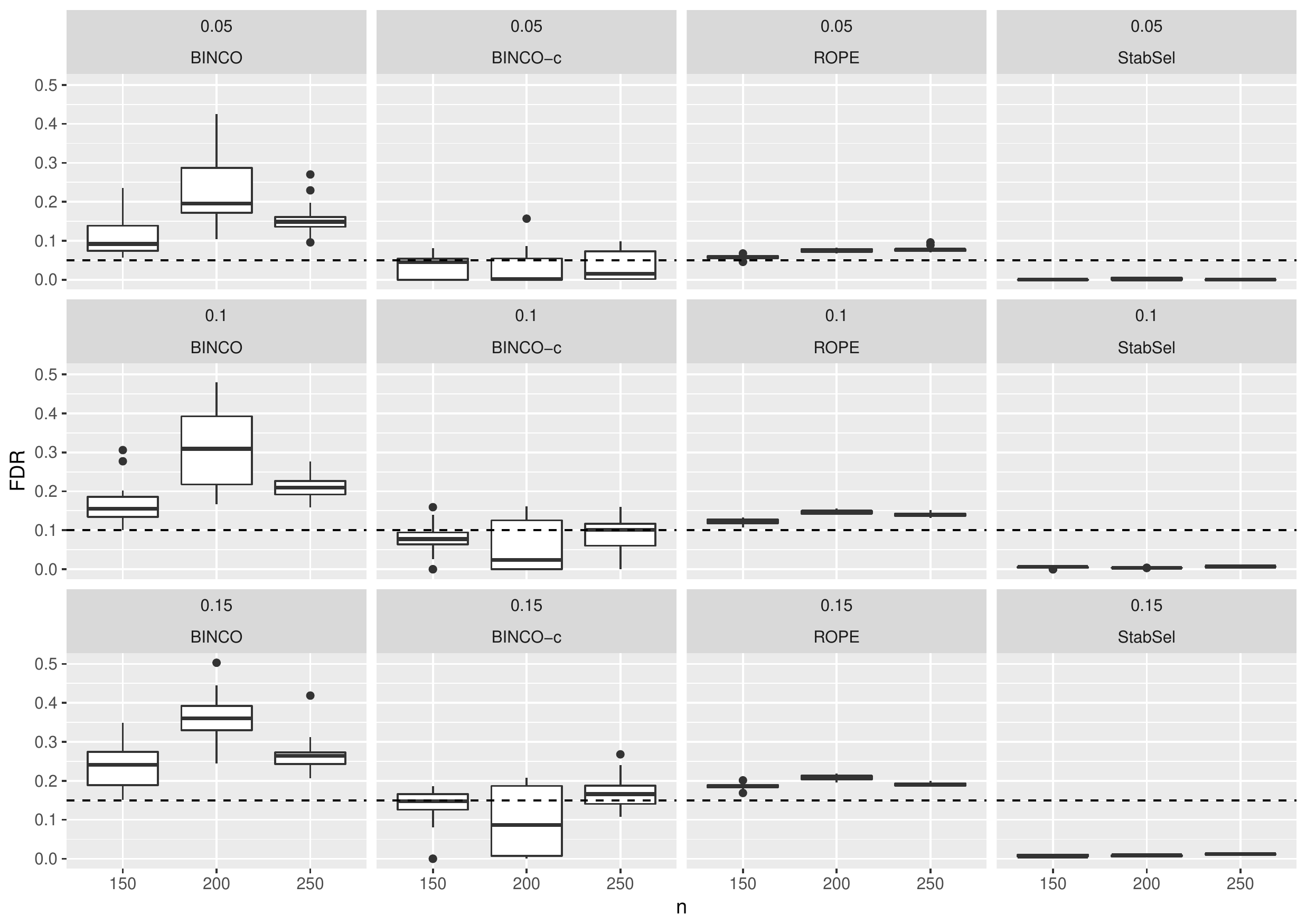}}
\caption{Network topology: chain, $B=500$, steps: 15, weakness: 0.8, facet titles: target FDR and method.}
\label{sim5}
\end{figure}

\begin{figure}[p]
\centerline{\includegraphics[width=0.9\linewidth]{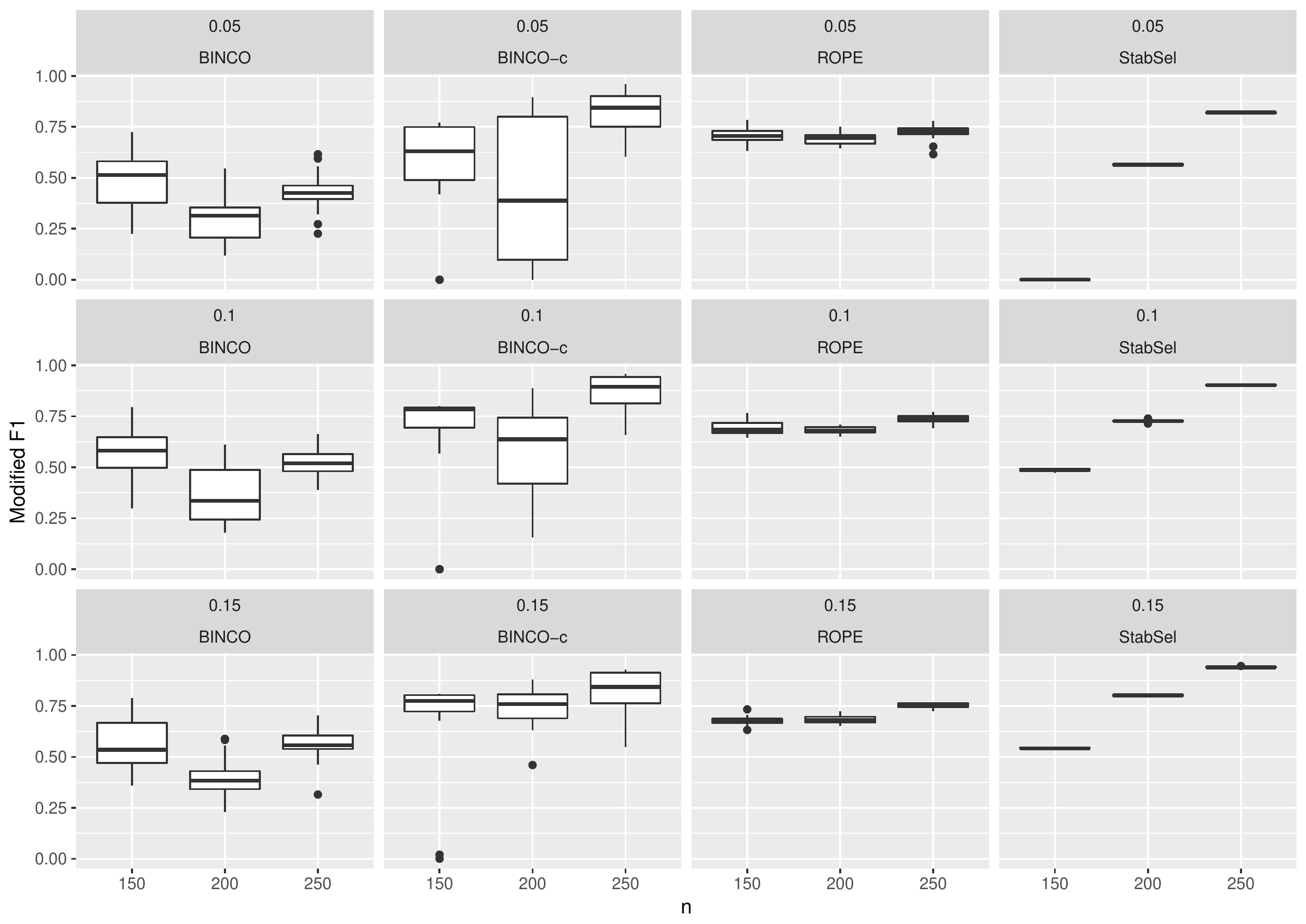}}
\caption{Network topology: chain, $B=500$, steps: 15, weakness: 0.8, facet titles: target FDR and method.}
\label{sim6}
\end{figure}

\begin{figure}[p]
\centerline{\includegraphics[width=0.9\linewidth]{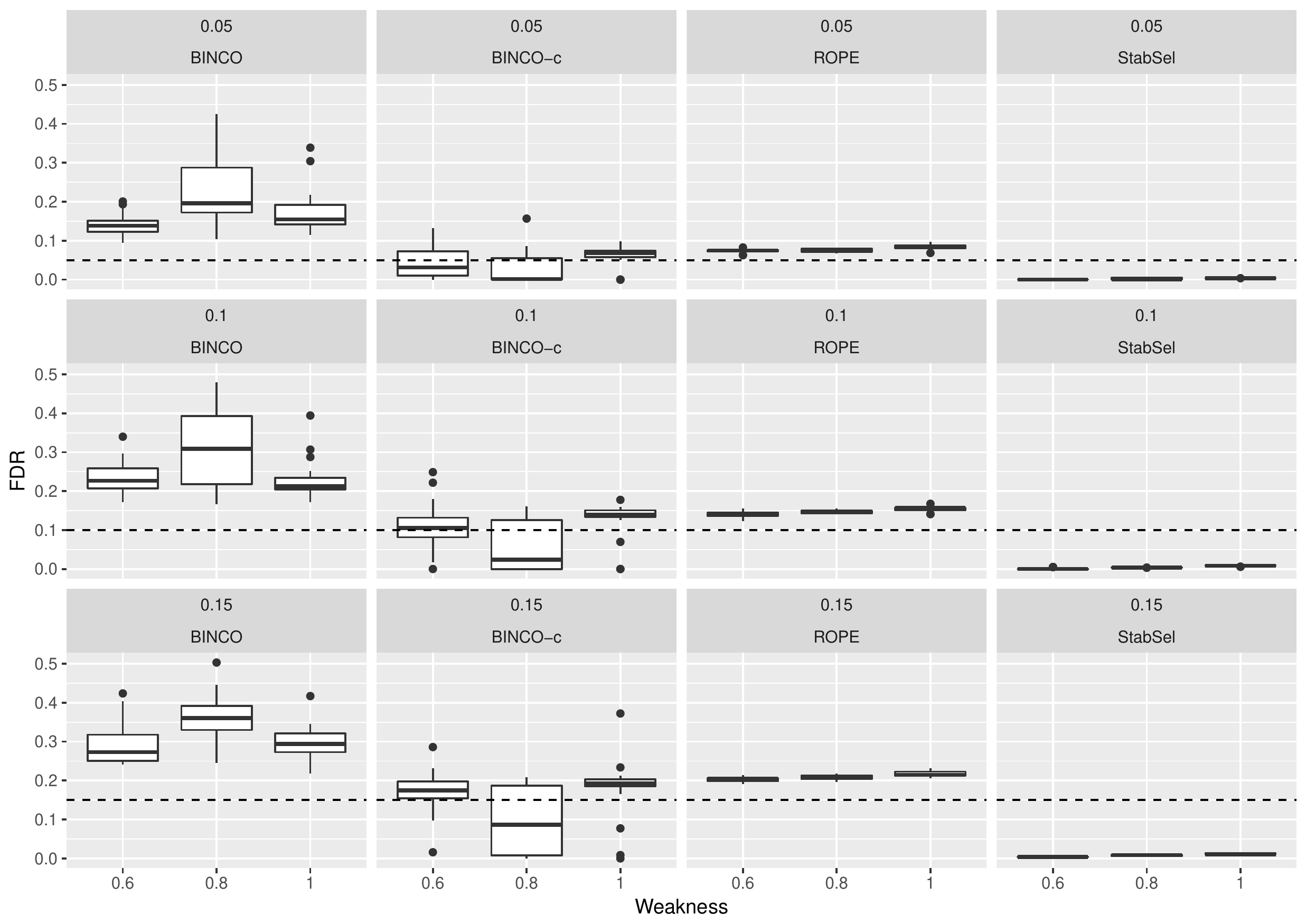}}
\caption{Network topology: chain, $B=500$, steps: 15, $n=200$, facet titles: target FDR and method.}
\label{sim7}
\end{figure}

\begin{figure}[p]
\centerline{\includegraphics[width=0.9\linewidth]{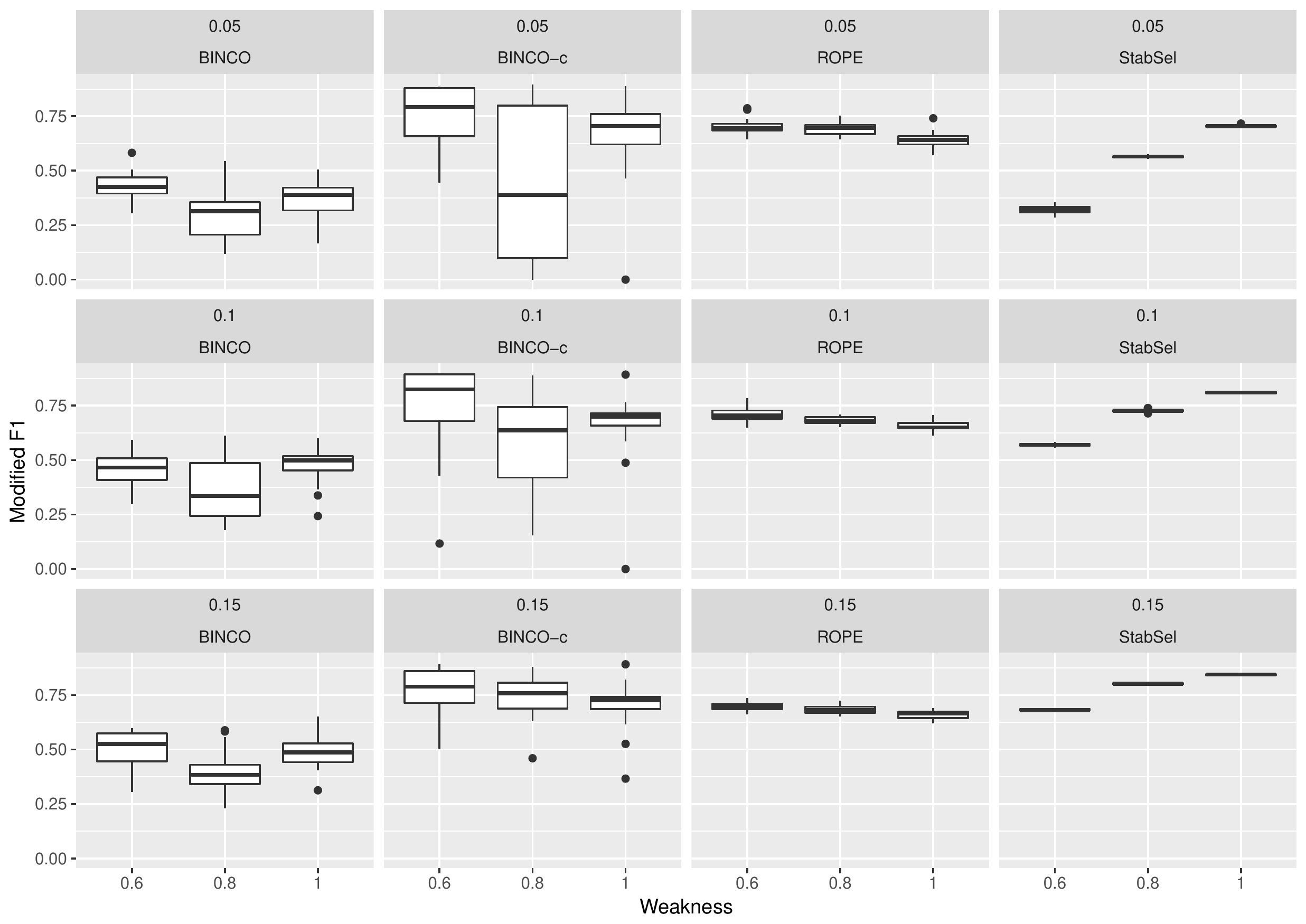}}
\caption{Network topology: chain, $B=500$, steps: 15, $n=200$, facet titles: target FDR and method.}
\label{sim8}
\end{figure}

\begin{figure}[p]
\centerline{\includegraphics[width=0.9\linewidth]{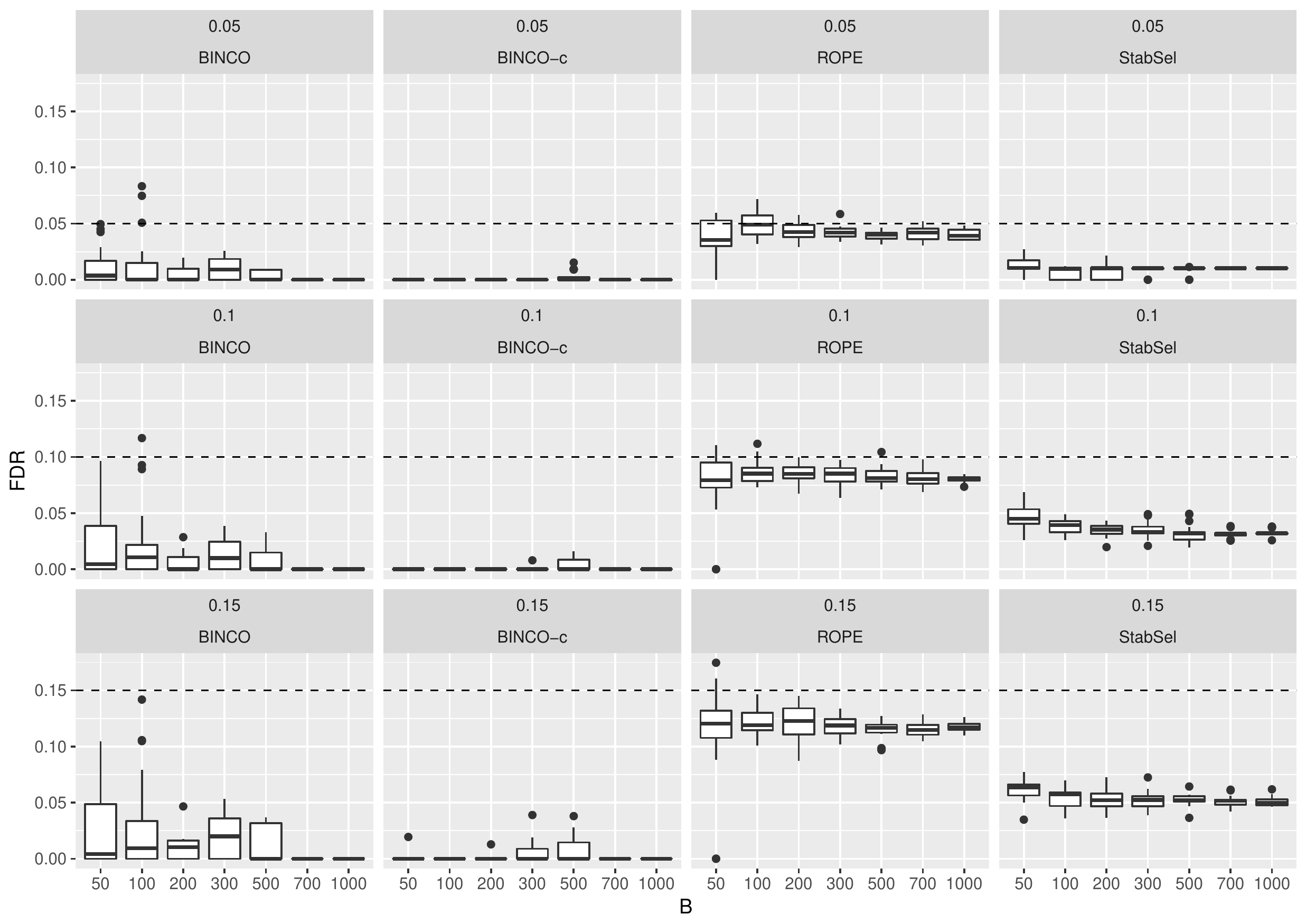}}
\caption{Network topology: dense, steps: 15, $n=200$, weakness=0.8, facet titles: target FDR and method.}
\label{sim9}
\end{figure}

\begin{figure}[p]
\centerline{\includegraphics[width=0.9\linewidth]{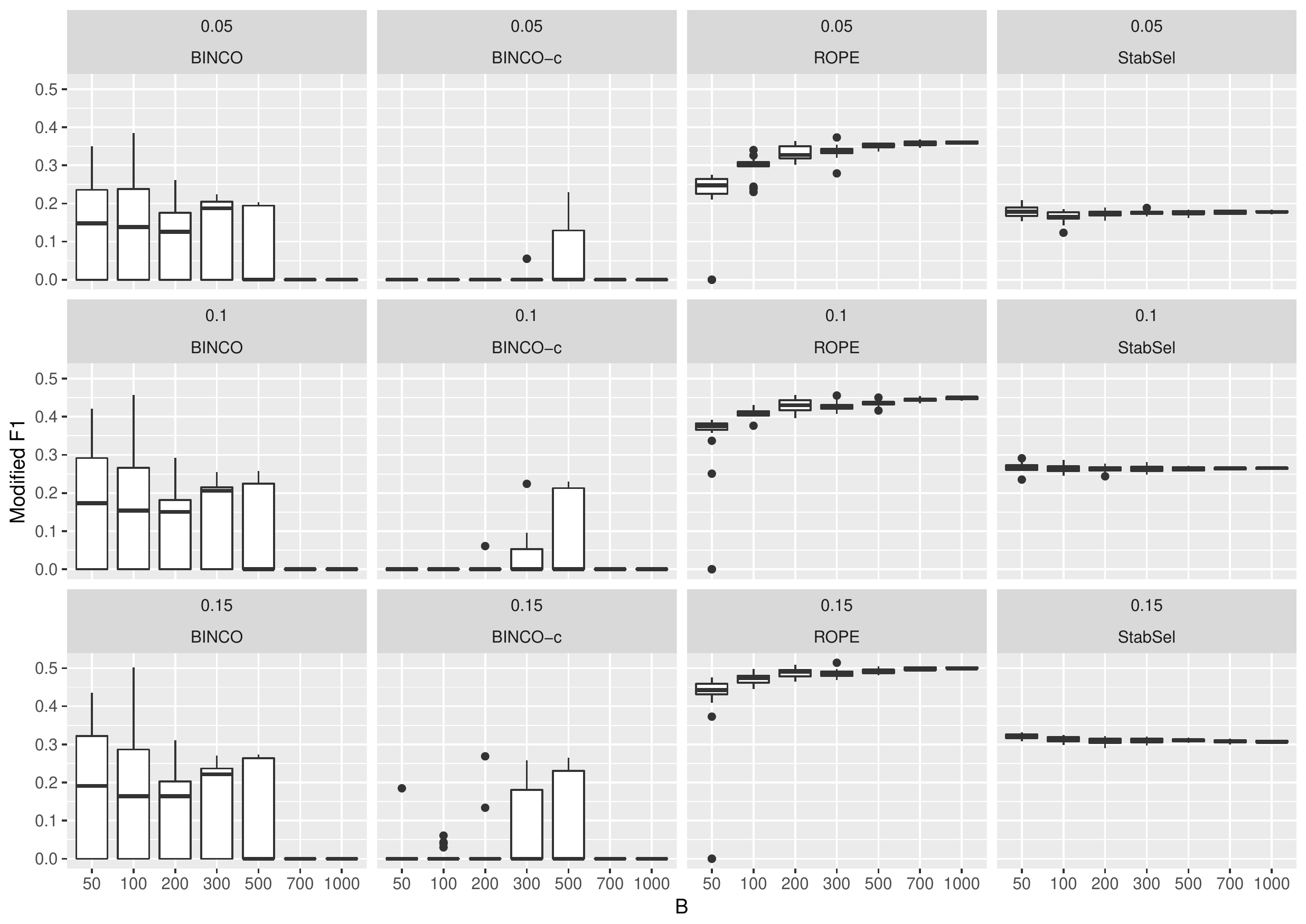}}
\caption{Network topology: dense, steps: 15, $n=200$, weakness=0.8, facet titles: target FDR and method.}
\label{sim10}
\end{figure}

\begin{figure}[p]
\centerline{\includegraphics[width=0.9\linewidth]{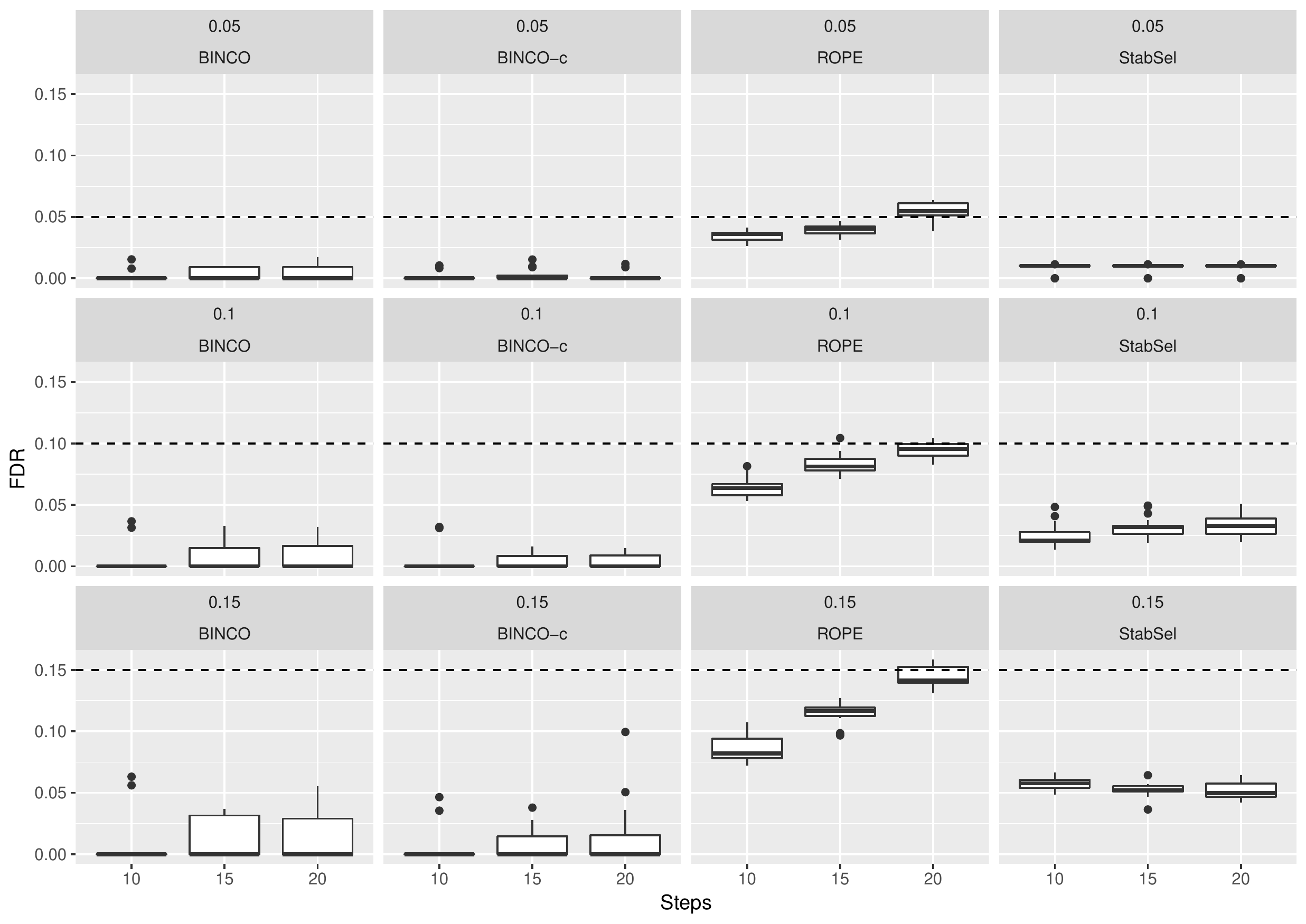}}
\caption{Network topology: dense, $B=500$, $n=200$, weakness=0.8, facet titles: target FDR and method.}
\label{sim11}
\end{figure}

\begin{figure}[p]
\centerline{\includegraphics[width=0.9\linewidth]{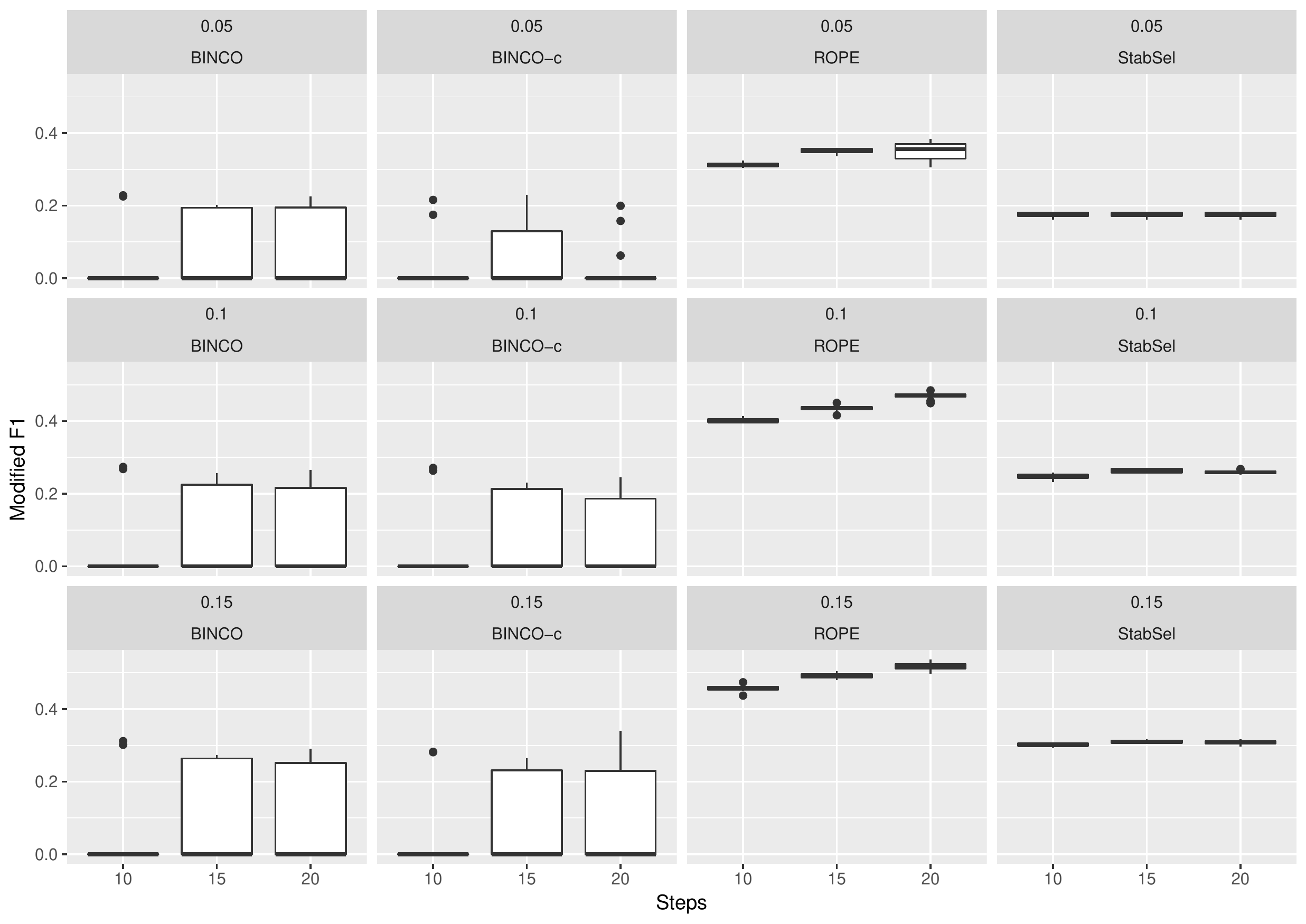}}
\caption{Network topology: dense, $B=500$, $n=200$, weakness=0.8, facet titles: target FDR and method.}
\label{sim12}
\end{figure}

\begin{figure}[p]
\centerline{\includegraphics[width=0.9\linewidth]{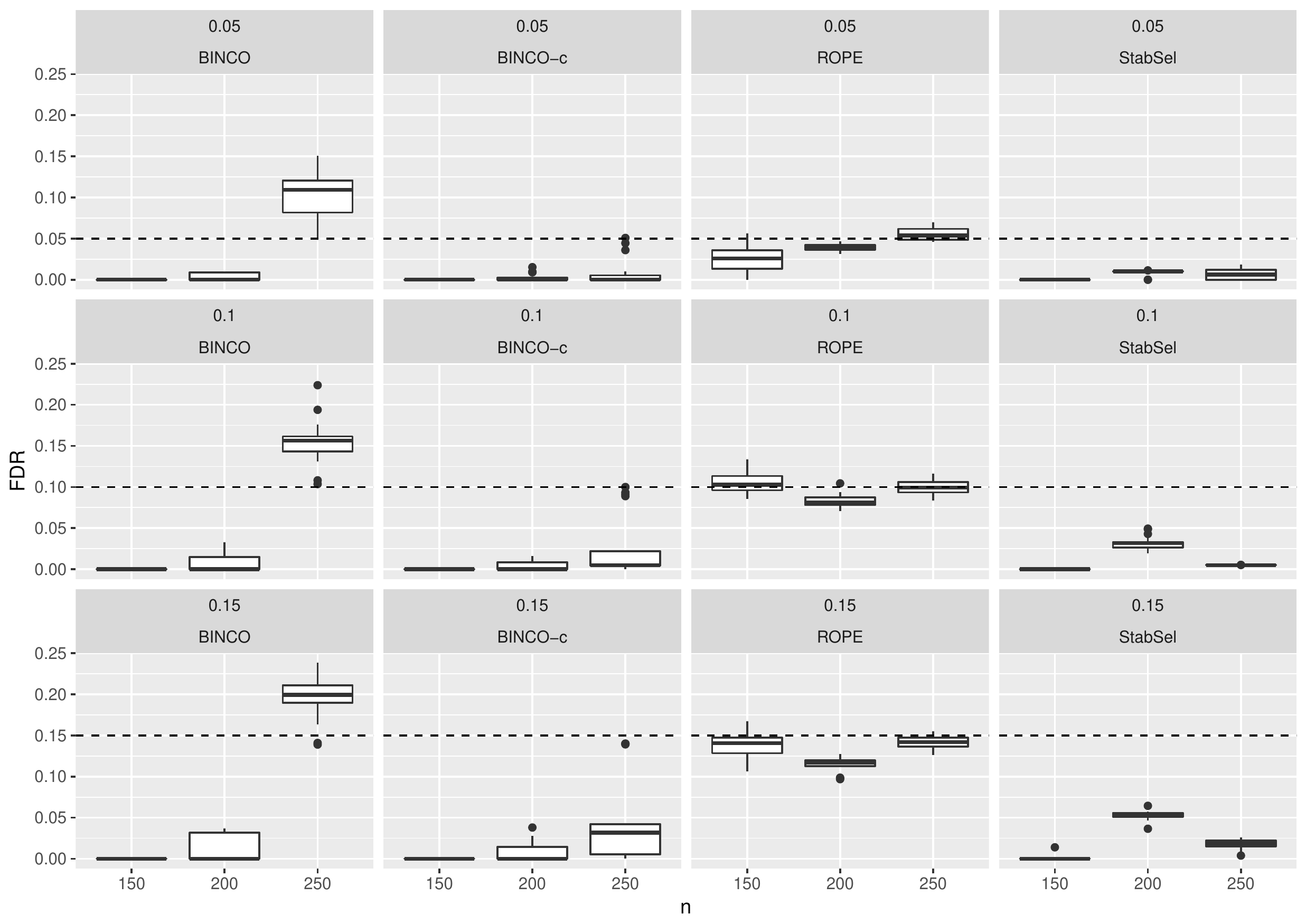}}
\caption{Network topology: dense, $B=500$, steps:15, weakness=0.8, facet titles: target FDR and method.}
\label{sim13}
\end{figure}

\begin{figure}[p]
\centerline{\includegraphics[width=0.9\linewidth]{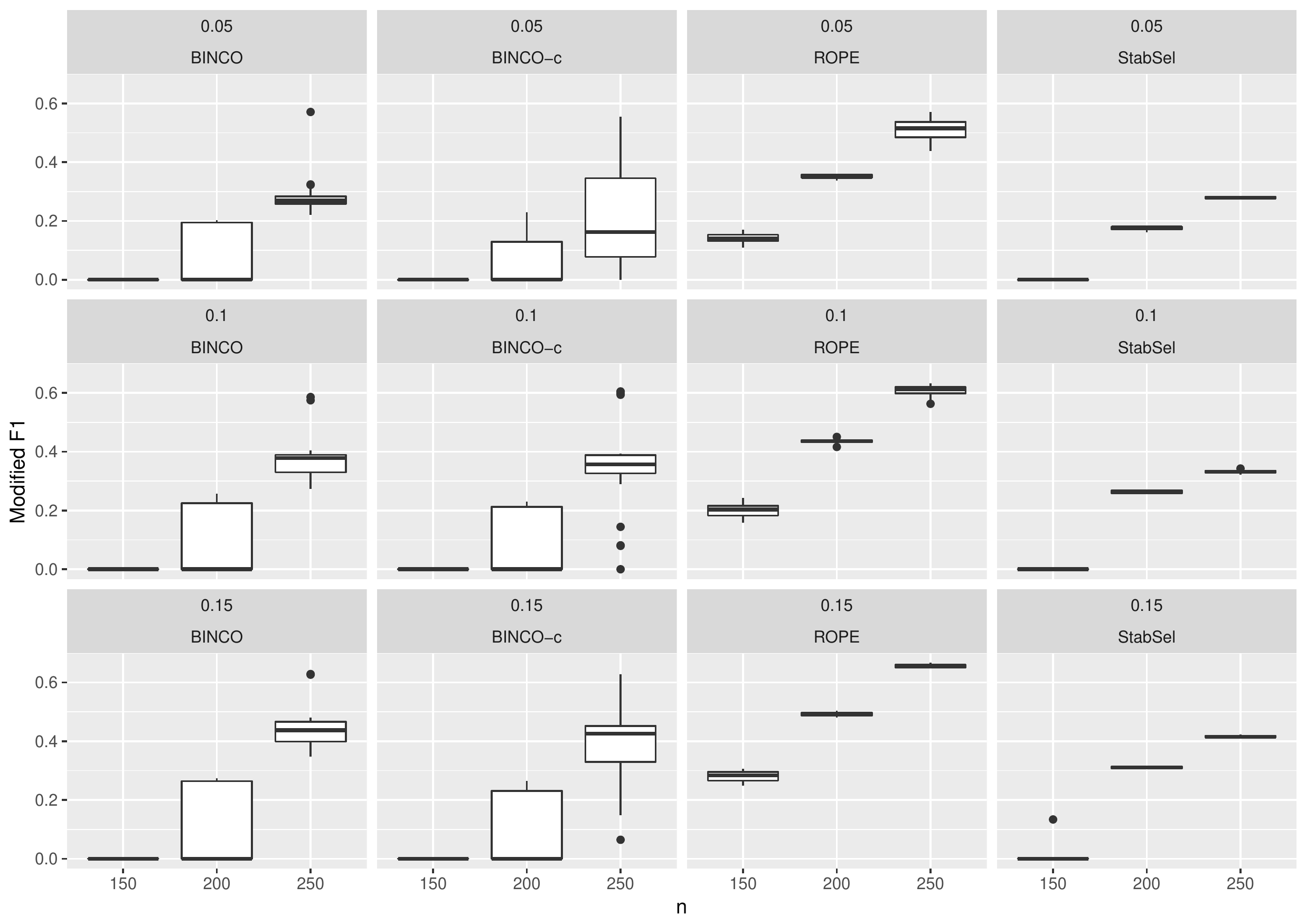}}
\caption{Network topology: dense, $B=500$, steps:15, weakness=0.8, facet titles: target FDR and method.}
\label{sim14}
\end{figure}

\begin{figure}[p]
\centerline{\includegraphics[width=0.9\linewidth]{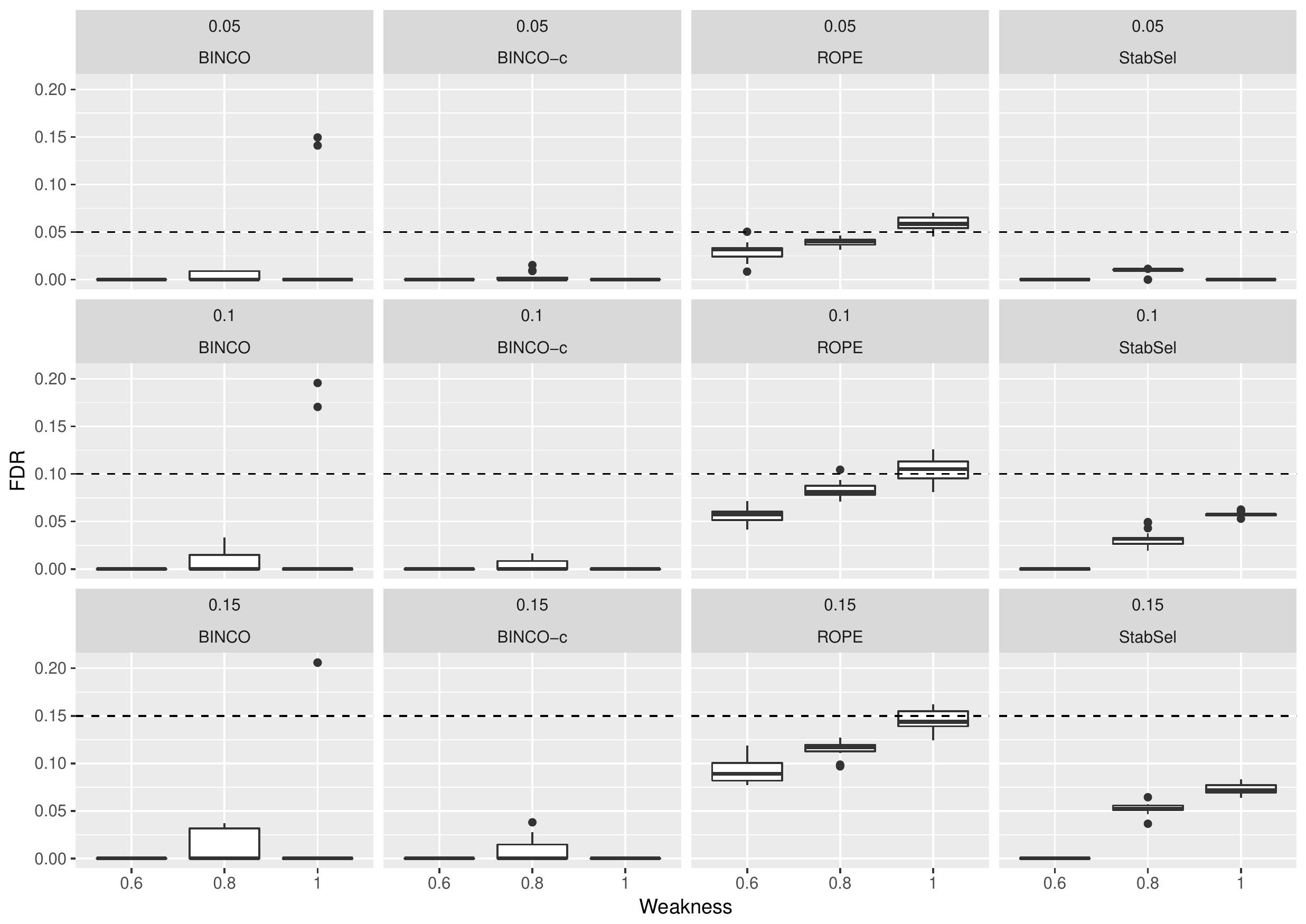}}
\caption{Network topology: dense, $B=500$, steps:15, $n=200$, facet titles: target FDR and method.}
\label{sim15}
\end{figure}

\begin{figure}[p]
\centerline{\includegraphics[width=0.9\linewidth]{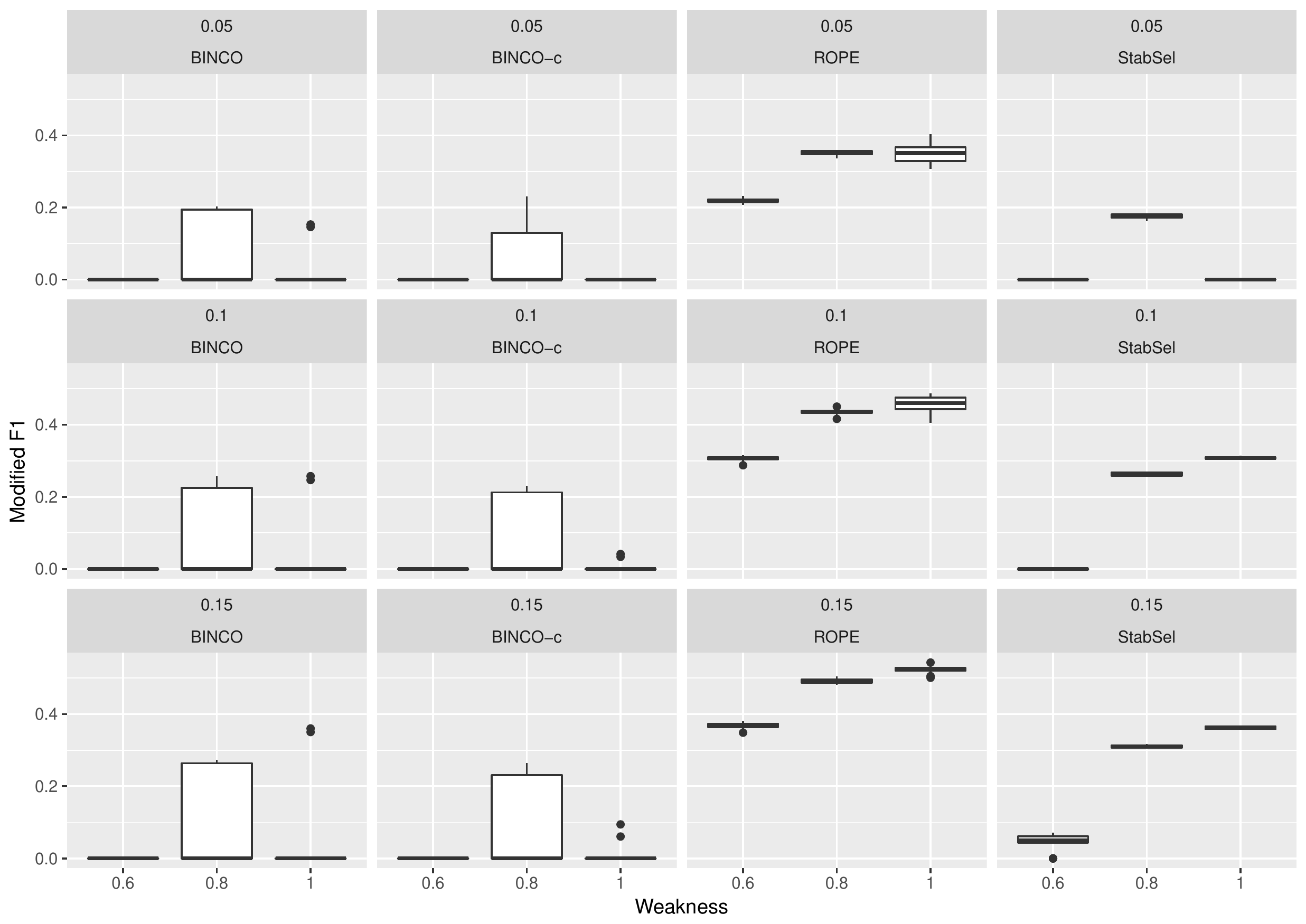}}
\caption{Network topology: dense, $B=500$, steps:15, $n=200$, facet titles: target FDR and method.}
\label{sim16}
\end{figure}

\begin{figure}[p]
\centerline{\includegraphics[width=0.9\linewidth]{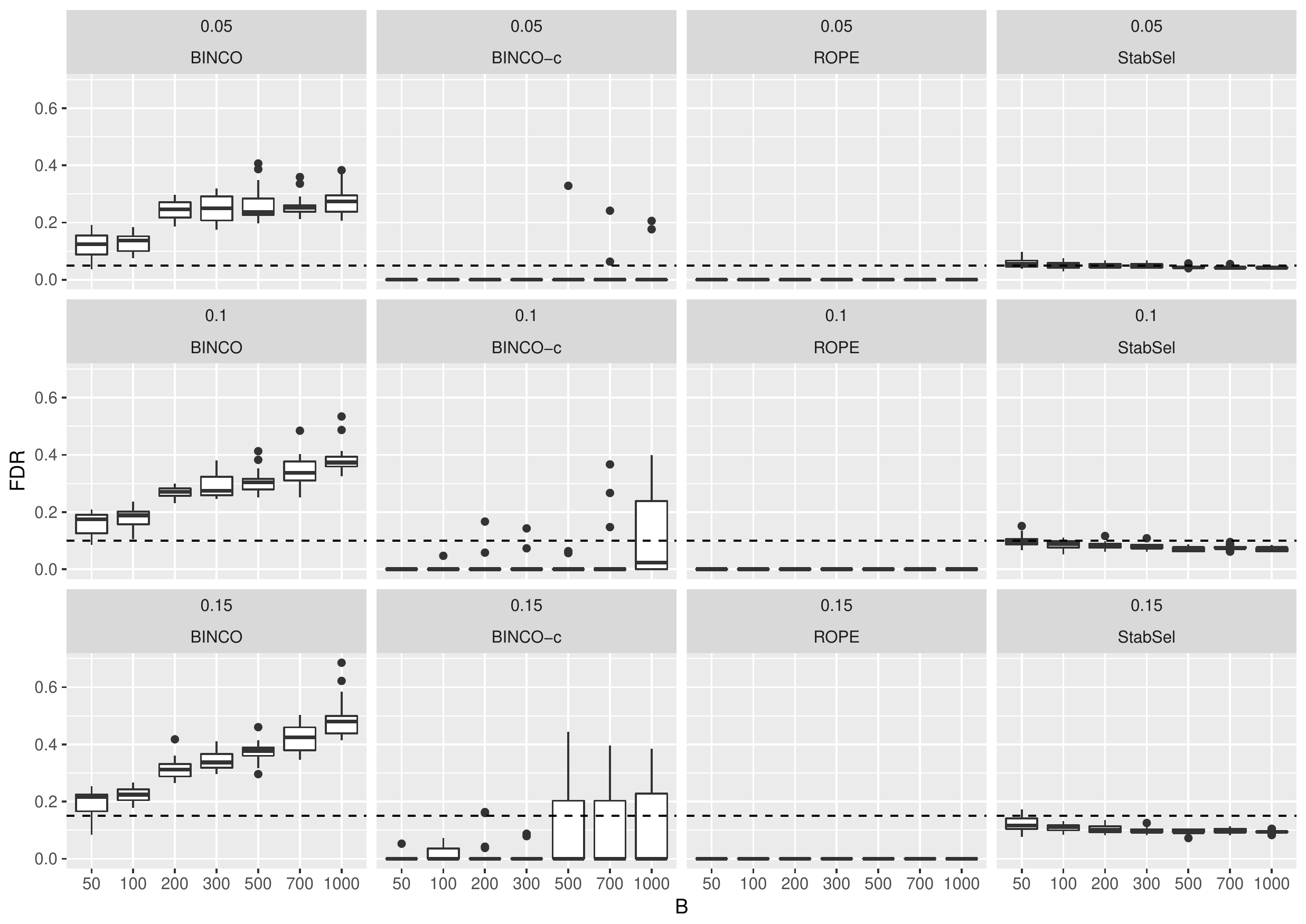}}
\caption{Network topology: hubby, steps:15, $n=200$, weakness=0.8, facet titles: target FDR and method.}
\label{sim17}
\end{figure}

\begin{figure}[p]
\centerline{\includegraphics[width=0.9\linewidth]{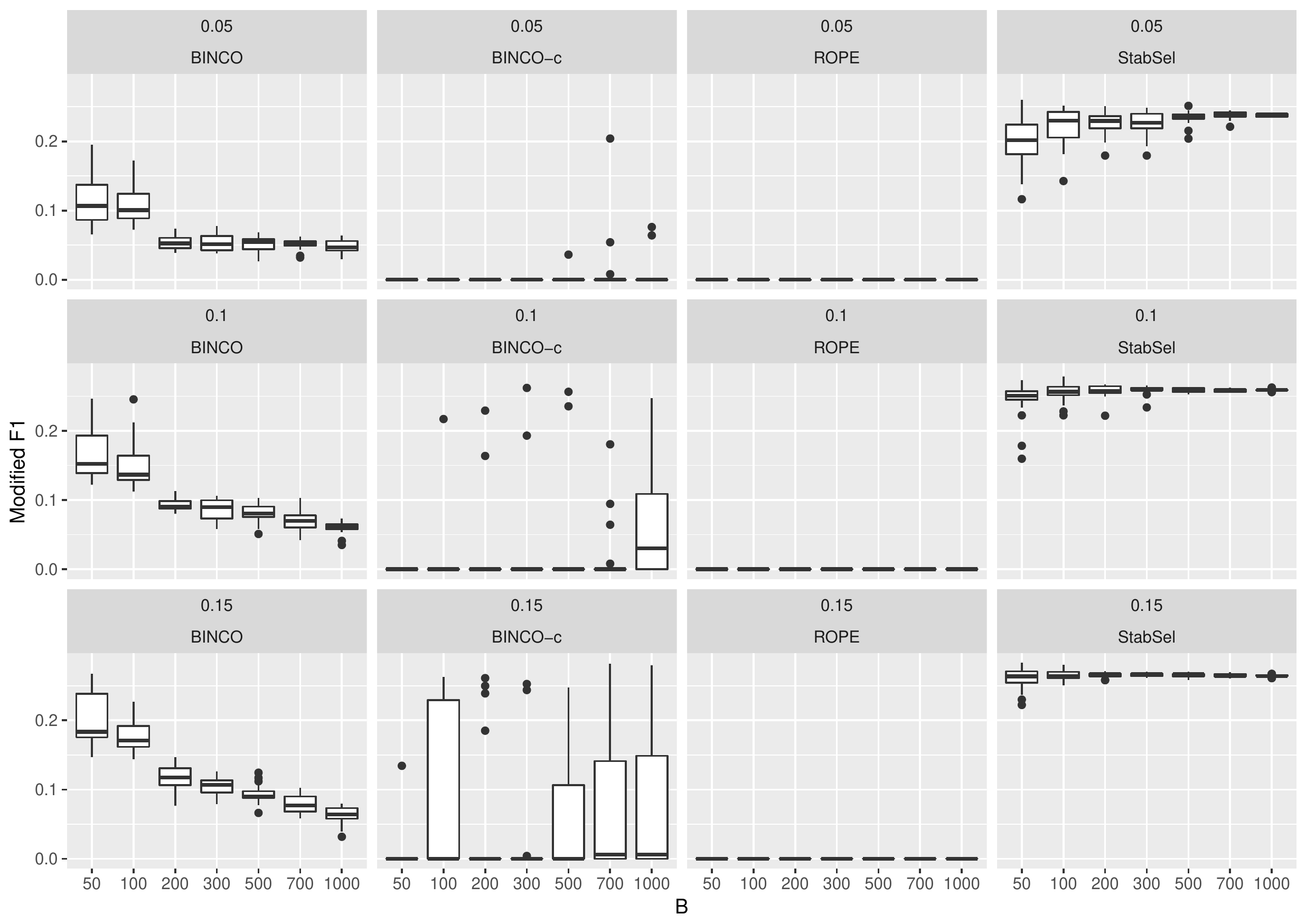}}
\caption{Network topology: hubby, steps:15, $n=200$, weakness=0.8 facet titles: target FDR and method.}
\label{sim18}
\end{figure}

\begin{figure}[p]
\centerline{\includegraphics[width=0.9\linewidth]{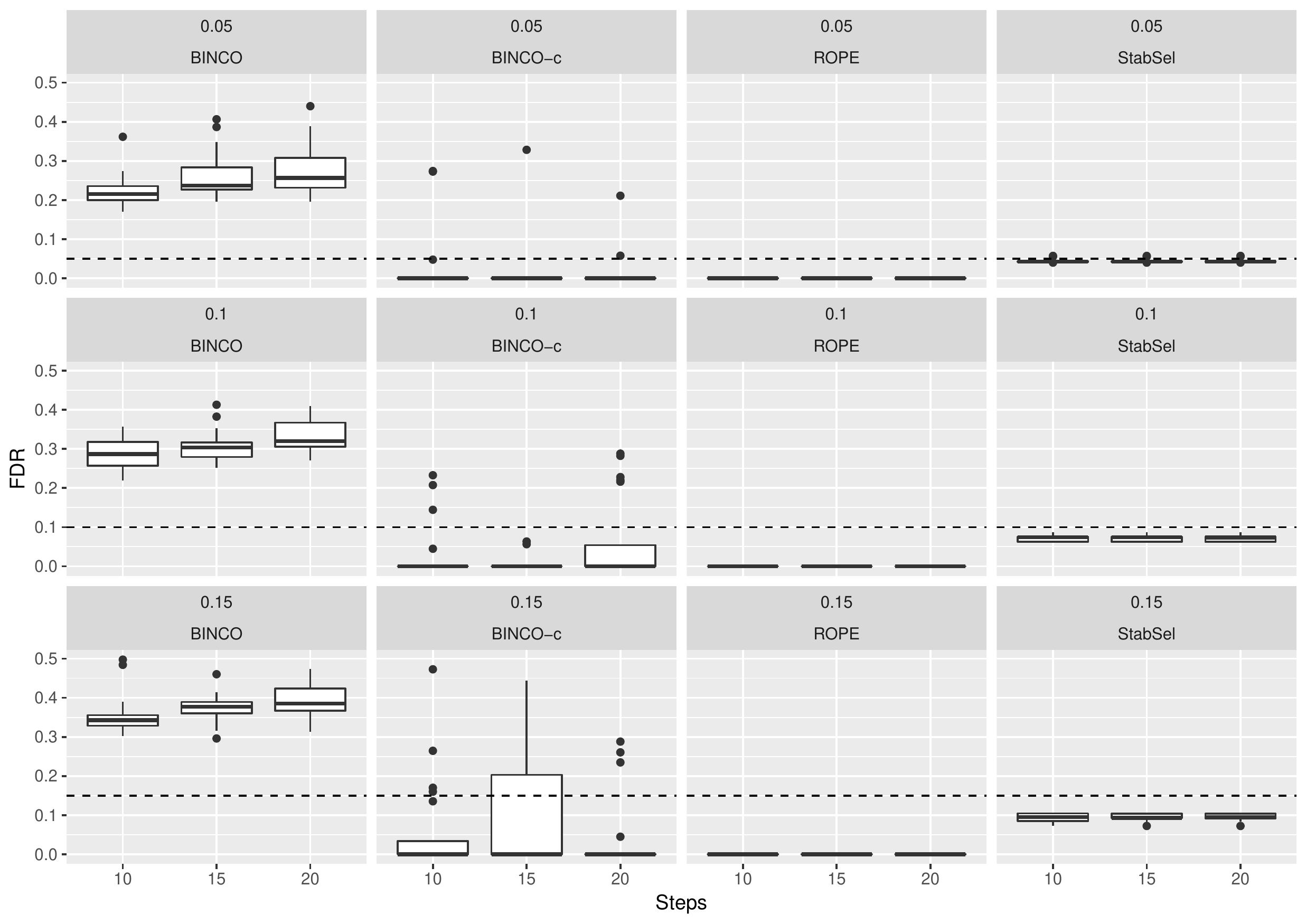}}
\caption{Network topology: hubby, $B=500$, $n=200$, weakness=0.8, facet titles: target FDR and method.}
\label{sim19}
\end{figure}

\begin{figure}[p]
\centerline{\includegraphics[width=0.9\linewidth]{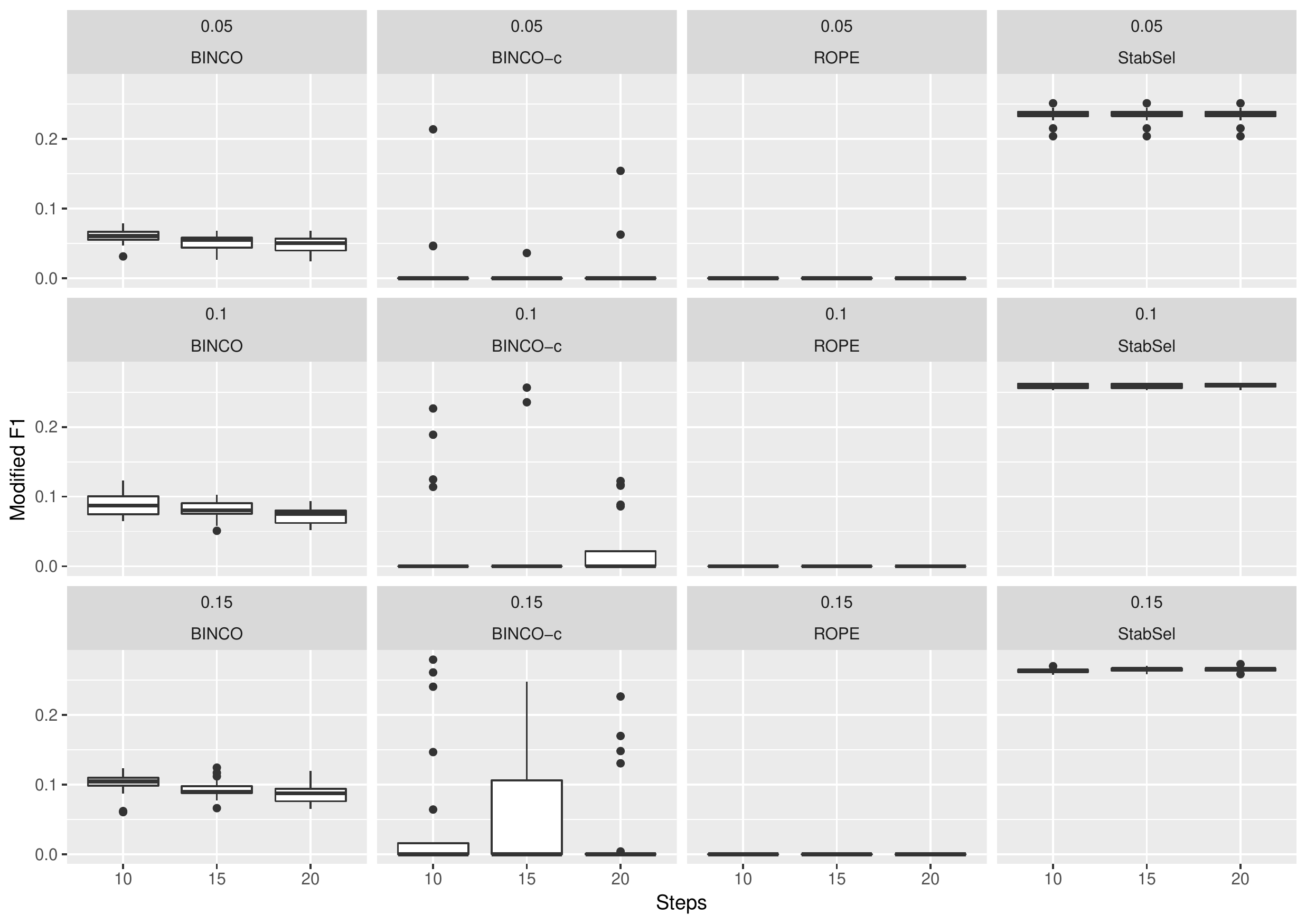}}
\caption{Network topology: hubby, $B=500$, $n=200$, weakness=0.8 facet titles: target FDR and method.}
\label{sim20}
\end{figure}

\begin{figure}[p]
\centerline{\includegraphics[width=0.9\linewidth]{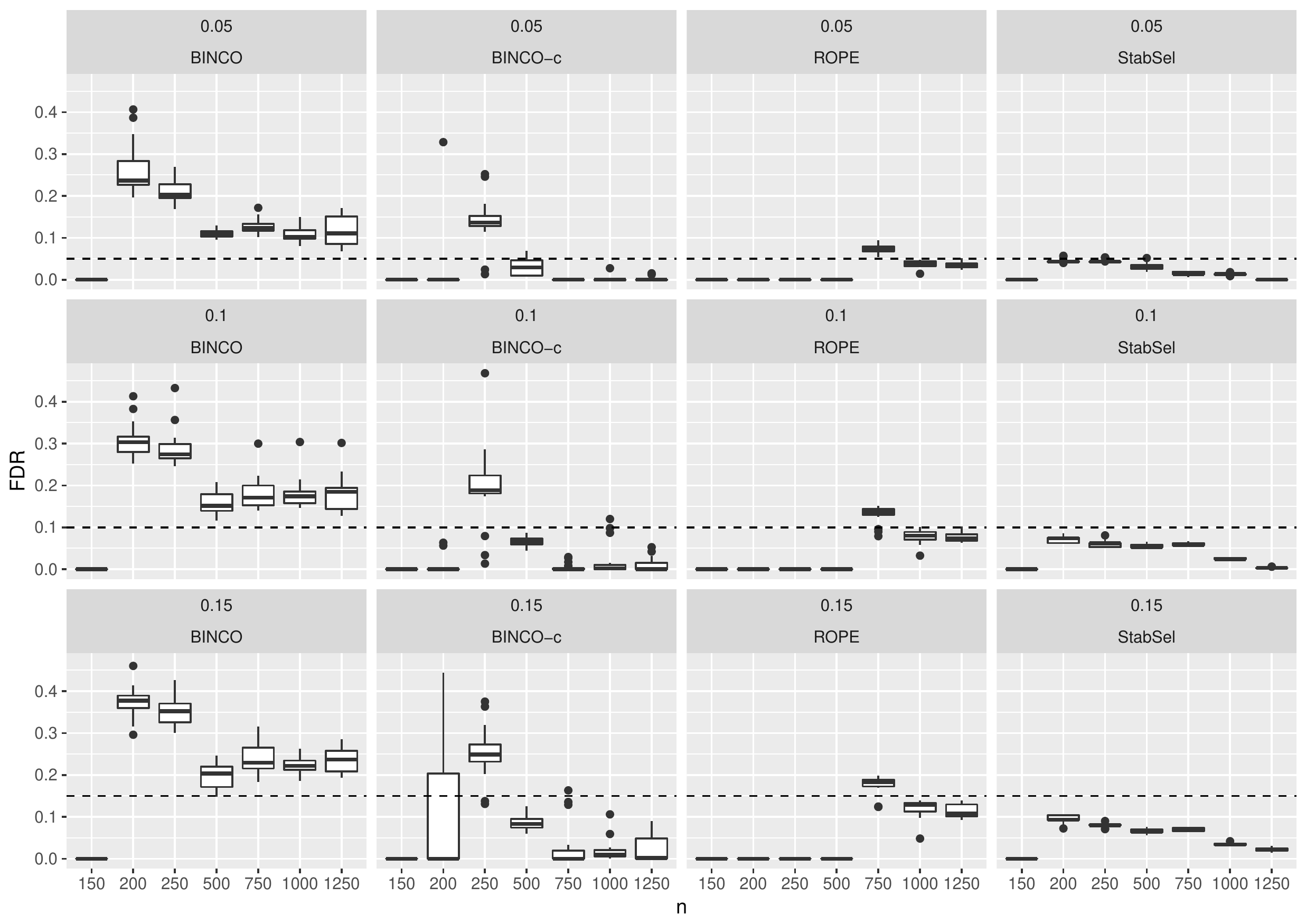}}
\caption{Network topology: hubby, $B=500$, steps:15, weakness=0.8, facet titles: target FDR and method.}
\label{sim21}
\end{figure}

\begin{figure}[p]
\centerline{\includegraphics[width=0.9\linewidth]{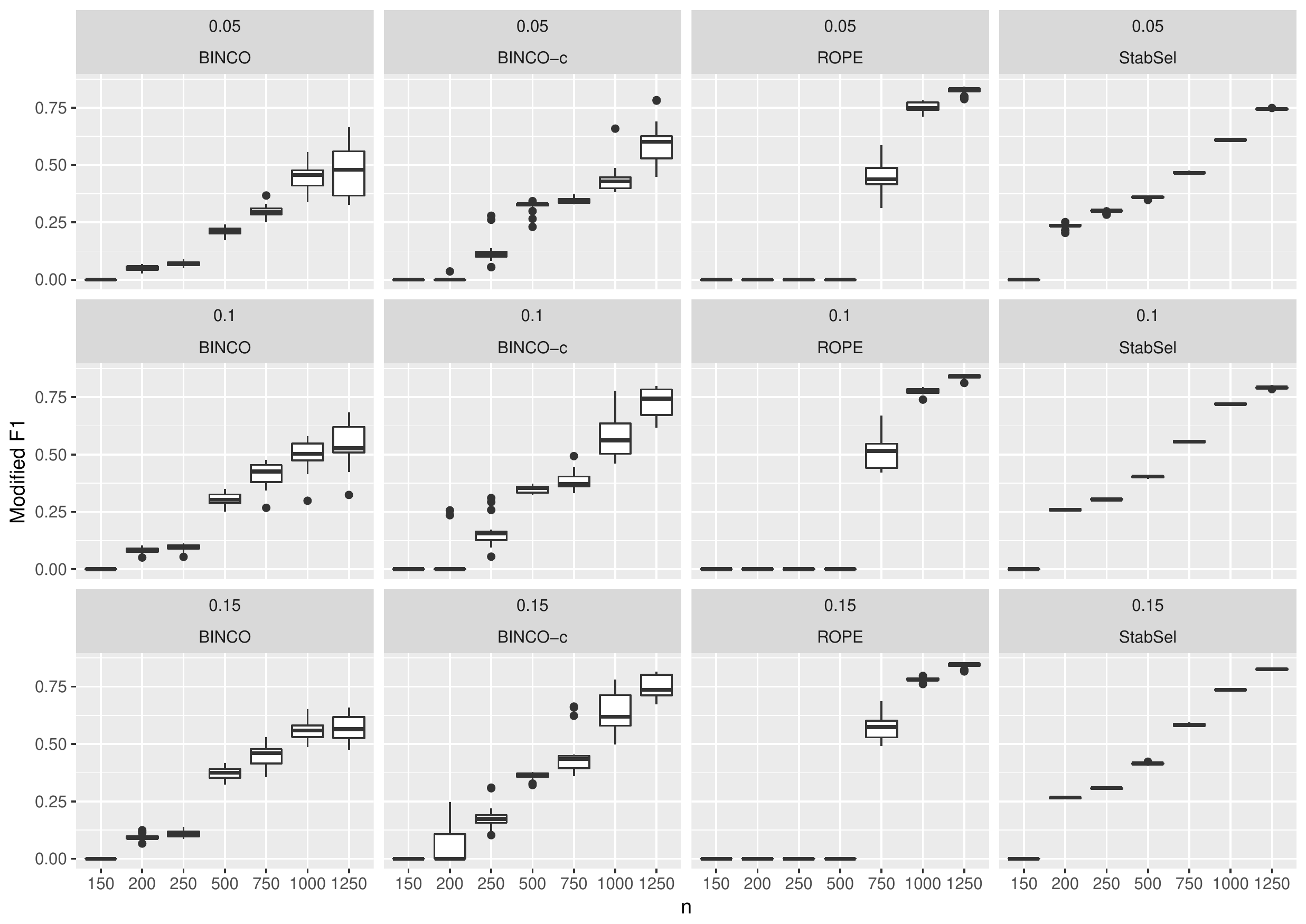}}
\caption{Network topology: hubby, $B=500$, steps:15, weakness=0.8, facet titles: target FDR and method.}
\label{sim22}
\end{figure}

\begin{figure}[p]
\centerline{\includegraphics[width=0.9\linewidth]{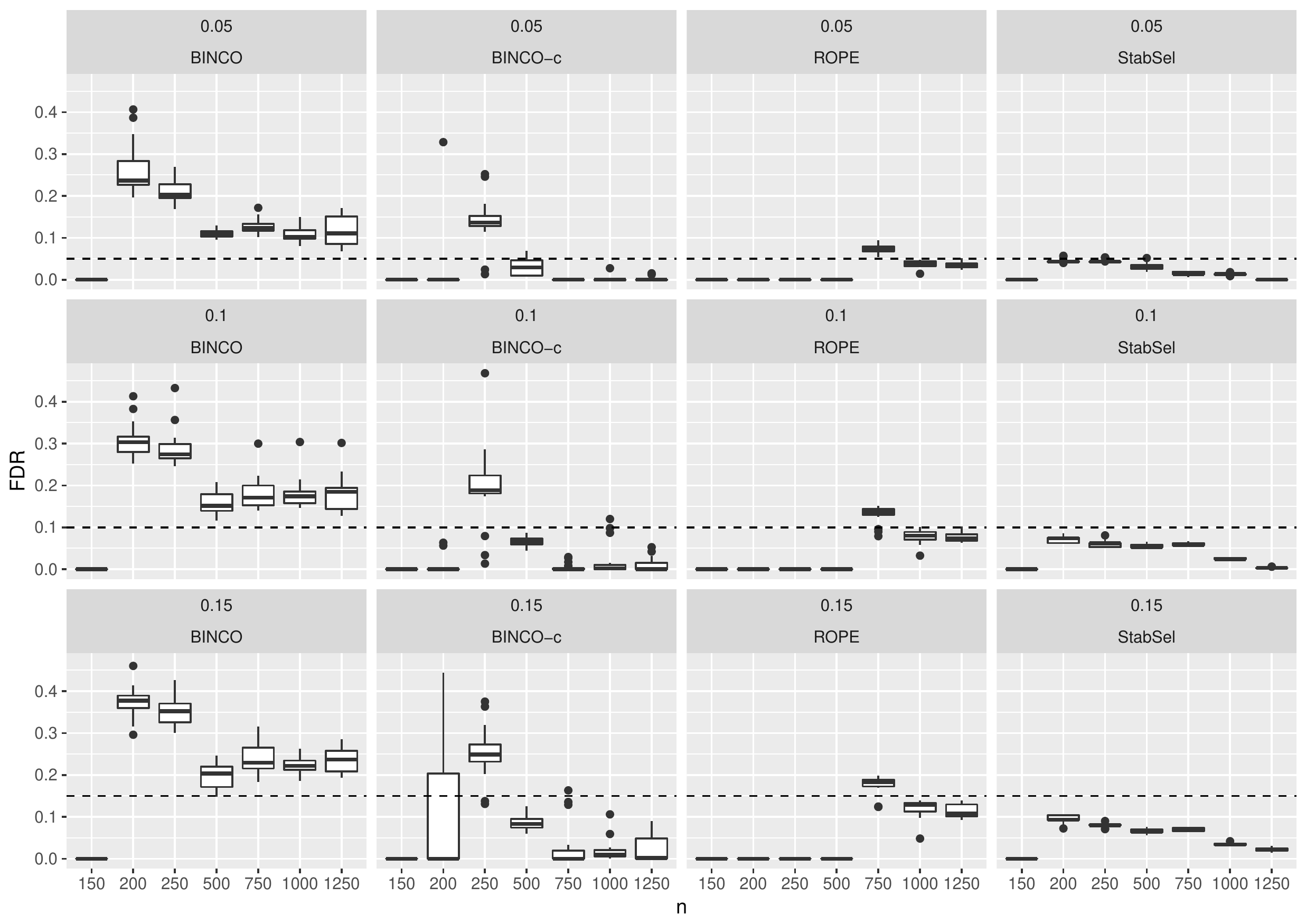}}
\caption{Network topology: hubby, $B=500$, steps:15, $n=200$, facet titles: target FDR and method.}
\label{sim23}
\end{figure}

\begin{figure}[p]
\centerline{\includegraphics[width=0.9\linewidth]{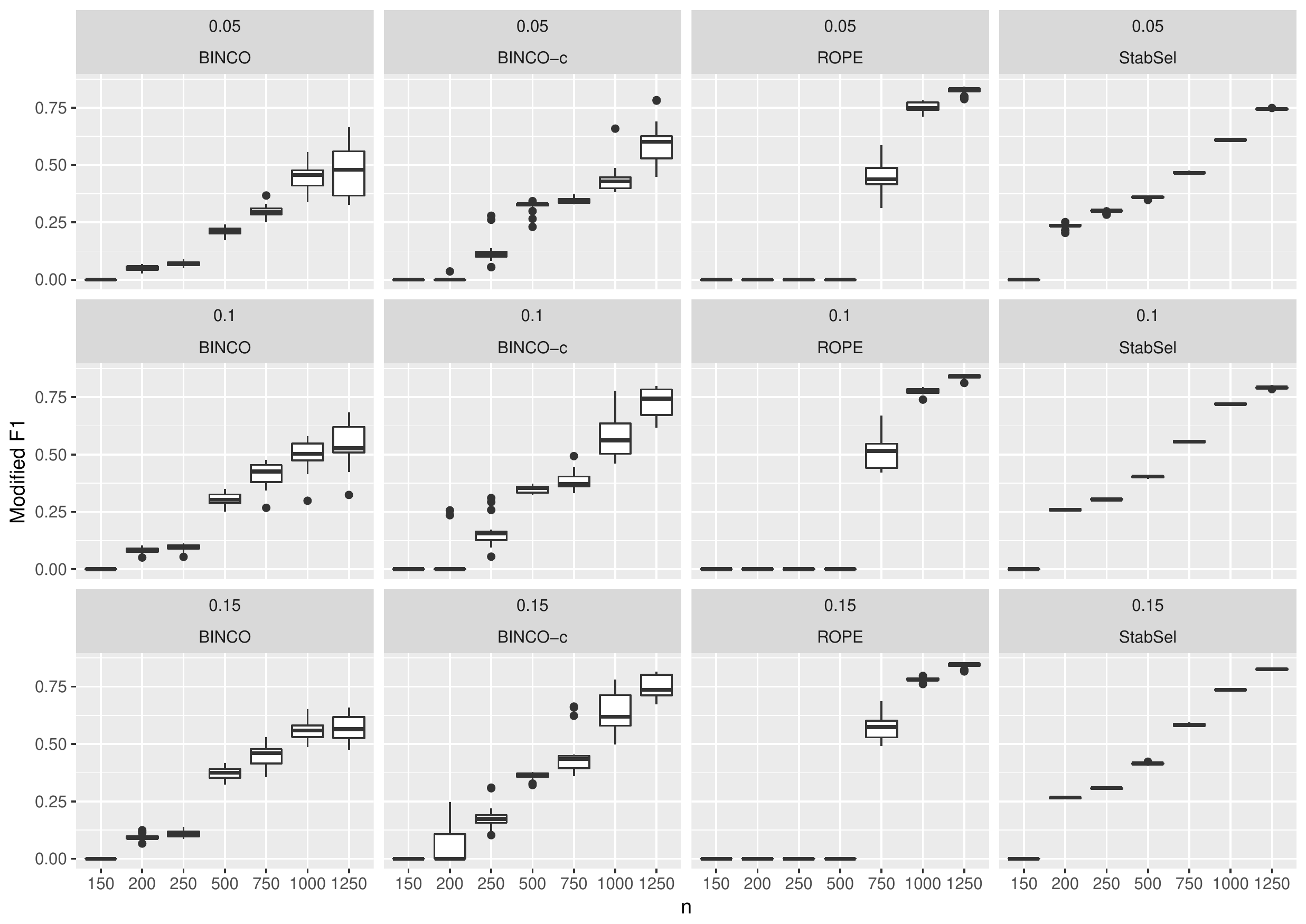}}
\caption{Network topology: hubby, $B=500$, steps:15, $n=200$, facet titles: target FDR and method.}
\label{sim24}
\end{figure}

\begin{figure}[p]
\centerline{\includegraphics[width=0.9\linewidth]{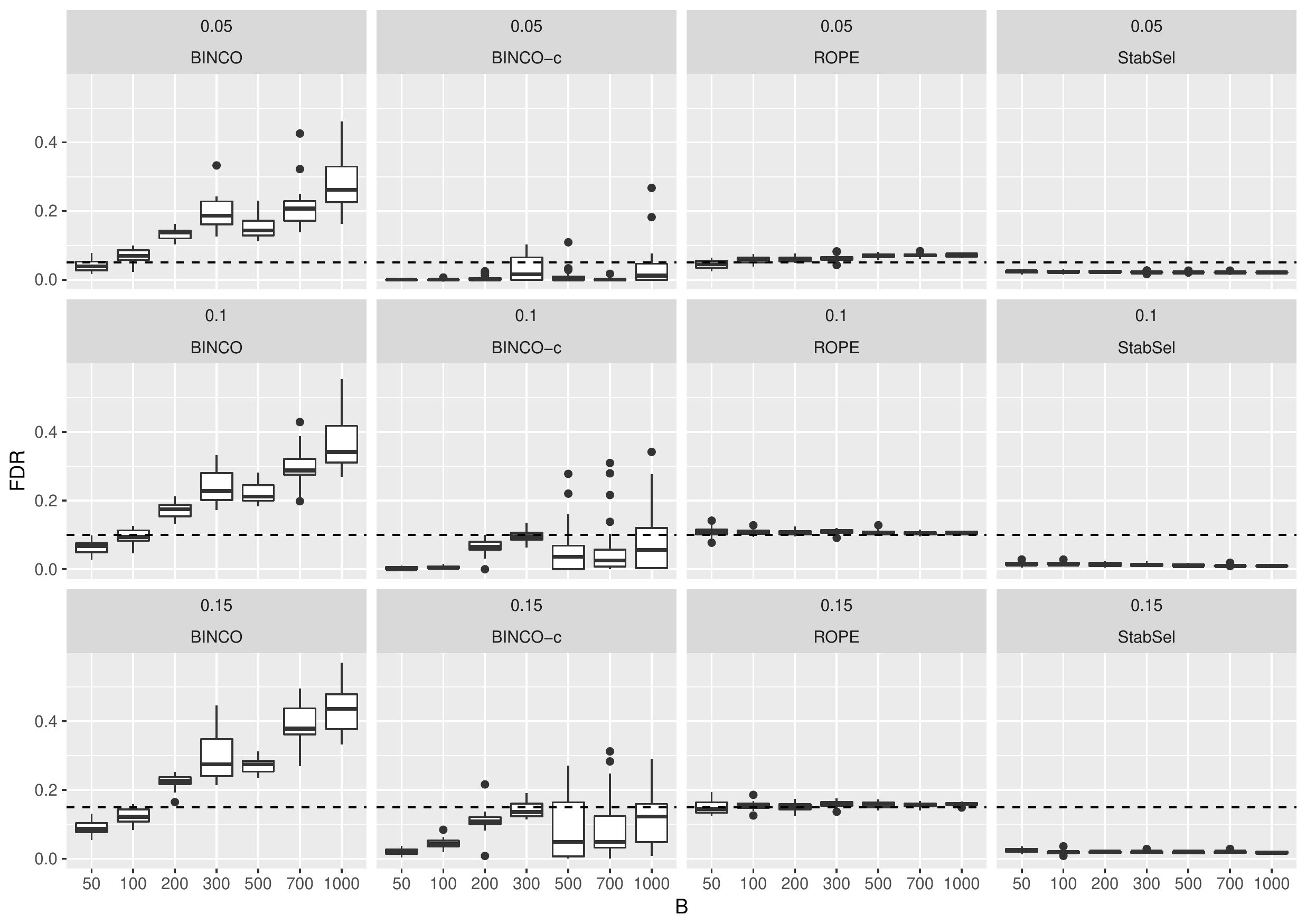}}
\caption{Network topology: scale-free, steps:15, $n=200$, weakness=0.8, facet titles: target FDR and method.}
\label{sim25}
\end{figure}

\begin{figure}[p]
\centerline{\includegraphics[width=0.9\linewidth]{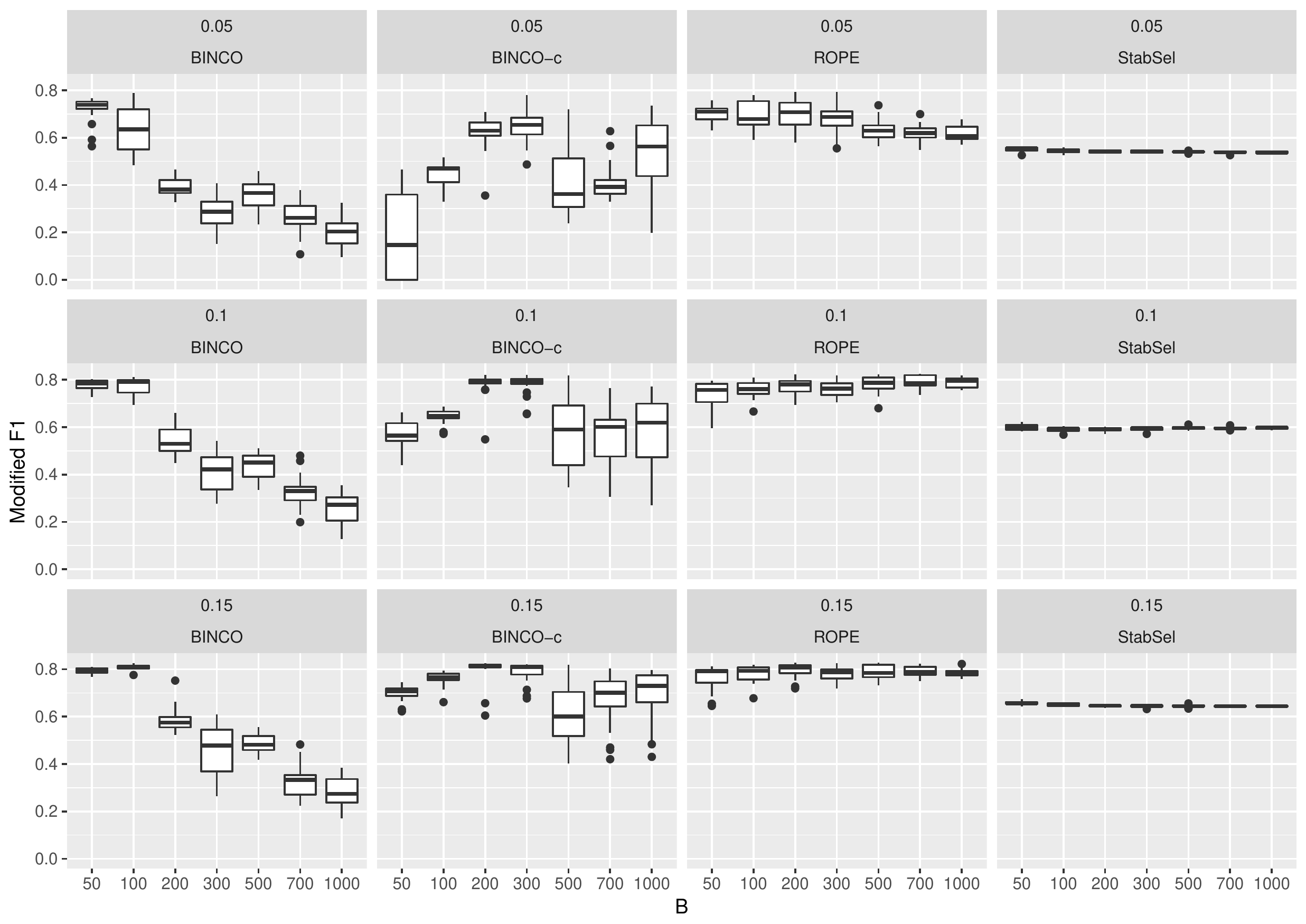}}
\caption{Network topology: scale-free, steps:15, $n=200$, weakness=0.8, facet titles: target FDR and method.}
\label{sim26}
\end{figure}

\begin{figure}[p]
\centerline{\includegraphics[width=0.9\linewidth]{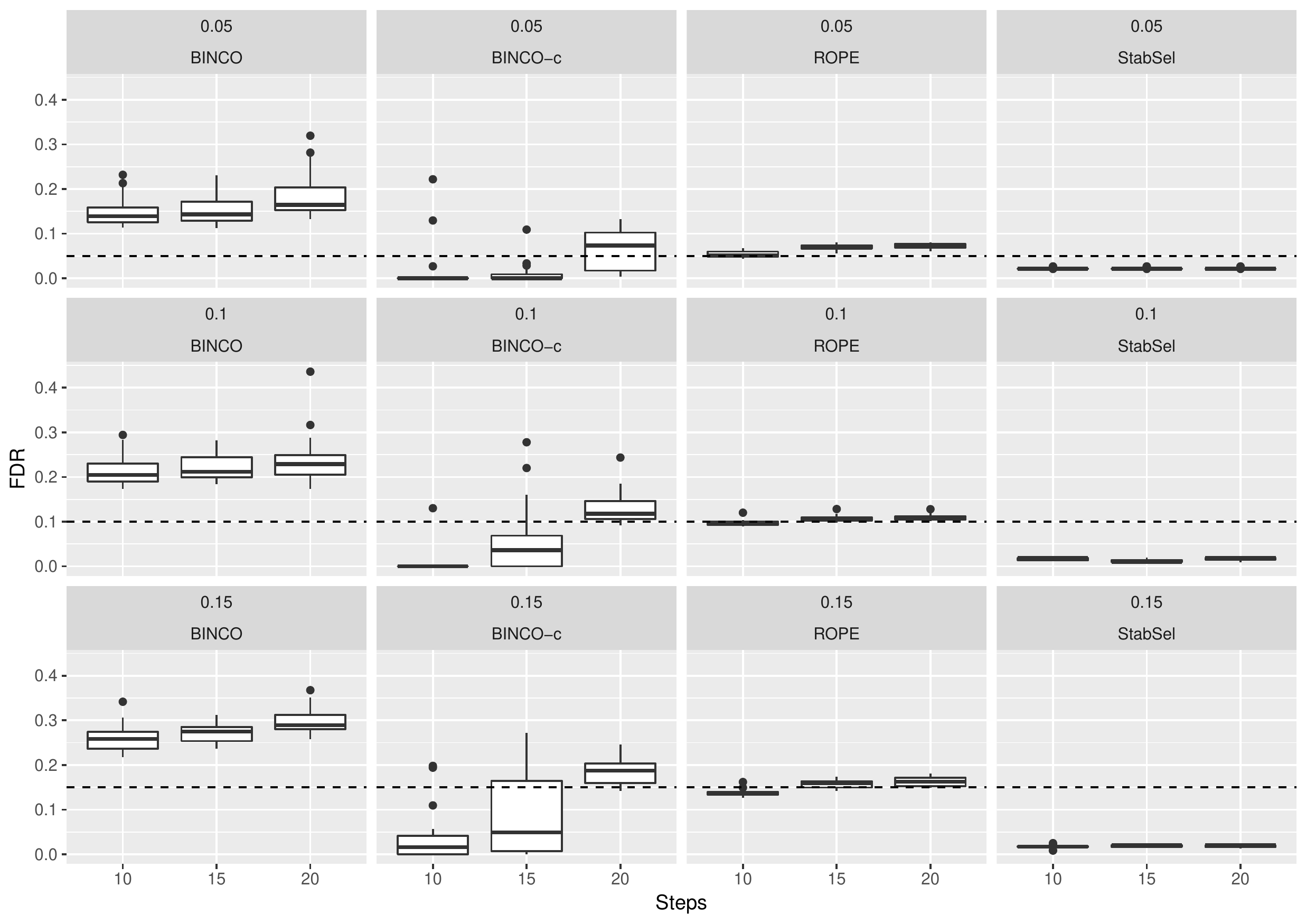}}
\caption{Network topology: scale-free, $B=500$, $n=200$, weakness=0.8, facet titles: target FDR and method.}
\label{sim27}
\end{figure}

\begin{figure}[p]
\centerline{\includegraphics[width=0.9\linewidth]{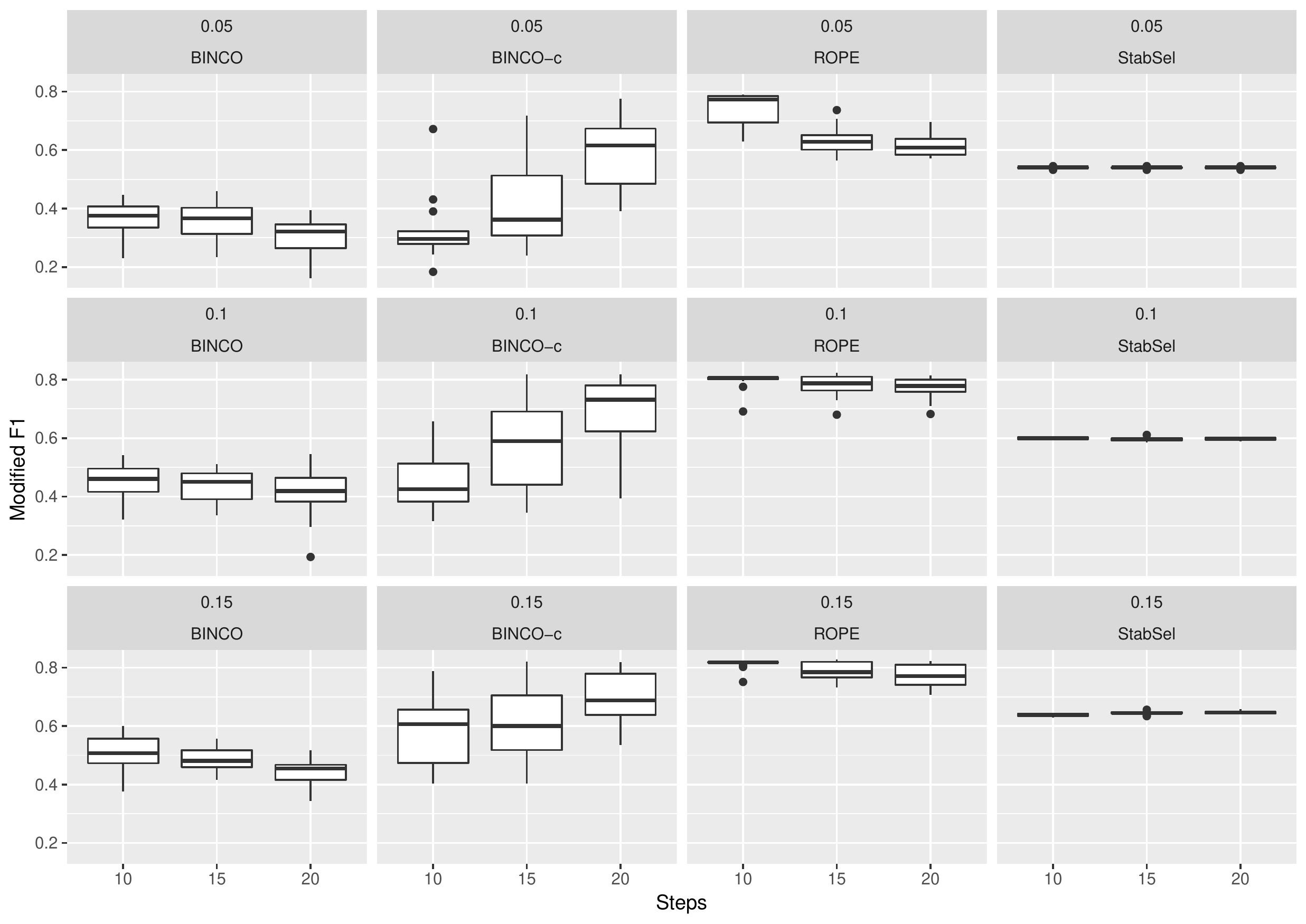}}
\caption{Network topology: scale-free, $B=500$, $n=200$, weakness=0.8, facet titles: target FDR and method.}
\label{sim28}
\end{figure}

\begin{figure}[p]
\centerline{\includegraphics[width=0.9\linewidth]{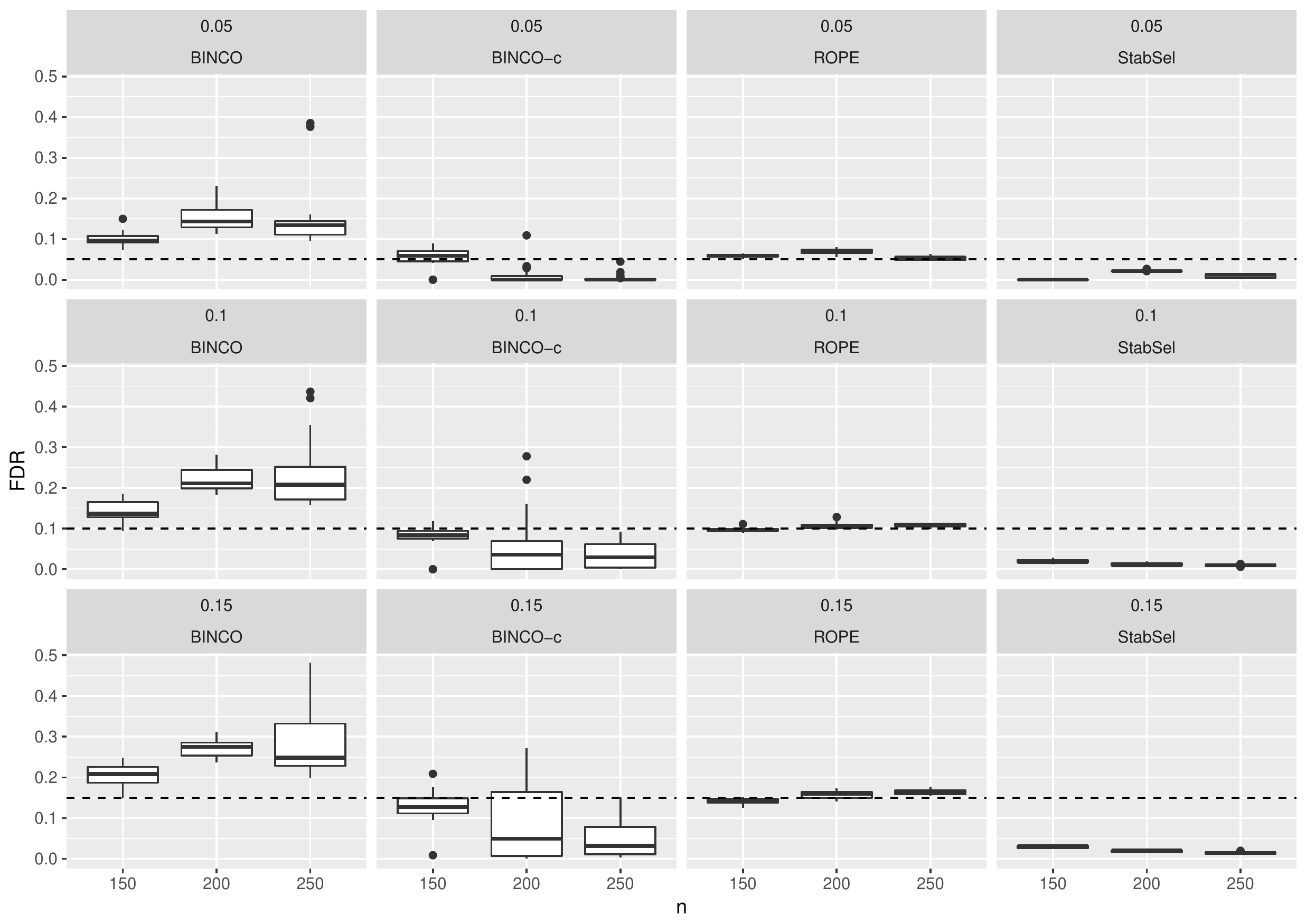}}
\caption{Network topology: scale-free, $B=500$, steps:15, weakness=0.8, facet titles: target FDR and method.}
\label{sim29}
\end{figure}

\begin{figure}[p]
\centerline{\includegraphics[width=0.9\linewidth]{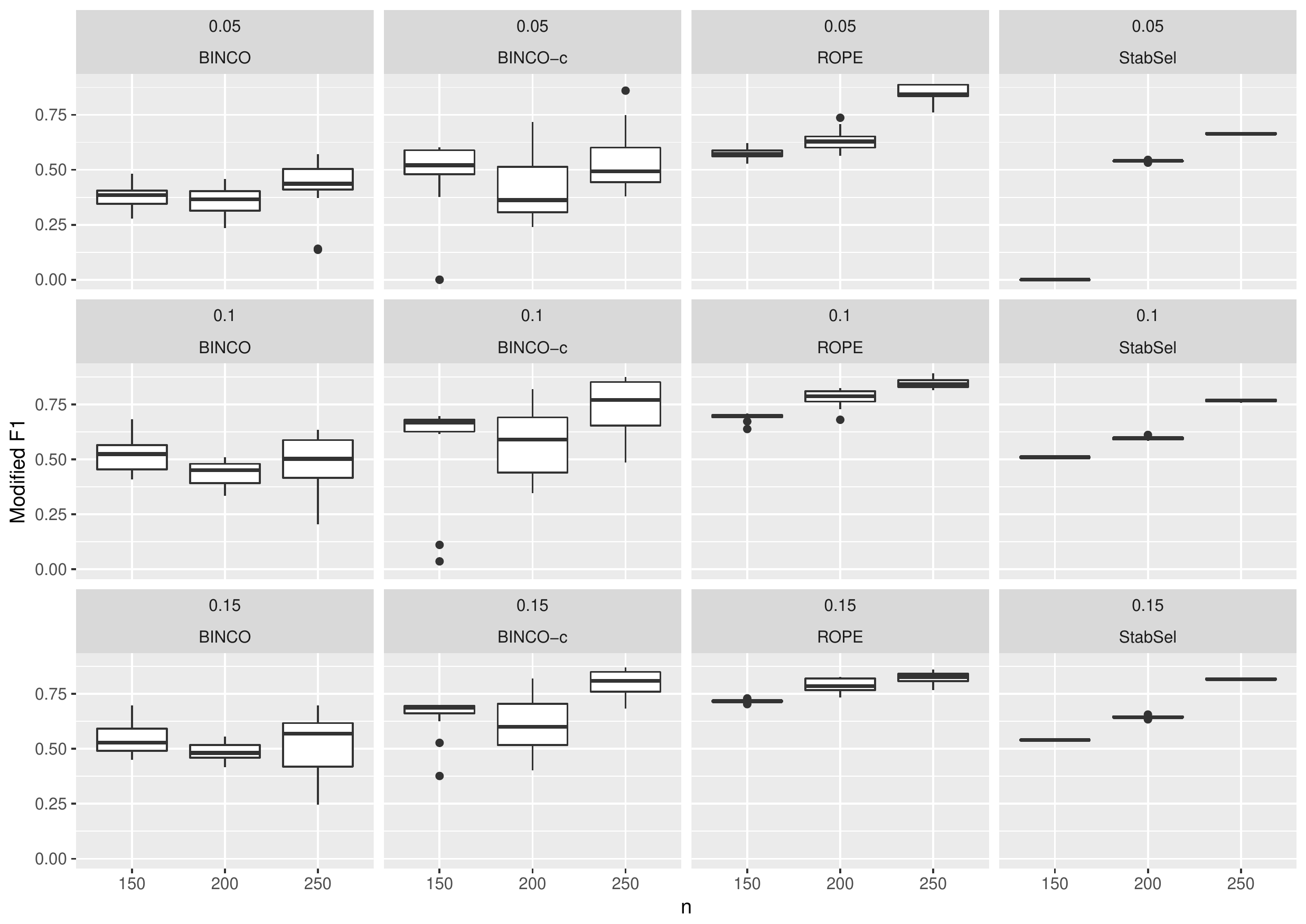}}
\caption{Network topology: scale-free, $B=500$, steps:15, weakness=0.8, facet titles: target FDR and method.}
\label{sim30}
\end{figure}

\begin{figure}[p]
\centerline{\includegraphics[width=0.9\linewidth]{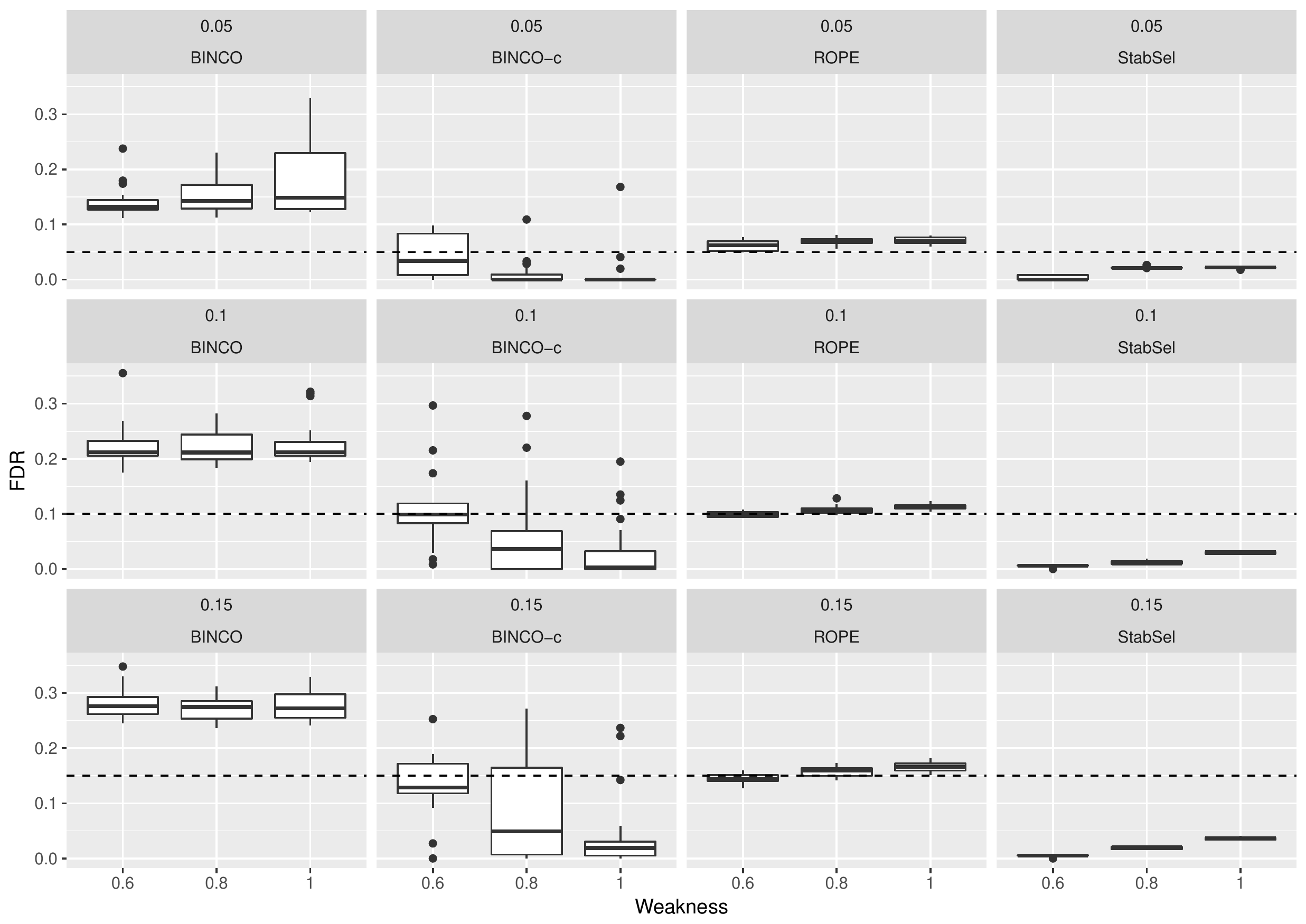}}
\caption{Network topology: scale-free, $B=500$, steps:15, $n=200$, facet titles: target FDR and method.}
\label{sim31}
\end{figure}

\begin{figure}[p]
\centerline{\includegraphics[width=0.9\linewidth]{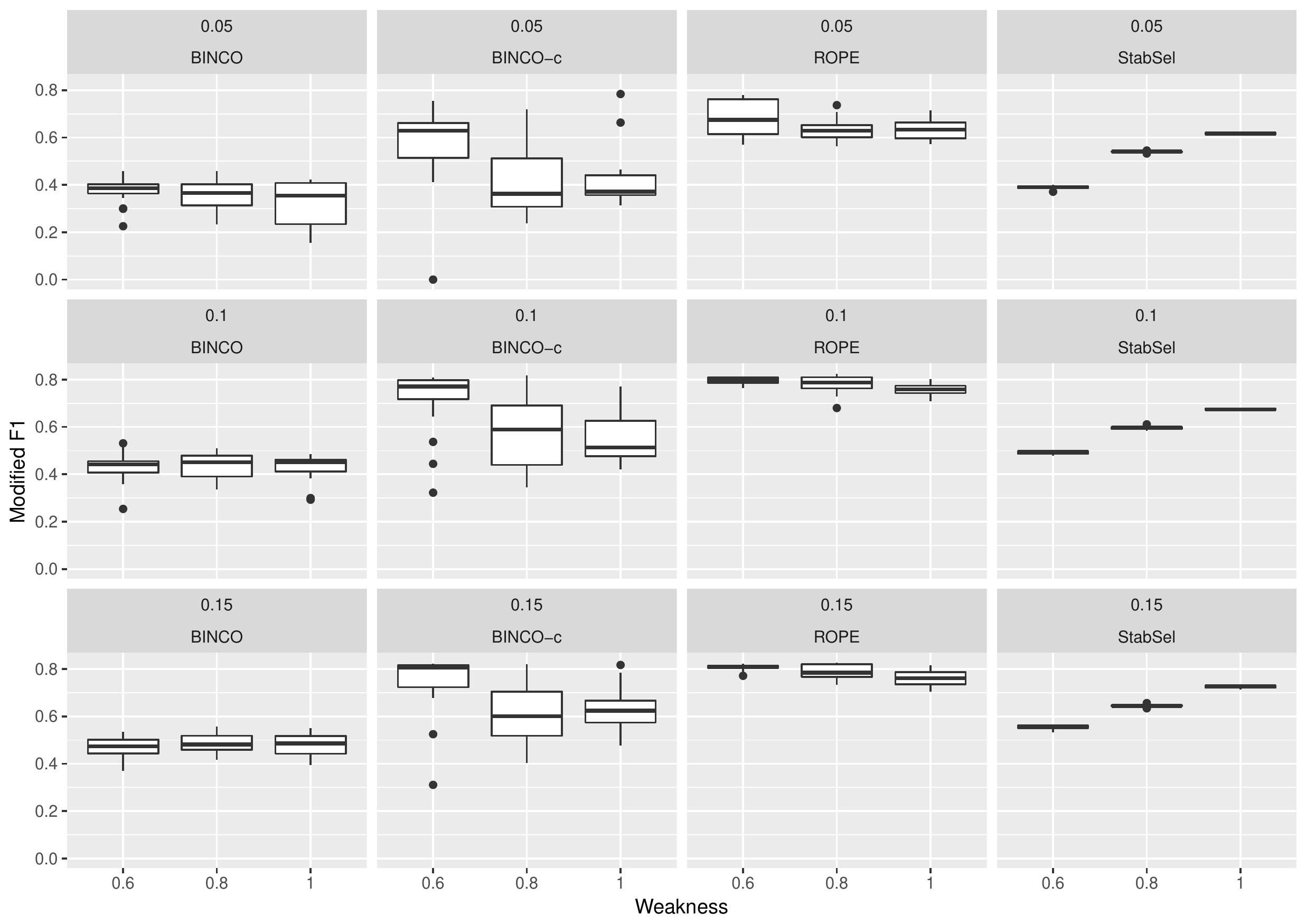}}
\caption{Network topology: scale-free, $B=500$, steps:15, $n=200$, facet titles: target FDR and method.}
\label{sim32}
\end{figure}

\begin{figure}[p]
\centerline{\includegraphics[width=0.9\linewidth]{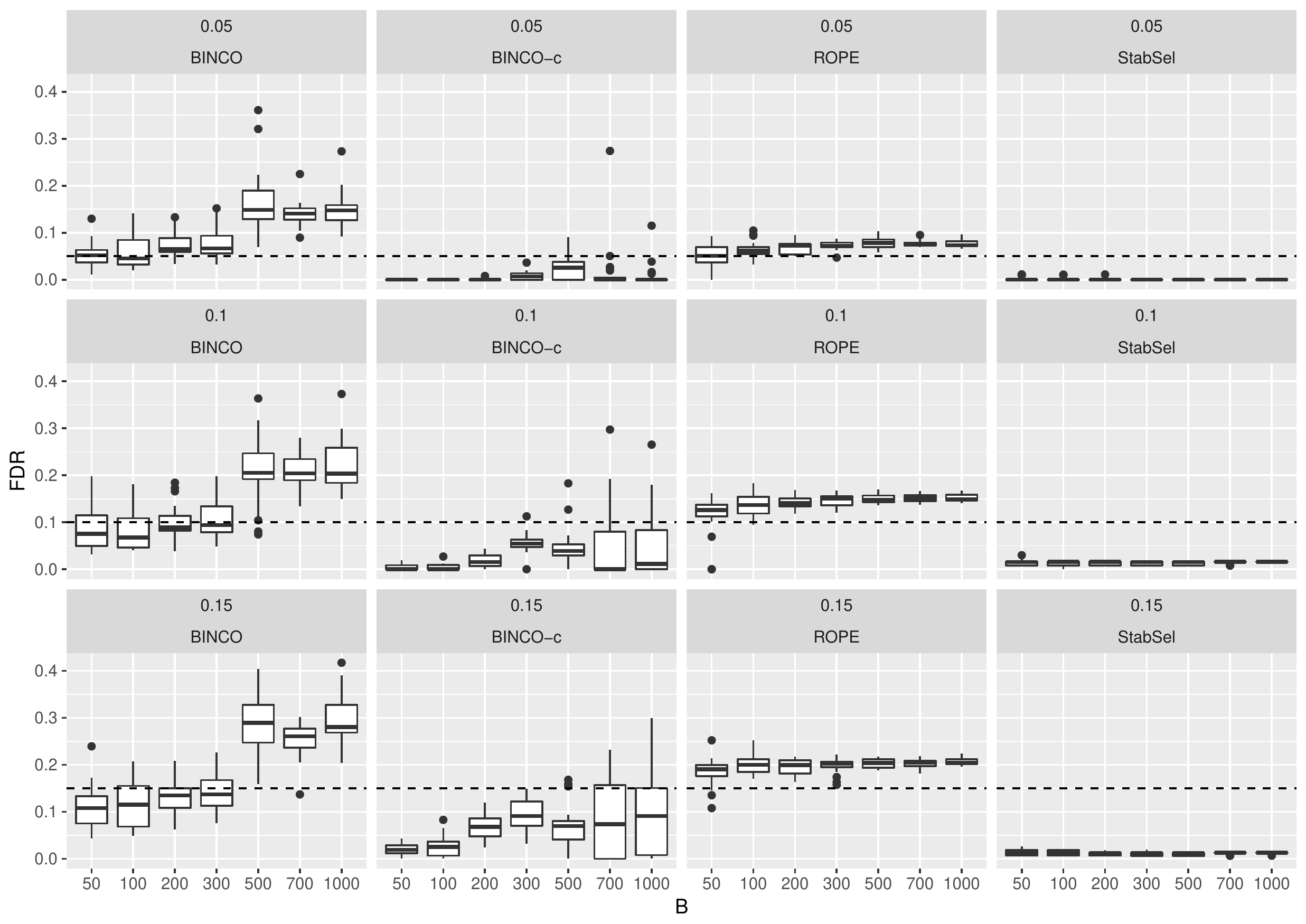}}
\caption{Network topology: small,  steps:15, $n=200$, weakness=0.8, facet titles: target FDR and method.}
\label{sim33}
\end{figure}

\begin{figure}[p]
\centerline{\includegraphics[width=0.9\linewidth]{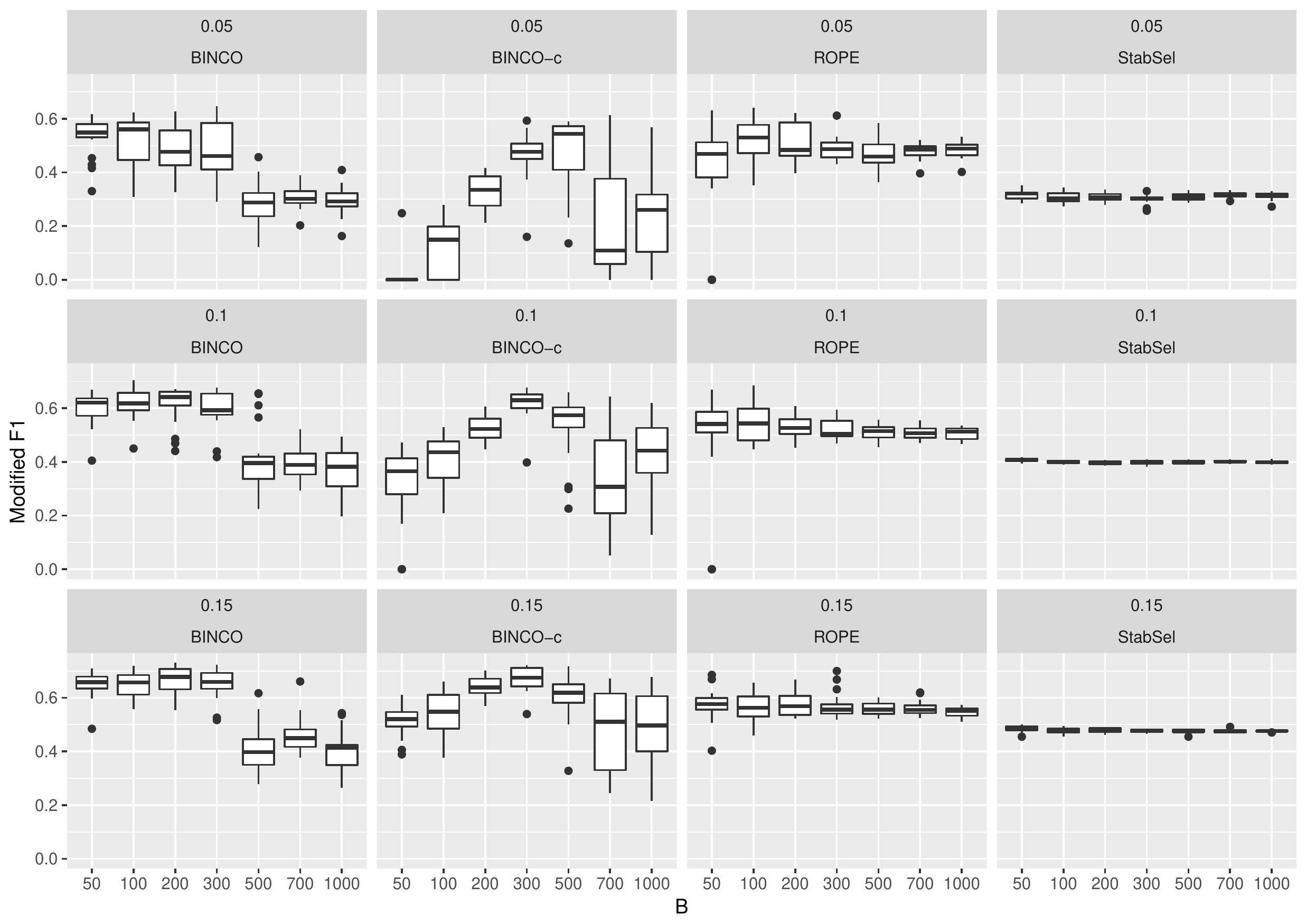}}
\caption{Network topology: small,  steps:15, $n=200$, weakness=0.8, facet titles: target FDR and method.}
\label{sim34}
\end{figure}

\begin{figure}[p]
\centerline{\includegraphics[width=0.9\linewidth]{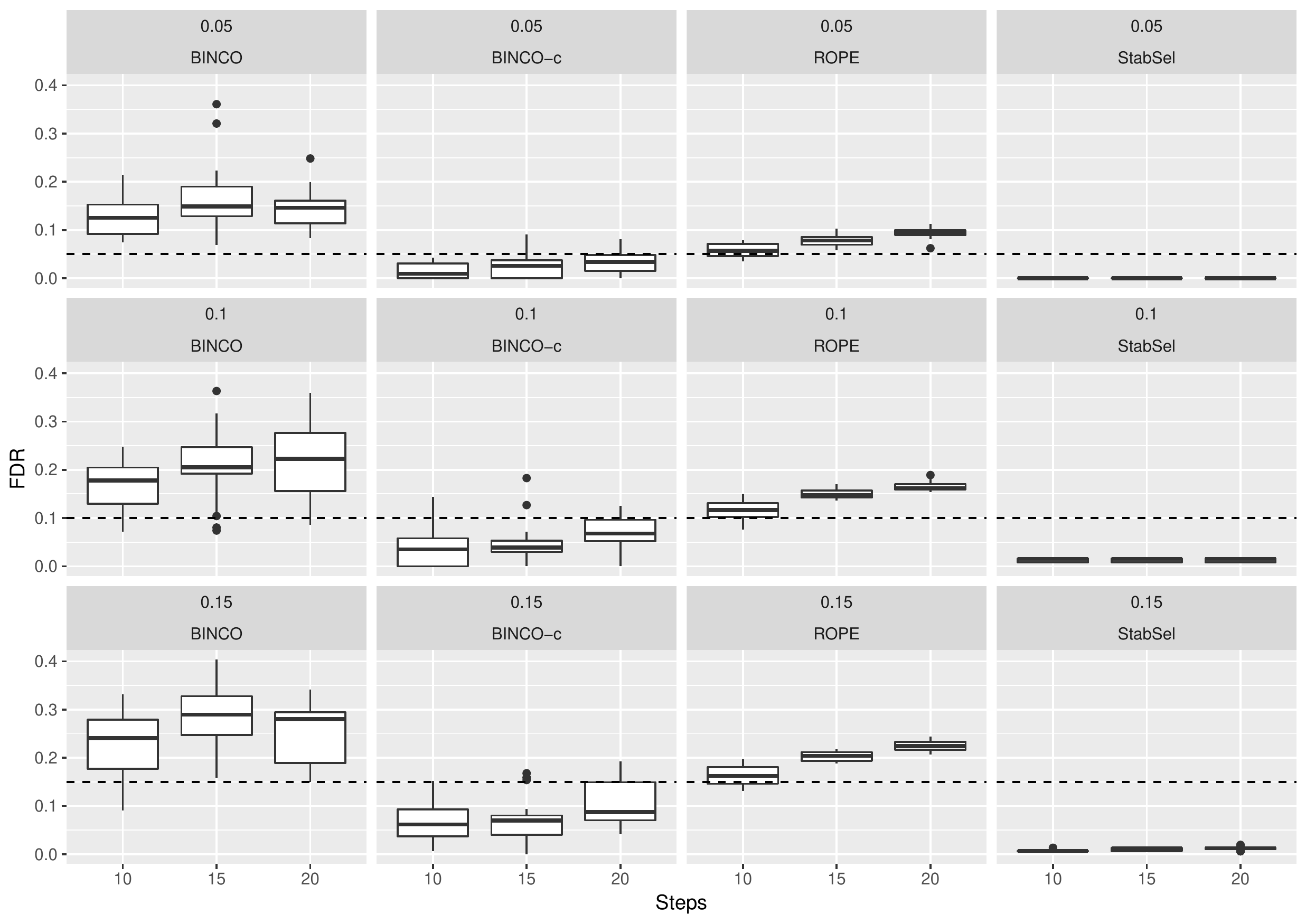}}
\caption{Network topology: small, $B=500$, $n=200$, weakness=0.8, facet titles: target FDR and method.}
\label{sim35}
\end{figure}

\begin{figure}[p]
\centerline{\includegraphics[width=0.9\linewidth]{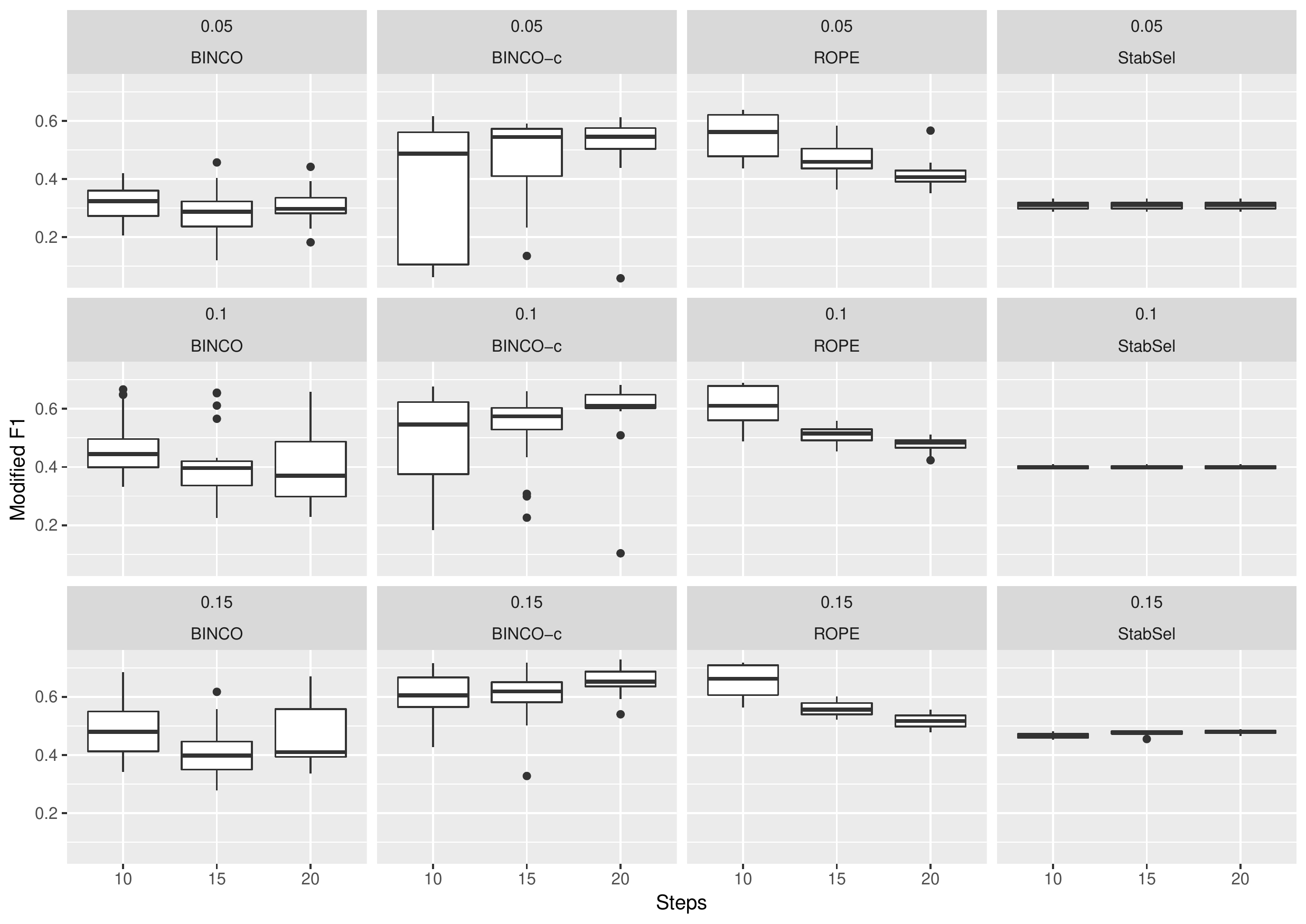}}
\caption{Network topology: small,  $B=500$,  $n=200$, weakness=0.8, facet titles: target FDR and method.}
\label{sim36}
\end{figure}

\begin{figure}[p]
\centerline{\includegraphics[width=0.9\linewidth]{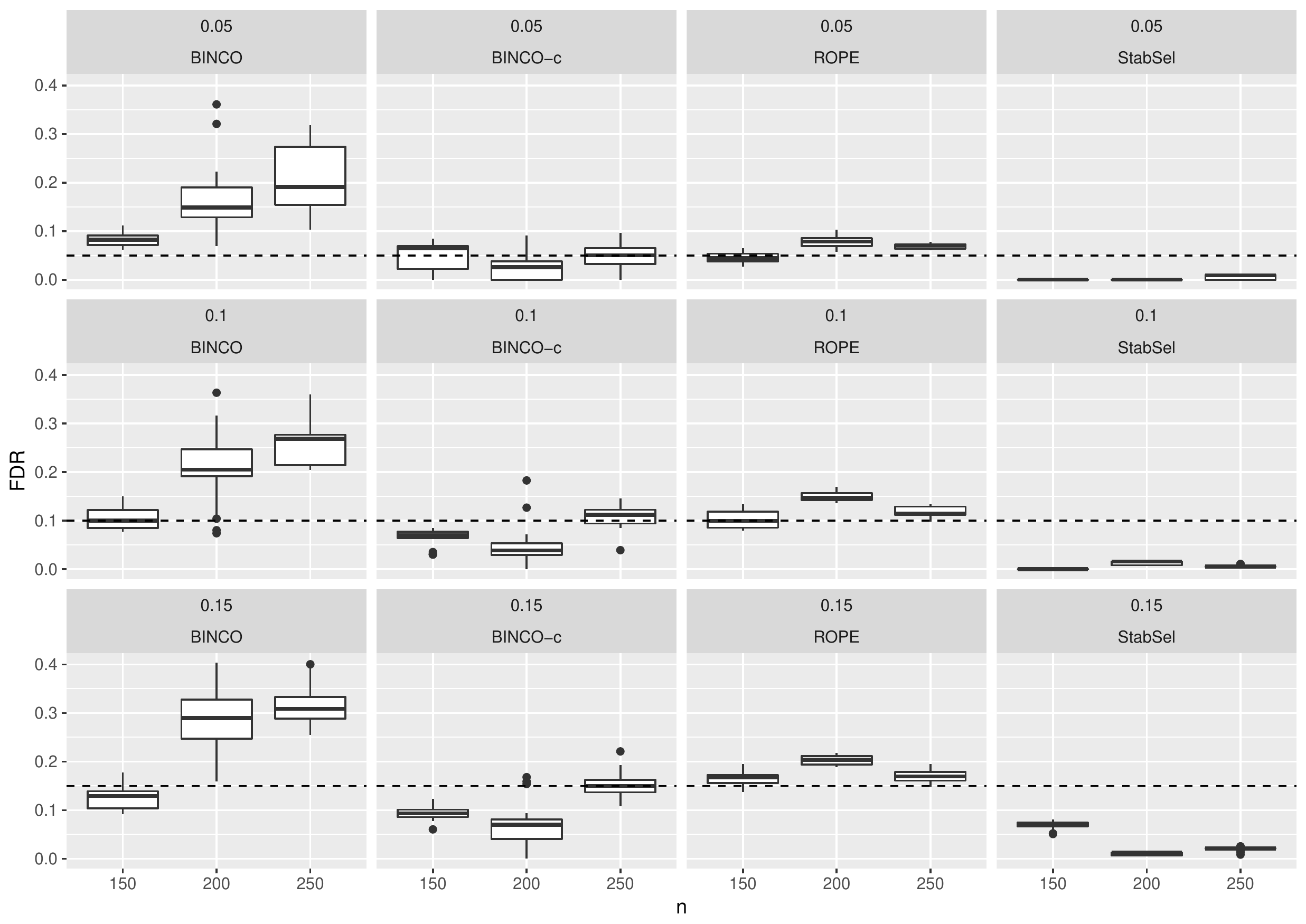}}
\caption{Network topology: small, $B=500$, steps:15, weakness=0.8, facet titles: target FDR and method.}
\label{sim37}
\end{figure}

\begin{figure}[p]
\centerline{\includegraphics[width=0.9\linewidth]{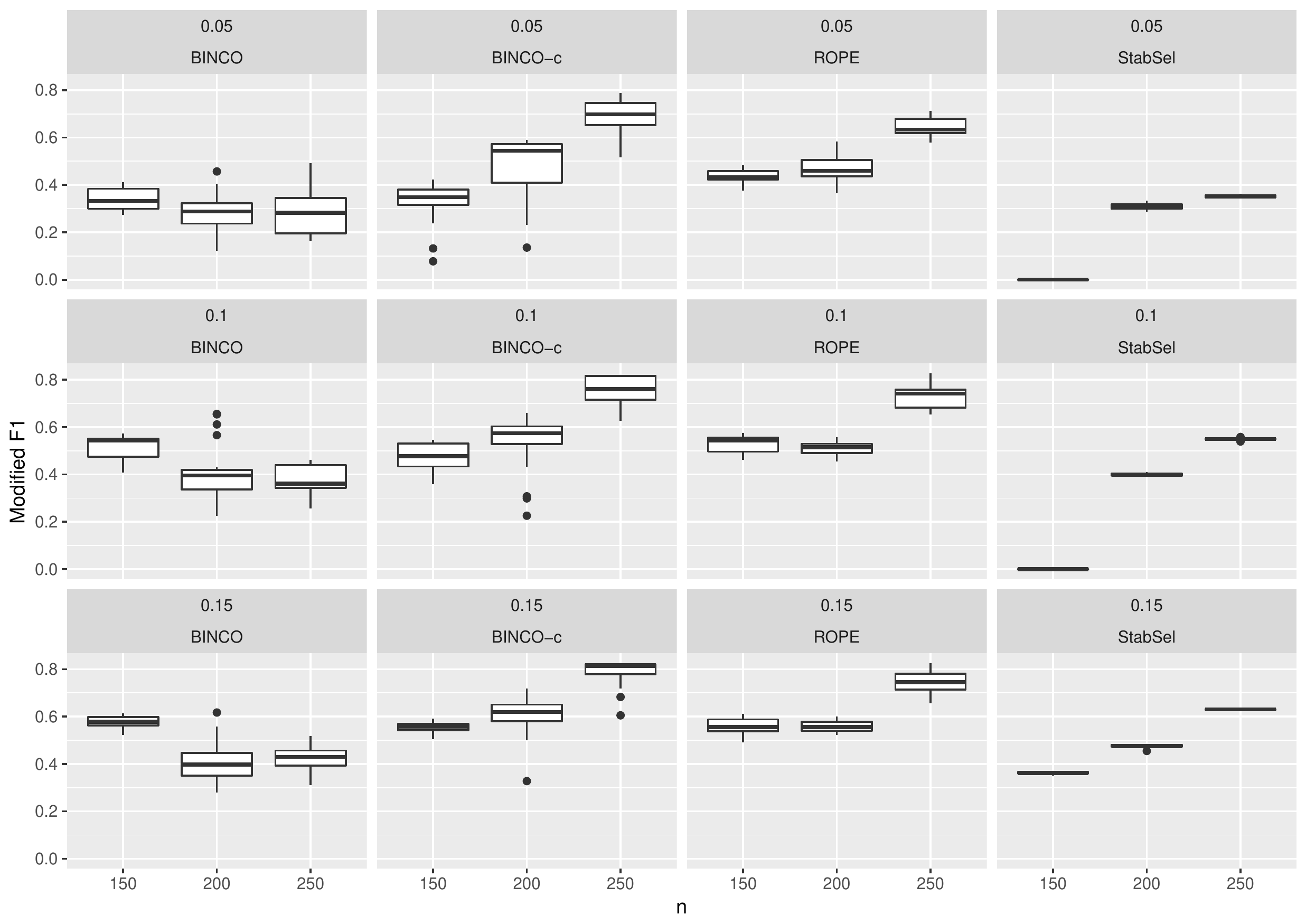}}
\caption{Network topology: small,  $B=500$,  steps:15, weakness=0.8, facet titles: target FDR and method.}
\label{sim38}
\end{figure}

\begin{figure}[p]
\centerline{\includegraphics[width=0.9\linewidth]{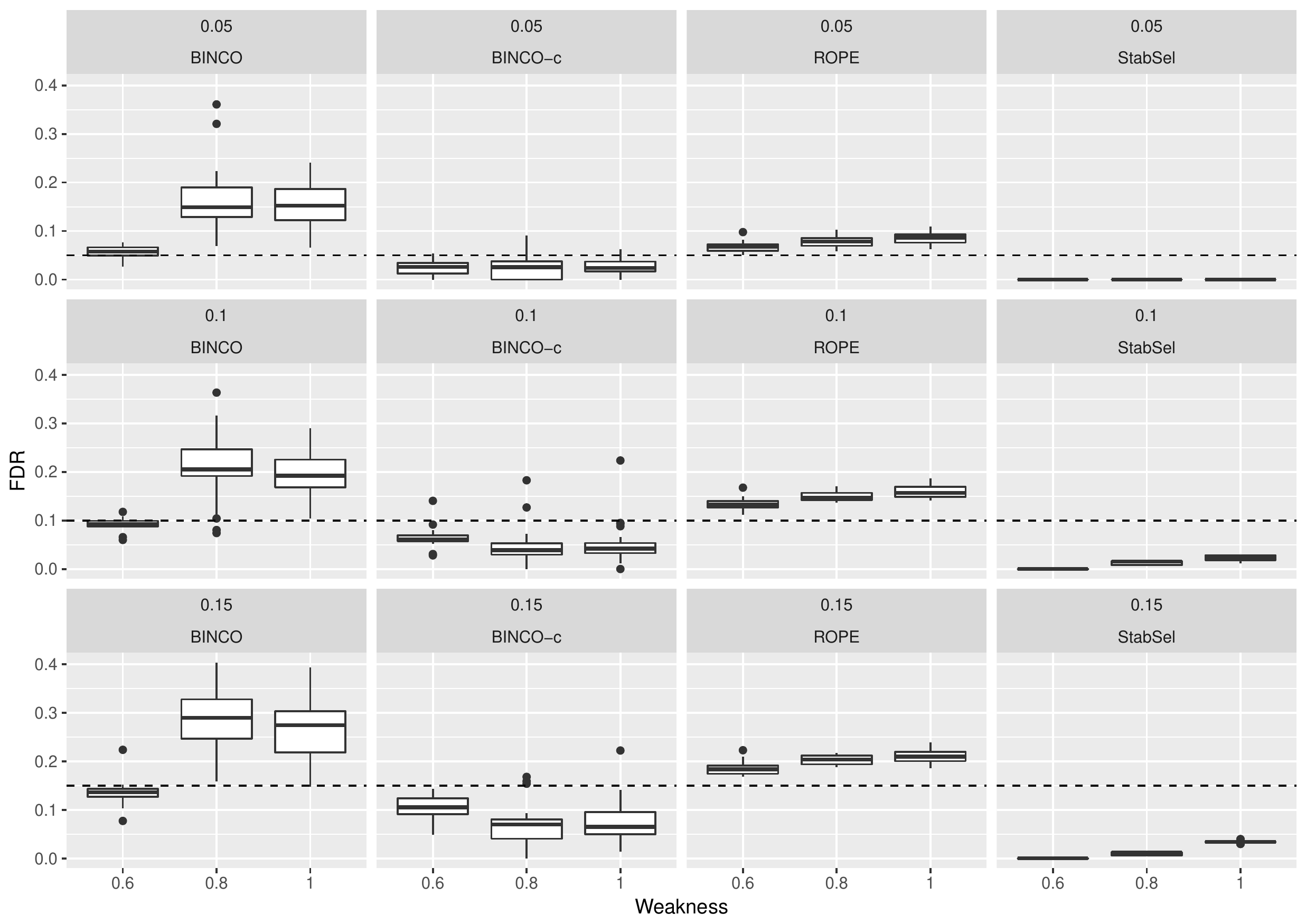}}
\caption{Network topology: small, $B=500$, steps:15, $n=200$, facet titles: target FDR and method.}
\label{sim39}
\end{figure}

\begin{figure}[p]
\centerline{\includegraphics[width=0.9\linewidth]{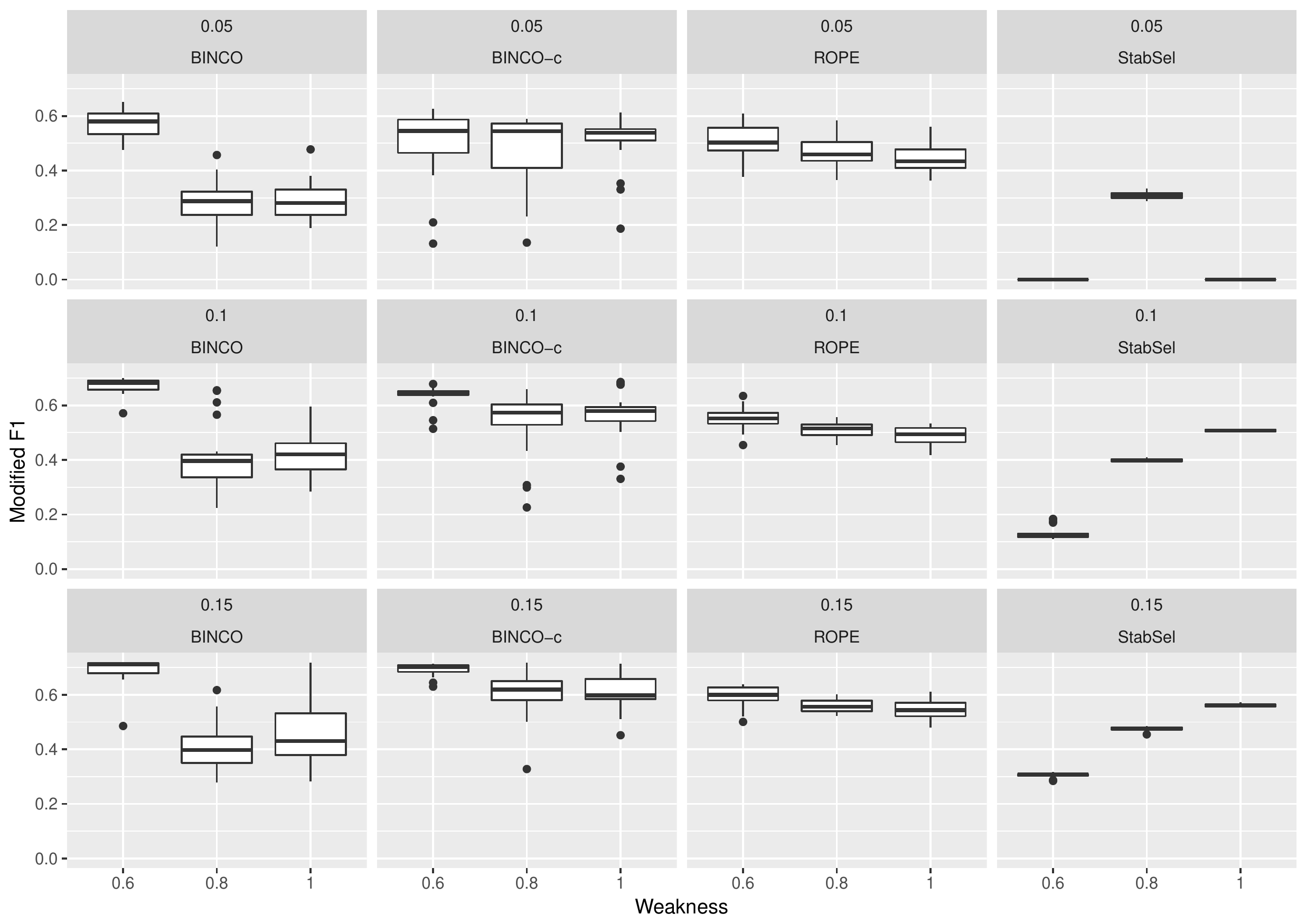}}
\caption{Network topology: small,  $B=500$,  steps:15, $n=200$, facet titles: target FDR and method.}
\label{sim40}
\end{figure}

\begin{figure}[p]
\centerline{\includegraphics[width=0.9\linewidth]{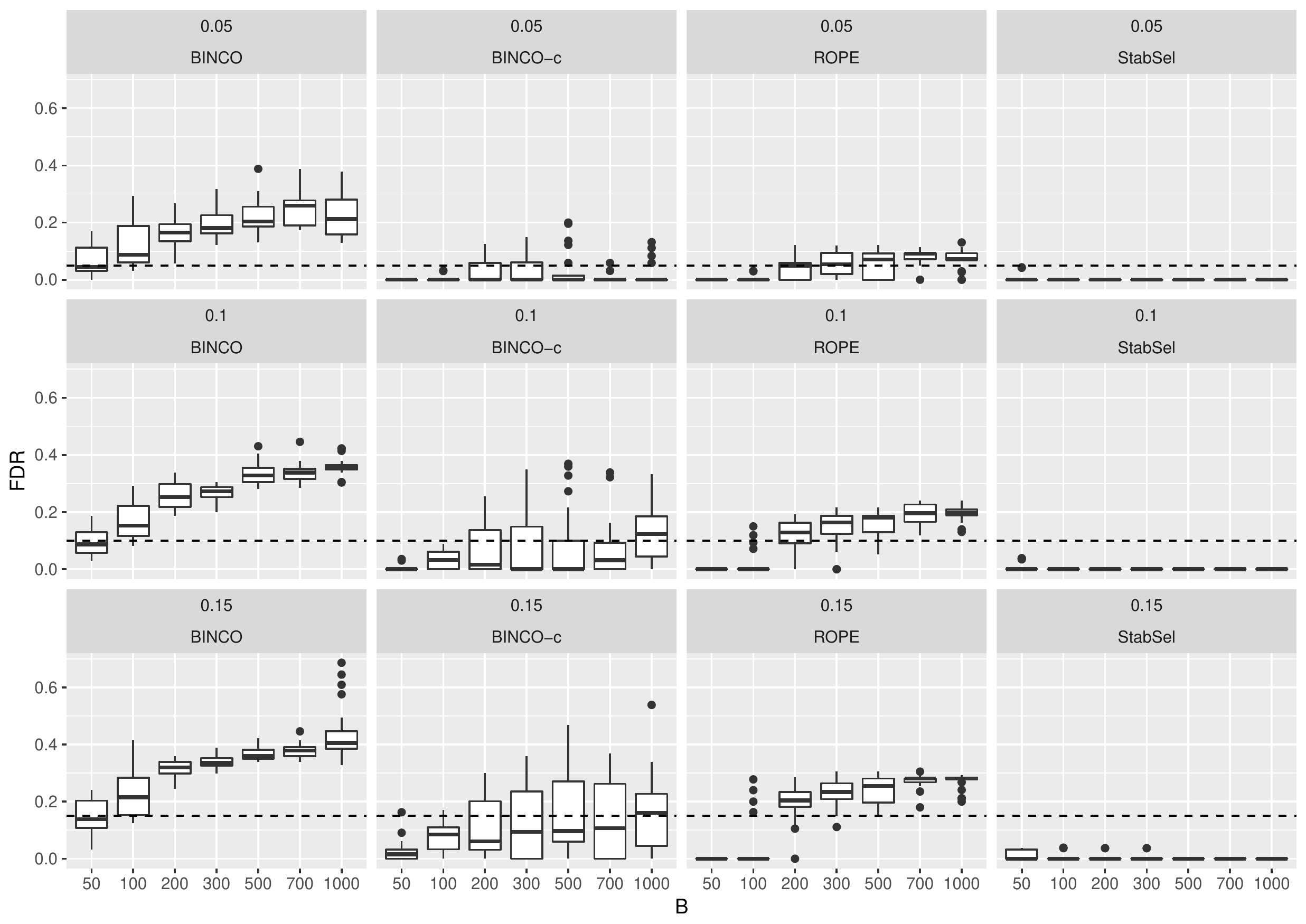}}
\caption{Network topology: sparse, steps:15, $n=200$, weakness=0.8, facet titles: target FDR and method.}
\label{sim41}
\end{figure}

\begin{figure}[p]
\centerline{\includegraphics[width=0.9\linewidth]{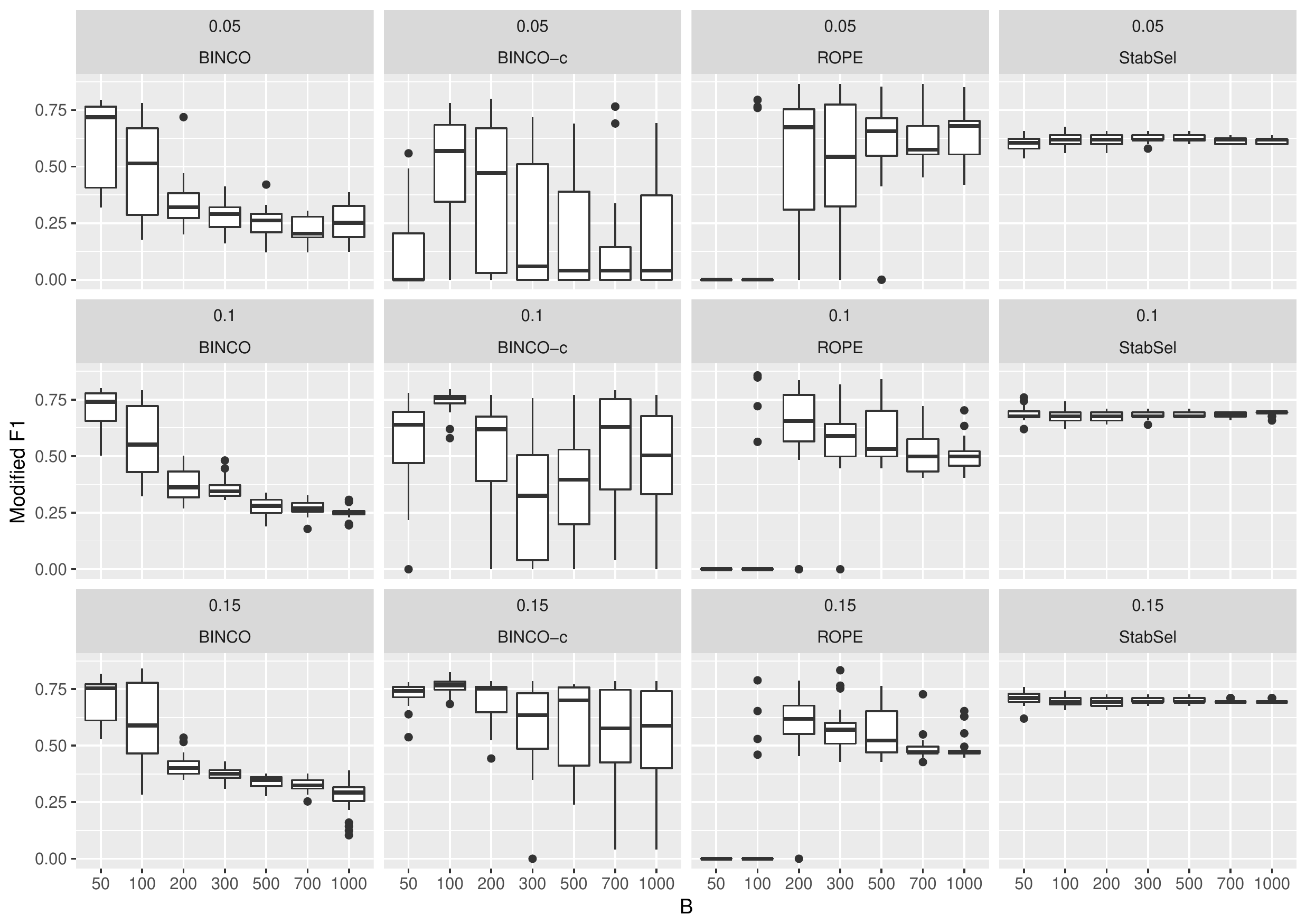}}
\caption{Network topology: sparse,  steps:15, $n=200$, weakness=0.8, facet titles: target FDR and method.}
\label{sim42}
\end{figure}

\begin{figure}[p]
\centerline{\includegraphics[width=0.9\linewidth]{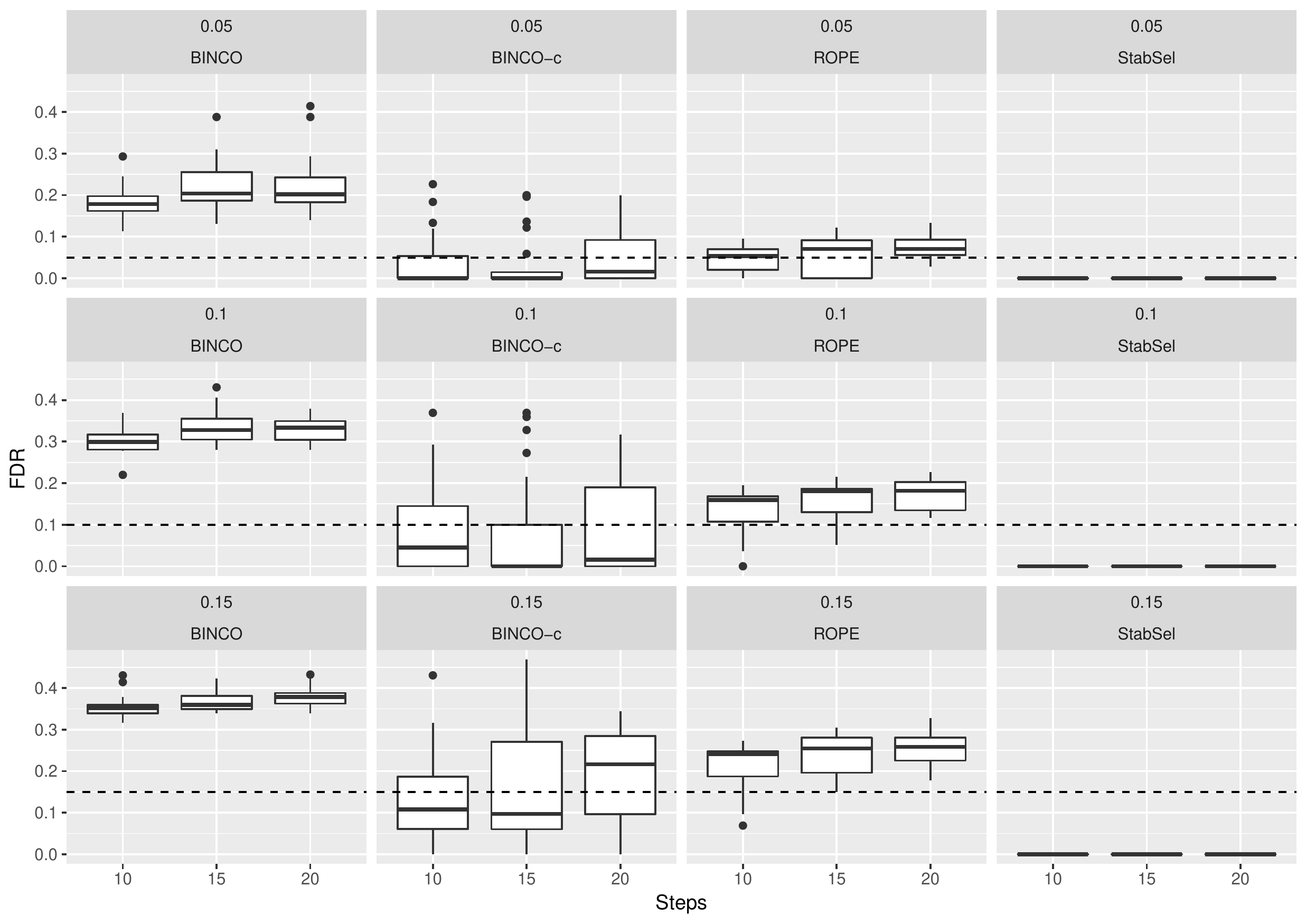}}
\caption{Network topology: sparse, $B=500$, $n=200$, weakness=0.8, facet titles: target FDR and method.}
\label{sim43}
\end{figure}

\begin{figure}[p]
\centerline{\includegraphics[width=0.9\linewidth]{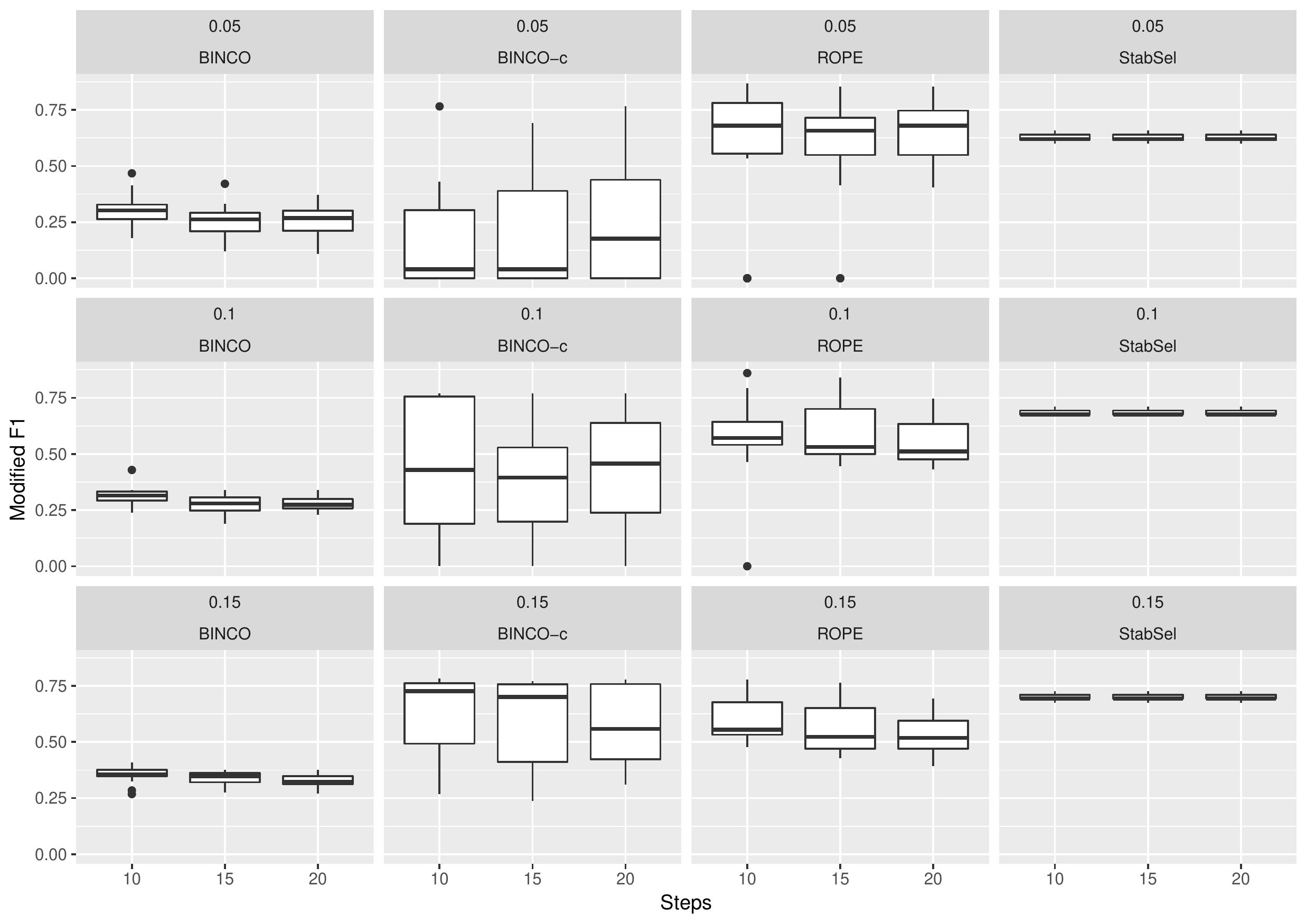}}
\caption{Network topology: sparse,  $B=500$, $n=200$, weakness=0.8, facet titles: target FDR and method.}
\label{sim44}
\end{figure}

\begin{figure}[p]
\centerline{\includegraphics[width=0.9\linewidth]{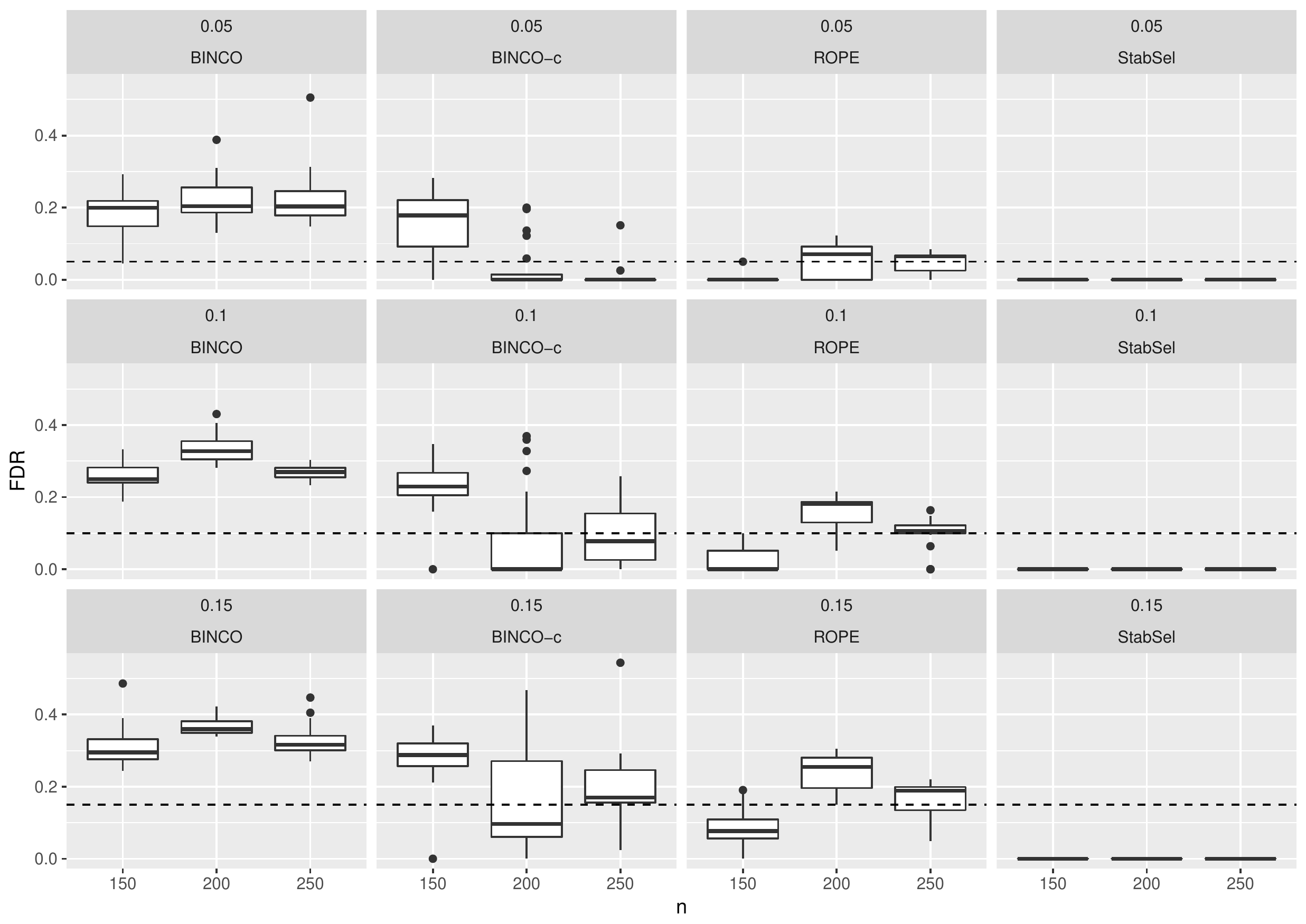}}
\caption{Network topology: sparse, $B=500$, steps:15, weakness=0.8, facet titles: target FDR and method.}
\label{sim45}
\end{figure}

\begin{figure}[p]
\centerline{\includegraphics[width=0.9\linewidth]{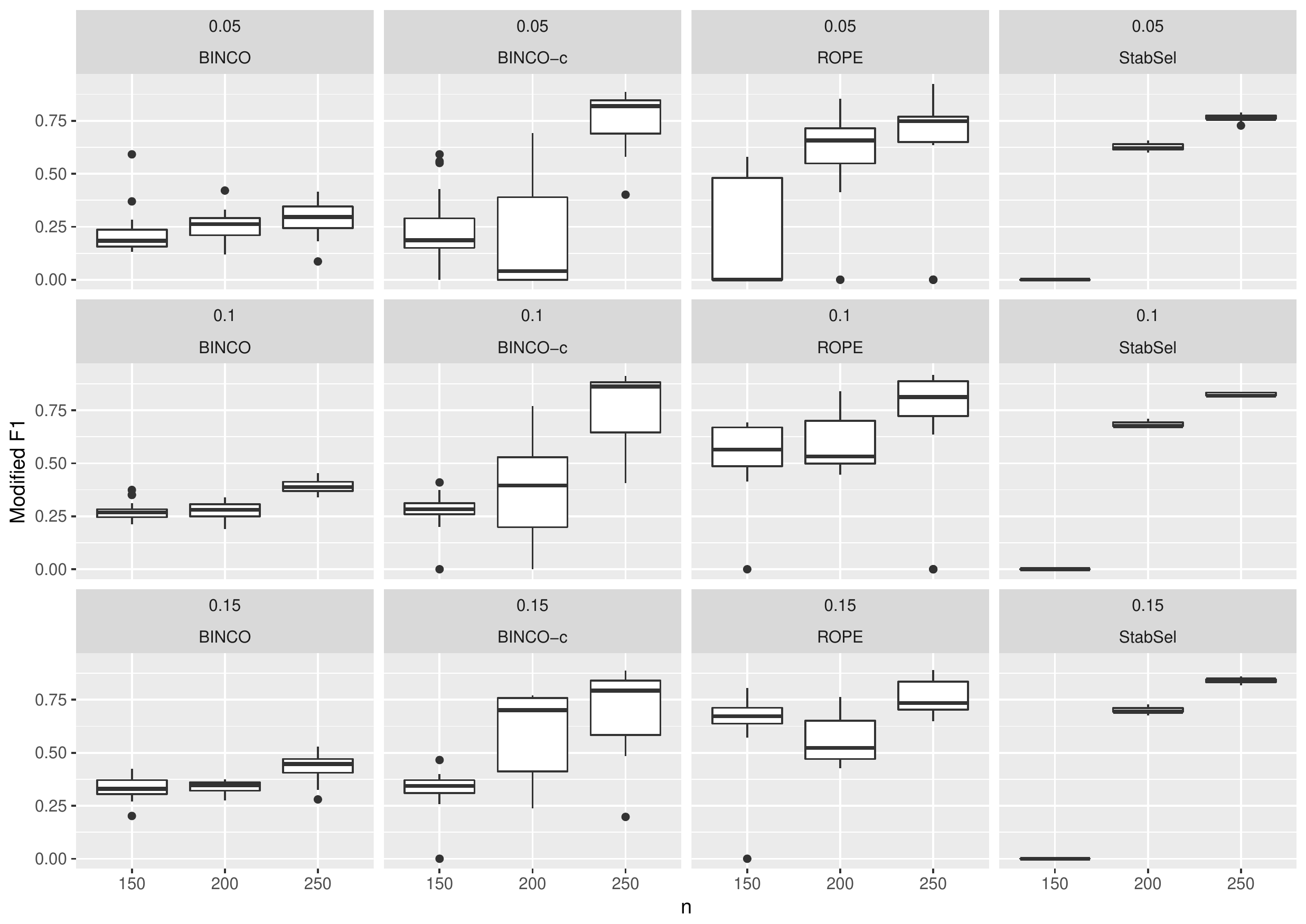}}
\caption{Network topology: sparse,  $B=500$, steps:15, weakness=0.8, facet titles: target FDR and method.}
\label{sim46}
\end{figure}

\begin{figure}[p]
\centerline{\includegraphics[width=0.9\linewidth]{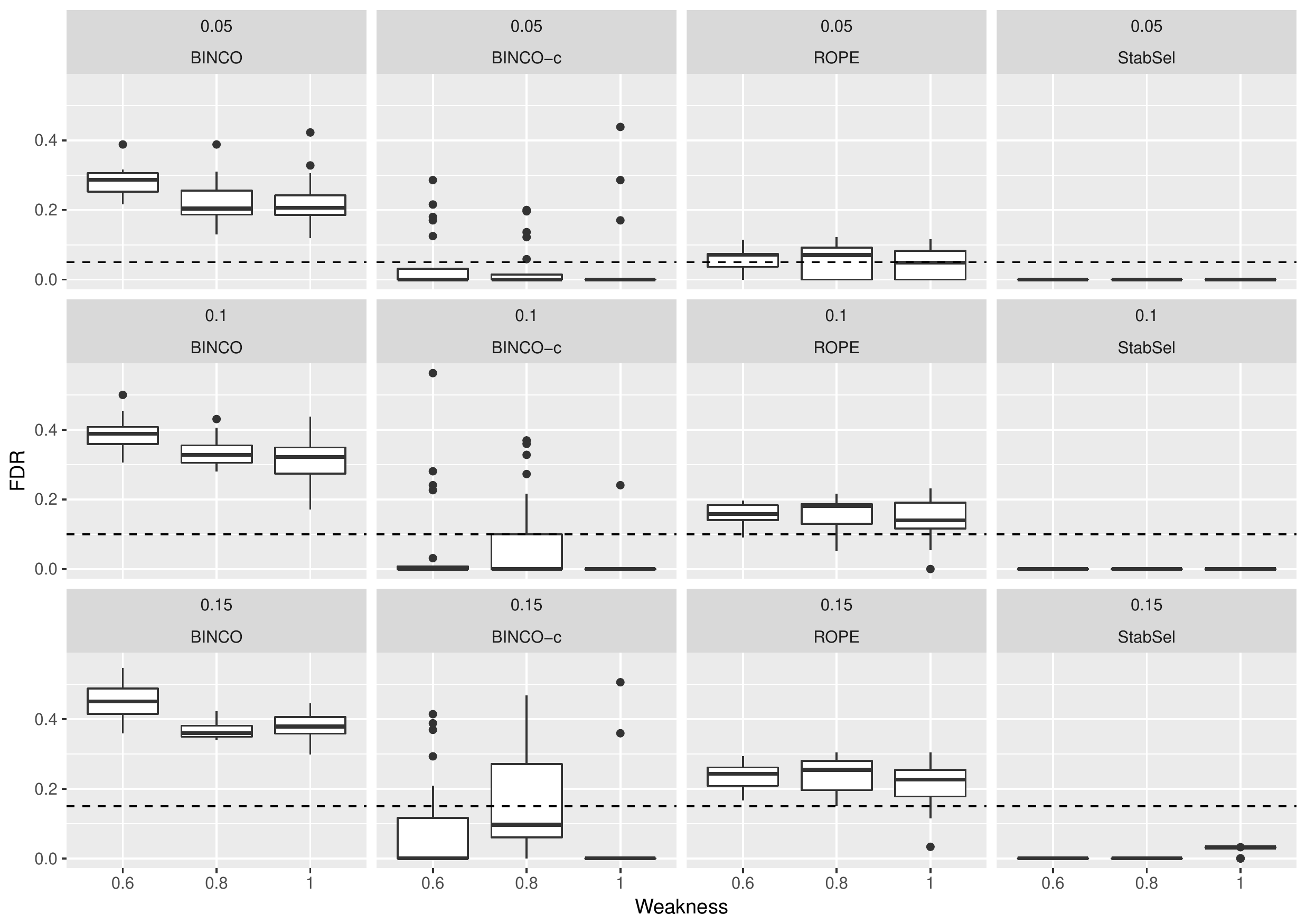}}
\caption{Network topology: sparse, $B=500$, steps:15, $n=200$, facet titles: target FDR and method.}
\label{sim47}
\end{figure}

\begin{figure}[p]
\centerline{\includegraphics[width=0.9\linewidth]{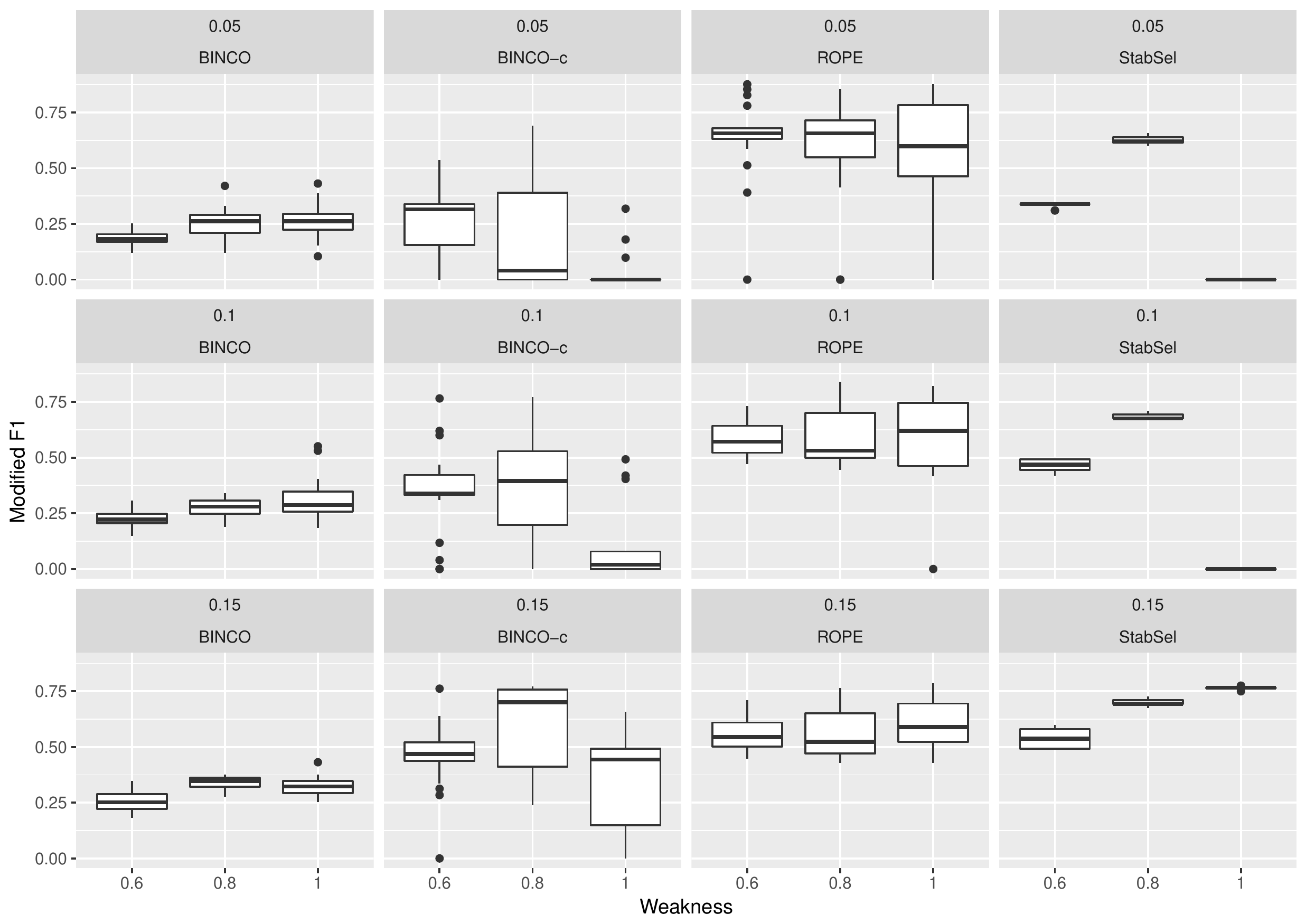}}
\caption{Network topology: sparse,  $B=500$, steps:15, $n=200$, facet titles: target FDR and method.}
\label{sim48}
\end{figure}

\end{document}